\documentclass[11pt]{article}
\renewcommand{\textwidth}{6.35 in}

\usepackage{pdflscape}

\usepackage{caption}
\usepackage{graphicx}
\usepackage{amsmath}
\usepackage{mathrsfs}
\usepackage{mathtools}
\usepackage{float}
\usepackage[pdftex]{color}

\usepackage{rotating}  
\usepackage{lipsum}    

\usepackage{textcomp}

\makeatletter
\def\normaljustify{%
  \let\\\@centercr\rightskip\z@skip \leftskip\z@skip%
  \parfillskip=0pt plus 1fil}
\makeatother

\usepackage[T1]{fontenc}
\usepackage[utf8]{inputenc}
\usepackage{tabularx,ragged2e,booktabs,caption}
\newcolumntype{C}[1]{>{\Centering}m{#1}}
\renewcommand\tabularxcolumn[1]{C{#1}}

\restylefloat{table}
\restylefloat{figure}

\begin{document}

\pagestyle{empty}

\long\def\symbolfootnote[#1]#2{\begingroup%
\def\thefootnote{\fnsymbol{footnote}}\footnote[#1]{#2}\endgroup}

\hfuzz 50pt

\rightline{Current draft:  January 17, 2024}
\rightline{First draft: February 1, 2012}

\begin{center}

\vskip 1.4cm

{\bf An Empirical Assessment of Characteristics and Optimal Portfolios}

\vskip .25in

Christopher G. Lamoureux\symbolfootnote[1]{Department of Finance, The University
of Arizona, Eller College of Management, Tucson, 85721,
520--621--7488, lamoureu@arizona.edu.}

{\em and}

Huacheng Zhang\symbolfootnote[2]{Department of Accounting and Finance,
University of Edinburgh Business School, UK,
zhanghuacheng1@gmail.com.

\noindent
We are grateful to the editor, Clemens Sialm and two anonymous referees for helping us to improve our paper.  We retain responsibility for all errors.}
\end{center}

\small

\vskip .25in


\baselineskip 15pt

\parindent 18pt

\vskip .7in

\normalsize

\begin{center}
{\bf Abstract}
\end{center}

\noindent
We implement a dynamically regularized, bootstrapped two-stage out-of-sample parametric portfolio policy to evaluate characteristics' efficacy in the conditional stock
return generating process in the metric of expected power utility.
Traditional characteristics, such as momentum and size afforded large utility gains before 1999.  These opportunities have since vanished.
Overfitting--imprecision in weight estimation--is correlated with the optimal portfolio's variance.
Therefore, it is not a problem for power utility investors with coefficients of relative aversion greater than four.  For more risk-tolerant investors,
we successfully reduce estimation error by increasing the curvature of the loss function relative to the investor's utility function.

\vskip 1.25in


\newpage

\setcounter{footnote}{0}

\baselineskip 19pt

\parindent 18pt

\pagestyle{plain}

\setcounter{page}{1}

\leftline{{\bf 1. Introduction}}

Much of the empirical research in asset pricing over the past forty years examines the predictive content of measurable stock characteristics.
We are interested in the question of whether
risk-averse investors who
care about all moments of the return distribution can optimally exploit this predictability.  Furthermore, if the original findings of predictability are
robust and economically significant for such an investor, have they vanished in recent years, as investors have learned of the predictability and their capabilities
of exploiting that predictability have increased?

We answer these questions in the metric of power utility functions with a bootstrapped out of sample approach.  We use Brandt, Santa-Clara, and Valkanov's
(2009) parametric portfolio policy (PPP) to build characteristic-based portfolios that maximize in-sample power utility.
We evaluate these portfolios
out of sample.  We confront estimation risk and model selection with a two-stage dynamically regularized out-of-sample design.  
We use the first out-of-sample stage to construct density
functions of all of the optimal portfolios' certainty equivalent returns.  We use a minmax procedure to select the optimal portfolio policy.  We evaluate this
optimal portfolio policy in a second out-of-sample stage.
Since both the in- and out-of-sample periods are bootstrapped we construct (small sample) empirical distributions and report confidence intervals for functions of 
interest, such
as portfolio alpha, Sharpe ratio, and certainty equivalent return.\footnote{Lewellen, Nagel, and Shanken (2010) stress the importance of presenting confidence intervals
since sample statistics in asset pricing are often biased and skewed.}
Evaluating characteristics' predictive efficacy with this loss function, in the expected utility
metric, where we consider all moments of
portfolios' return distributions, addresses concerns about the statistical robustness and economic relevance of the return predictability.
Furthermore, expected return predictability may persist in equilibrium if it is subject to large outliers and/or
negative skewness.
Barroso and Santa-Clara (2015a)  and Kadan and Liu (2014) show that
characteristic-based portfolios that generate a high alpha and Sharpe ratio may come at the cost of negative skewness.
Nagel (2021, p. 33) notes, ``whether methods that deliver the most accurate
return forecasts at the individual stock level also automatically give us the best performing portfolio once we aggregate across stocks is an
open question that does not have an obvious answer.''

We find that over the period 1955-1998, all six of the characteristics that we consider: size, the book-to-market ratio, momentum, average same-month
return, residual volatility, and beta have economically meaningful predictive content for the purpose of forming optimal portfolios from a power utility investor's perspective.
Our primary performance metric is the portfolio's certainty equivalent return to a power utility investor with
coefficient of relative risk aversion, $\gamma = 2$.  In the out-of-sample period, 1990 - 1998,
this investor's regularized dynamically optimized optimal portfolio's certainty equivalent return has a 95\% confidence interval
of $(329 \, , \, 529)$ basis points per month compared to the market portfolio's $(121 \, , \, 137).$

We analyze overfitting in Brandt, Santa-Clara, and Valkanov's (2009) PPP which
has been used successfully in a variety of applications.\footnote{DeMiguel, Plyakha, Uppal, and Vilkov (2013) use PPP to examine the predictive content of
option-implied moments in a mean-variance setting.  Faias and Santa-Clara (2017) analyze optimal option portfolios.  Kroencke, Schindler, and Schrimpf (2014) and
Barroso and Santa-Clara (2015b) consider foreign exchange portfolio strategies, including the carry trade.
Barroso, Reicheneker, Reicheneker, and Rouxelin (2023) optimize jointly over global equity and currency exposure.}
PPP is parsimonious and avoids the first step in traditional portfolio selection -- estimating, or even taking a stand, on the
conditional distribution of returns--given measurable characteristics.  A\"it-Sahalia and Brandt (2001, p. 1299) characterize
this first step as the ``Achilles' heel of conditional portfolio choice because although the moments are predictable, this
predictability is for some moments quite tenuous.''   They argue that not specifying
a likelihood (i.e., the conditional
return distribution) avoids ``introducing additional noise and potential misspecifications through the intermediate, but unnecessary,
estimation of the return distribution.''

Best and Grauer (1991) stress that overfitting causes the documented poor out-of-sample performance of optimal mean-variance portfolios.  Optimization
amplifies estimation errors.  The literature suggests three approaches to mitigate overfitting in portfolio optimization settings.  These regularization procedures include: 
constraining the weights or estimators, Bayesian priors, and machine learning.  We use machine learning to manage estimation risk.
Jagannathan and Ma (2003) demonstrate that there is a duality between constraining the weights, for example with short-selling constraints, and
shrinking moment estimators in the mean-variance optimization context.
Bayesian approaches establish a prior using economic theory.
P\'{a}stor (2000) and P\'{a}stor and Stambaugh (2000) use asset pricing models to form the prior.  MacKinlay and P\'{a}stor (2000) impose moment restrictions
according to a factor model.
Kan and Zhou (2007) derive the expected loss function from using sample
(rather than true) moments when returns
are normally distributed.  They show that estimation risk can be diversified by holding a minimum variance portfolio
in addition to the estimated tangency portfolio.  These solutions which effectively reduce portfolio variance relative to the population solution
suggest an approach to overfitting more generally by increasing the shadow cost of the return variance (and kurtosis) in terms of mean return (and skewness).
DeMiguel, Garlappi, and Uppal (2009) show that in general these attempts to mitigate estimation error in (mean-variance) 
portfolio selection are dominated by an equally-weighted benchmark (the $\frac{1}{N}$ rule).  Barroso, Reicheneker, Reicheneker, and Rouxelin (2023) consider benchmark
constraints which limit the deviation of weights from standard equally-weighted and value-weighted benchmark portfolios.  Since such constraints tend to reduce portfolio
variance they mitigate estimation risk.

Machine learning approaches such as Lasso introduce a hyperparameter, or tuning parameter, to manage estimation risk.
For example, DeMiguel, Mart\'{i}n-Utrera, Nogales, and Uppal (2020) use a PPP
algorithm to maximize the Sharpe ratio (i.e., they specify a quadratic utility function), with transactions costs.  They impose an L1-norm
penalty on the parameter space, and demonstrate that it does better out of sample than a non-regularized optimization.  Freyberger, Neuhirl, and Weber (2020) use a group Lasso
procedure to shrink the model and manage overfitting.
Ao, Li, and Zheng (2019) develop a Lasso-type estimator to deal with a large cross-section specifically designed to address the out-of-sample deterioration of the Sharpe ratio.
However, Kozak, Nagel, and Santosh (2020, p.274) note that
such a penalty has poor statistical properties when the characteristics are correlated, and it lacks economic motivation.
We regularize the PPP by separating the curvature
of the loss function that links portfolio weights directly to characteristics from the investor's utility function.
Under a power utility function, the coefficient of relative risk aversion, $\gamma$ is effectively the shadow cost of variance relative to expected returns.
We expand the parameter space to allow a power
utility investor with coefficient of relative risk aversion $\gamma$ to increase this shadow cost in sample by maximizing expected utility with coefficient of relative
aversion $\gamma^* = \gamma + \lambda$.
The hyperparameter $\lambda > 0$ 
will reduce estimation risk, if present, to the extent that noise is positively linked to the variance of the conditional return generating process.

We find that for mid-levels of relative risk aversion PPP does not suffer from estimation risk, lending credence to the
claim that estimating moments of the conditional return distribution is the source of much overfitting (A\"it-Sahalia and Brandt 2001).
However, estimation risk is a serious problem for PPP to our power utility investor with a coefficient of relative risk aversion of two.
Since the PPP is agnostic with respect to the conditional return generating process, we cannot appeal to Bayes' Theorem to manage estimation risk.
Instead, we rely on the multiprior decision theory of Gilboa and Schmeidler (1989).
Gilboa and Schmeidler (1989 p. 142) consider uncertainty--as distinct from risk-- where ``there is too little information to form a prior.''
They show that uncertainty aversion means that the agent should optimize over all feasible states and choose that rule
which produces the best outcome under the worst possible state of nature.

Because there is no likelihood we use the bootstrap to construct the sampling distribution of out-of-sample portfolio properties for each model configuration.
A configuration consists of the curvature of the loss function used to estimate an optimal portfolio rule in-sample ($\lambda$) and the (sub)set of
measurable characteristics.  With 6 characteristics there are $63 = \sum\limits_{j=1}^6 \frac{6!}{(6-j)! \cdot j!}$ unique combinations.  We consider 14 values of $\lambda$.
So we evaluate all $63 \times 14 = 882$ alternative configurations at the beginning of each year in our out-of-sample periods. 
After (minimally) 15 years
of out of sample data we evaluate the utility function of these out-of-sample returns.
We now have a bootstrapped sampling distribution of the utility function
of the out-of-sample returns from each configuration.
Confronted with a finite sample, the investor seeks to maximize expected utility in the worst-case scenario (i.e., maxmin).
We select that configuration with the highest 1\%ile value of the loss function (certainty equivalent return) at the beginning of each
year in the second-stage out-of-sample period to construct the bootstrap distribution of the returns on the dynamically optimal portfolio policy.\footnote{We
use the bootstrap 1\%ile as the
``worst-case scenario'' to accommodate numerical issues and link to statistics.  Our results do not change in any qualitative way if we use the 2.5\%ile instead of
the 1\%ile certainty equivalent to select the optimal out-of-sample configuration.}  This is linked to statistical assessment of the portfolio.
We consider that Portfolio A dominates (i.e., is statistically significantly strictly preferred to) Portfolio B if:
a) A's 2.5\%ile certainty equivalent return is greater than B's 97.5\%ile certainty equivalent return; and/or b) A's 2.5\%ile certainty equivalent return is positive but
B's 2.5\%ile certainty equivalent return is negative.

Our use of the maxmin criterion on the bootstrapped out-of-sample certainty equivalent returns relates to the literature in decision making under uncertainty with
machine learning.  We learn from specifications' out-of-sample performance, as in Barroso and Saxena (2022) and Freyberger, Neuhierl, and Weber (2020).
Gilboa, Postlewaite, and
Schmeidler (2008) provide an overview and survey of the problem of decision making under uncertainty, and the
``multiple prior'' approach.  A\"{i}t-Sahalia and Brandt (2001) suggest the use of maxmin for a CRRA investor in the
case where the (conditional) return distribution is unknown.
This approach has been extended broadly to dynamic
optimization by Hansen and Sargent (2008).  There is related work in operations research and machine learning on {\em robust optimization}, where ``it is assumed that the
decision maker has no distributional knowledge about the underlying uncertainty except for its support, and the model minimizes the worst-case cost over an uncertainty
set,'' (Rahimian and Mehrotra 2019, p. 1).  Bertsimas, Gupta, and Kallus (2018) characterize this approach as ``data-driven robust optimization.''

Nagel (2021, p. 48) notes that, ``shrinkage can improve portfolio performance if there is heterogeneity in the covariates' relative
contribution to moments and estimation error.  Shrinkage must reduce undesirable contributions (estimation error, variance, and kurtosis) more than
desirable ones (return mean and skewness).''  We regularize or shrink estimation by disentangling the loss function maximized on the data to obtain portfolio weights from the investor's
utility function.
Our two findings, that $\lambda > 0$ greatly reduces estimation risk for
our primary (relatively risk-tolerant) investor and $\lambda = 0$ for more risk-averse investors,
suggest that estimation
risk, the tendency to find spurious patterns in a sample, is related to the variance of the portfolio return distribution.  This is fully consistent with all of the literature
that demonstrates that constraining portfolio leverage (hence variance) serves to mitigate estimation risk. 

We find that characteristics' predictive usefulness for portfolio selection is temporally unstable, and has vanished post-1998.  A jump in $\lambda$ under the rolling protocol
prior to 2001 accompanies this structural break.  Our minmax regularization suggests more conservative portfolios following out-of-sample results that are disappointing compared to
prior expectations.  This result is consistent with recent research.  For example, Martin and Nagel (2022) provide a learning framework in which an
econometrician can detect such {\em cross-sectional} predictability using standard tests, but the predictability is no longer present out of sample.
They motivate the use of an out-of-sample test design--noting that we should expect to find evidence of predictability {\em in sample} in a high dimensional
highly complex environment.  In this setting, the usual implications of informationally efficient markets place testable restrictions on {\em out-of-sample} predictability.
Green, Hand, and Zhang (2017) document a significant drop in characteristics' predictive content in 2003, which they attribute to institutional changes that reduce trading 
frictions.\footnote{These changes stem from both regulations: Regulation FD (2000) and Sarbanes-Oxley (2002); and technological advances: decimalization (2001) 
and enhanced autoquote (2003).}  McLean and Pontiff (2016) also document a drop in the return to trading on anomalies documented in the literature subsequent to publication.
Our result, that measurable
characteristics did have economically and statistically predictive content for portfolio construction prior to 1999, but no longer do, complements this research.

We consider power utility investors with higher aversions to risk, $\gamma$ values of 5 and 8.  The corresponding
tables and figures are provided in an Internet Appendix.  
We confirm the findings that CRRA investors could use PPP to exploit the predictive content in characteristics 
prior to 1999, but not since then.  As noted, while overfitting is a severe problem for the more risk-tolerant ($\gamma = 2$) investor
in this first subperiod it is not for the more risk-averse investors.  
In the first subperiod optimal portfolio variance declines statistically significantly in risk aversion.
These results are consistent with the hypothesis that estimation risk shrinks in portfolio variance.
However, optimal portfolio Sharpe ratio, skewness, and kurtosis are flat in risk aversion.

We complement studies by Lewellen (2015), Green, Hand, and Zhang (2017), and Freyberger, Neuhirl, and Weber (2020)
by showing {\em how} the set of characteristics affects optimal portfolios' factor exposures in this first subperiod.  Our most risk-tolerant investor
uses the characteristics to short the market, and get positive exposures to the Fama-French value, size, and momentum factors.  
Roughly half of the portfolio's excess mean
return and return variance come from outside the span of the six Fama-French factors.
The portfolio
has a small positive exposure to their RMW factor, and a statistically significant  negative loading on their CMA factor.  
While the size of characteristic weight tilts diminish in risk aversion,
the percentages of portfolio mean returns and return variance within the span of the Fama-French six factor model are stable in $\gamma$.  

\newpage

\leftline{\bf 2. Portfolio Selection}

\vskip .2 in

\leftline{\em 2.1 Algorithm}

In Brandt, Santa-Clara, and Valkanov's (2009) algorithm, the vector $\theta$ is estimated to maximize a concave loss function over $M$ periods: 
\begin{eqnarray}
\max_\theta \sum\limits_{m=0}^{M-1}\frac{(1 + r_{p,m+1})^{1-\gamma^*}}{1 - \gamma^*} \left( \frac{1}{M} \right)
\end{eqnarray}
by allowing portfolio weights to depend on observable stock characteristics:
\begin{eqnarray}
r_{p,m+1} = \sum\limits_{i=1}^{N_m} \left(\overline{\omega}_{i,m} + \frac{1}{N_m} \theta^{'} x_{i,m} \right) \cdot r_{i,m+1} \, ,
\end{eqnarray}
where: $x_{i,m}$ is the $K$-vector of cross-sectionally standardized characteristics on firm $i$, measurable at month $m$; $\overline{\omega}_{i,m}$ is
the weight of stock $i$ in the (value-weighted) market portfolio at month $m$; and $N_m$ is the number of stocks in
the sample in month $m$.  Conditioning only on information that is available to investors at the time the portfolios
are formed avoids the overconditioning bias analyzed by Boguth, Carlson, Fisher, and Simutin (2011).
Unlike Brandt, Santa-Clara, and Valkanov or  DeMiguel, Mart\'{i}n-Utrera, Nogales, and Uppal (2020), we
do not identify $\gamma^*$, the parameter used to generate a feasible portfolio strategy in (1) with the relevant statistical loss function
(i.e., a specific investor's utility function).
Instead, we consider a statistical loss function (alternatively ``an investor'') that takes the same form as (1)
indexed by $\gamma$ (the curvature of the statistical loss function which is pre-determined and fixed).
From this perspective, $\gamma^*$ is a choice variable -- a tool to manage estimation risk.  Letting $\gamma^* \equiv \gamma + \lambda$, 
$\lambda$ is a hyperparameter, or tuning parameter of shrinkage or regularization.\footnote{This algorithm is general and can be used with
many alternative specifications.  One extension that Brandt, Santa-Clara, and Valkanov (2009, Section 1.3.3) mention is allowing weights to
be nonlinear functions of characteristics.  Freyberger, Neuhirl, and Weber's (2020) finding that non-linear functions of characteristics improve
the Sharpe ratio relative to a linear specification rationale for such an extension.}

Our primary focus is on a power utility investor with coefficient of relative risk aversion, $\gamma = 2$.
We also consider the effects of increasing risk aversion on estimation risk and 
the predictive content of characteristics for portfolio formation.
Our interest is in out-of-sample statistical comparisons across portfolios generated by various feasible portfolio rules--from 
the perspective of this nonlinear statistical loss function.  An eligible portfolio rule is a configuration consisting of a subset of characteristics
and the increased penalty on variance and kurtosis, expressed in utility terms, $\lambda$.
By definition, the optimal {\em in-sample} portfolio is obtained using all characteristics and by setting $\lambda = 0$, 
(i.e., $\gamma^* = \gamma$).
Whether this is also true out of sample is an empirical question as it depends on the unmodeled relationship between estimation error, 
the characteristics and the loss function.
If there is a positive relationship between an optimal portfolio's in-sample variance and its noise
(i.e., higher {\em out-of-sample} variance) then using a more concave loss function
than the investor's utility function ($\lambda > 0$), may generate portfolios that are preferred
to those obtained by constraining $\gamma^*$ to equal $\gamma$.

\vskip .2 in

\leftline{\em 2.2 Data}

Because our model selection stage uses out-of-sample analysis of expected utility from
hundreds of configurations we require a comparatively long time series.  
Therefore we use characteristics that can be directly computed from market prices, and the book value from the CRSP-Compustat merged file.
Our sample of returns and characteristics spans the years $Y_1 = 1960$ through $Y_{62} = 2021$, which means we use complete data starting in
January 1955 to obtain all measurable characteristics as of the start of 1960.  The initial in-period estimation uses the 180 months January, 1960 - December, 1974, 
and our initial out-of-sample validation period is 1975-1989. Our fully out-of-sample period comprises the 32 years 1990 -- 2021.
We use $Y_y$ to indicate year $Y,$ for $y = 1, \, \ldots, , \, 62$, and $m = 1, \, \ldots, , \, M = 744$ to denote the month in our sample.

For a stock to be eligible for investment in month $m$, we require five years of (non-missing) returns in months
$[m-60, \; m-1]$.  If the stock return is missing in month $m$, we look to the CRSP delisting return.  If that is missing, we substitute $-30\%$ for NYSE--
and AMEX--listed stocks and $-50\%$ for Nasdaq stocks.
Thus the stocks in the January 1960 sample have no missing data from January 1955 through
December 1959.  These requirements limit the sample.  For example, Brandt, Santa-Clara, and Valkanov (2009) report that the smallest cross-section in their study comprises 
1,033 stocks in February 1964.  Only 624 firms satisfy our data requirements in that month.  
Prior research suggests that it is important to exclude penny stocks and
stocks with low relative market capitalization.  To this end we add two additional criteria for a stock to be eligible for inclusion in month $m$.  
First, to exclude nano and small microcap stocks we impose a real dollar
minimum equity market capitalization in month $m-1$ of \$110 million in December 2021 dollars.  We obtain the US Consumer Price Index from the Federal Reserve 
(FRED).\footnote{We extracted the series CPIAUCSL: consumer price index for all urban consumers: all items.}  This restriction excludes stocks with market capitalization less 
than \$11.5 million in January 1960 and \$50 million in January 1990.  Second, we exclude the smallest 10\% of qualifying stocks in the months before Nasdaq stocks 
enter our sample, which is January 1978.  We exclude 20\% of eligible firms when Nasdaq stocks enter the sample.
In February 1964, the dollar criterion discards 39 of the eligible
624 stocks, and the percent exclusion discards another 58 stocks.  This leaves a final sample of 527 stocks -- 51\% of Brandt, Santa-Clara, and Valkanov's cross-section on
that date.  

Table IA-1 in the Internet Appendix provides details on the sample.  For each of the 744 months in the full sample, the table shows:
prior-month end, the number of stocks that meet the data availability requirement for month $m$ (at month $m-1$), the number excluded by the minimum equity market capitalization constraint, and the final sample size; along
with the minimum and median market capitalizations in the sample.
There are 411 (exclusively New York Stock Exchange) stocks in the final sample in
January 1960.  The sample size jumps in August, 1967, from 678 to 881  when stocks listed on the American Stock Exchange are eligible for inclusion
in our sample.  The largest jump in sample size is in January 1978 (from 1,001 to 1,420 stocks) when Nasdaq stocks are eligible to enter
our sample.  The maximum number of stocks is 2,290 in April, 2006.  There are 1,848 stocks in our sample in August 2008
and 1,693 stocks in the last month in our sample, December 2021.

We use the following six characteristics: momentum ($\zeta$), the book-to-market ratio (V), log size (S), beta ($\beta$), market model residual standard deviation ($\sigma_\epsilon$),,
and the average same-month return over the preceding five years ($\overline{r}$).\footnote{Same-month seasonality is analyzed in Heston and Sadka (2008).} 
Momentum is the stock's compounded return over the annual period $[m-13, \; m-2]$.  Equity market capitalization is the market value
of the company's outstanding shares (aggregated across all share classes)
at time $m-2$.   Book value is obtained from the Compustat database for the most recent fiscal year-end between $m-6$ and $m-18$.
We define the book-to-market ratio to be the natural log of one plus the ratio of book value to equity market capitalization.
Size is equity market capitalization.
We estimate beta and the residual standard deviation by regressing monthly returns over the period
$[m-60, \; m-1]$ on the CRSP value-weighted index.

We normalize and standardize the characteristics so that each characteristic has a cross-sectional mean of zero and unit standard deviation.
Inspection of (2) shows
that optimal portfolio weights will thereby sum to unity for any value of $\theta$.  This also means that the characteristics
are observationally equivalent to shrinkage values.   For example, let $\beta$ be a stock's OLS beta.
Consider a prior-weighted beta, such as $\beta^S = .5 \cdot \beta + .5 \cdot 1$.
The normalized $\beta^S$ are identical to the normalized $\beta$.
This implicit shrinkage mitigates the usual concerns about outliers in
characteristic space so we do not winsorize the characteristics.
A single observation ($\Psi_{i,m}$) comprises stock $i$'s return in month $m$, $r_{i,m}$, as well as the vector of characteristics,
measurable at month $m-1$, for stock $i$, $i = 1, \ldots, N_m$.  Importantly, $\Psi_{i,m}$ also includes stock $i$'s market capitalization at $m-1$,
since the passive portfolio at $m$ (i.e., the portfolio when the $\theta$-vector is zero) is the market-weighted portfolio at $m-1$.

We are interested in the optimal {\em sets} of characteristics so we consider all possible sets using these six characteristics.  There are 63 such sets including
each characteristic as a singleton and all six as a sextuplet.

\newpage

\leftline{\em 2.3 Bootstrap sample construction}

In light of the high noise-to-signal nature of stock return data, we develop and implement a bootstrap design for optimization,
regularization and to characterize the out-of-sample sampling distributions of portfolio properties,
such as certainty equivalent, portfolio loading on factors, portfolio skewness, etc.  
We stack the $744$ row-vectors $\Psi_{i,m}$ (of varying lengths) to form the array
\boldmath{$\Psi$}.  \unboldmath
This is not a traditional panel since stock $i$ at month $m$ is different from stock $i$ at any other time, and as noted, the number of columns is different for each row.

Our resampling is nonparametric cross-sectional bootstrap (see Kapetanios 2008), motivated by the (repeated) single period
optimization problem with out-of-sample cross-validation, regularization and inference.  It is nonparametric since we draw from the raw data.
It assumes that returns are temporally independent conditional on \boldmath{$\Psi$}\unboldmath.
As such, \boldmath{$\Psi^b$},\unboldmath is the $b^{th}$  resample from \boldmath{$\Psi$}\unboldmath , for $b = 1, \, \ldots, \, B = 10,000$.
Consistency holds under $N-$asymptotics (as the number of stocks in each period increases), since we can view $\hat{\theta}$ as a GMM estimator.
Preserving the original time
series in all resamples allows us to evaluate the effects of structural instability on out-of-sample portfolios, and also maintains the
cyclical (month of the year) and serial dependence patterns inherent in the design of \boldmath{$\Psi_m$}, \unboldmath for $m = 1, \, \ldots, \, M = 744$.
A resampled draw for month $m$ resamples (with replacement) an $N_m$-row vector from \boldmath{$\Psi_{m}$}\unboldmath.
Thus each pseudosample consists of the same number of observations in each period as the original sample and we preserve
the calendar structure of the original data.
We also maintain the temporal structure of the investment opportunity set.  In the original data \boldmath{$\Psi_m$} \unboldmath includes the raw values of the
characteristics as well as the market weight of the stock at $m-1$.  

Once we draw a resampled cross-section from \boldmath{$\Psi_{m} \,$}\unboldmath we
normalize (so that all characteristics have a zero mean in this month in this bootstrapped sample), standardize (so that all characteristics have unit variances), 
and construct the value-weighted market weights for each
resampled stock, based on the total market capitalization in this month in this bootstrapped sample.  
Thus our resampling design preserves the integrity of the investment opportunity set at each $m$.  This critically includes the
relationship between stocks' characteristics and sizes.  We also maintain the integrity of the seasonality in the data.
The month of January, for example occurs exactly once in any 12-month cycle, and its timing is deterministic.  The matrix of characteristics at $m$
is not independent of the size of each stock at $m$.
We manage concerns about unmodeled temporal dependencies by maintaining the out-of-sample test design with each bootstrapped sample.

\vskip .2in

\leftline{\em 2.4 Dynamic regularization}

We implement a dynamic two-stage bootstrapped out-of-sample regularization process using both updating and rolling construction.  We provide an appendix containing 
pseudo code for our approach.
Our first stage entails construction of the bootstrap 
distribution of out-of-sample returns from each of the 882 configurations.  
A configuration is a characteristic set and a value for $\lambda$ (or $\gamma^*$).  As noted above, with six characteristics, and 14 values of $\lambda = \{0, \; 1, \; \ldots, \; 11, \;  14, \;  20\}$,
we consider 882 configurations under each protocol.  In each of the $b = 1, \, \ldots, \, 10,000$ bootstrap samples \boldmath{$\Psi^b \, $}\unboldmath we estimate the parameter vector
$\hat{\theta}$ {\em in sample} by maximizing (1) over the $K-$vector $\theta$ for the in-sample period using a modified Newton method.\footnote{We use the model/trust region
algorithm of Gay (1983), and the analytical gradient and Hessian from (1).  Our 2-stage algorithm is numerically intensive. We maximize in-sample utility 414,540,000
times.}  The first in-sample period (under both protocols) is 1960 - 1974.  We use the
$\hat{\theta}$ estimated from this period to form the out-of-sample portfolio in each of the next 12 months (in 1975).  Under the updating protocol, then the 12 months
of 1975
are added to the sample, and $\hat{\theta}$ is estimated on this 192-month in-sample period.  This  $\hat{\theta}$ is used to construct 
out-of-sample portfolio returns in the 12 months of 1976.  Under the rolling protocol we drop
the first 12 months from the in-sample data so that the second year's $\theta$ is estimated using the 180-month in-sample period 1961 - 1975.  
At this point we have 10,000 bootstrap draws of 24 months of out-of-sample returns from each of the 882 configurations.

Our dynamic regularization selects the optimal configuration by looking back at the performance of prior {\em out-of-sample returns}.
We use at least 15 years of optimal portfolio out-of-sample returns for our second stage regularization, so the first year of second stage 
(i.e., dynamically regularized) out-of-sample returns is 1990.
At the end of 1989 we have 180 months of 10,000 draws of out-of-sample optimal portfolio returns from each of the 882 configurations.
We select that configuration whose bootstrap distribution of out-of-sample returns over the period 1975 - 1989 has the highest 1\%ile certainty equivalent
return as ex-post optimal, and use this to construct the dynamically optimized (and regularized) out-of-sample returns for the next 12 months (1990).
From this point until the end of the sample each additional year involves re-estimating the parameter vector using the additional year's worth of data 
(as in Stage 1 optimization).  Further, we (dynamically) regularize using the additional year of out-of-sample returns.
We proceed in this manner through the end of 2020 to 
construct the dynamically optimized (and regularized) out-of-sample returns for 2021.
The optimal configurations at the end of each year from both protocols are reported in Table 1, (and in Tables IA-2 and IA-3 for the dynamically optimized portfolios for the increasingly
more risk-averse power utility investors).
At any month in the second out-of-sample stage, the optimal configuration and its corresponding $\hat{\theta}$ vector are determined 
using information at the end of the previous year.
The stocks' characteristics (and market weights) are available prior to the start of the month.

As in Brandt, Santa-Clara, and Valkanov (2009), we estimate the $\hat{\theta}$ coefficients at the beginning of each year in the out-of-sample period.  
There are several reasons for this.  Since the updating protocol relies on the temporal structural stability of the return generating process, adding 12 months
of data increases the reliability and efficiency of the $\theta$ estimates.  By contrast, since the rolling protocol uses only the most recent 15 years of data,
if the conditional return generating process experiences structural breaks, that will be evident in material changes in $\hat{\theta}$ over time.
We similarly choose the optimal configuration using the minmax criterion at the beginning of each year in the second-stage out-of-sample period.  The rolling protocol
accommodates changes in the joint return, estimation error process.  This dynamic regularization is consonant with the annual updating of $\hat{\theta}$, and is consistent
with an investor updating her information set as she moves through time.  In this design that new information is used in two ways.  First to update the $\hat{\theta}$ 
vector, and second to select the optimal configuration.  As the Appendix shows, the sampling protocols differ in both the 
in-sample period used to obtain $\hat{\theta}$ and the period used in second stage model selection.

\vskip .4in

\leftline{\bf 3. Results}

Table 1 shows the ex-post optimal portfolio policy for the power utility investor with $\gamma = 2$ at the beginning of each year in the second-stage out-of-sample
period under both protocols.  This is the optimal configuration from the first-out-of-sample period that ends before the indicated year: the characteristic set (of the 63 possibilities)
and $\gamma^* = \gamma + \lambda$ (14 possibilities).  
The table provides sampling statistics (1\%ile value, mean, and standard deviation) of this configuration's certainty equivalent from this period.
Table 1 provides the same information about the sampling distribution of the two benchmarks' certainty equivalents  at the beginning of the second-stage
out-of-sample period.  The benchmarks are
the value-weighted portfolio of all eligible securities at the beginning of each month (VWI) and the equally-weighted portfolios of all eligible securities at the beginning of each month (EWI).  
The table also reports the certainty equivalent
sampling distribution for the configuration with the optimal characteristic set and $\lambda = 0$ at the end of the first out-of-sample period under both
protocols.  Figure 1 reports the sampling distributions (box and whiskers plots) of the $\theta$ coefficient on each of the six characteristics for each year in
the second-stage out-of-sample period from the rolling protocol.
The whiskers show the 95\% confidence interval on the estimate each year, and the box the interquartile range.  The median is the
bar inside the box.

Table 1 shows that $\lambda > 0$ is a necessary regularization for in-sample optimization to produce attractive out-of-sample returns for this investor under both
protocols.  This suggests that estimation risk is positively related to the variance of the conditional return distribution as $\lambda > 0$ penalizes variance (and kurtosis)
more than the investor's utility function.  This form of regularization has economic rationale--unlike penalizing the $\theta$ coefficients.
It is also more statistically appealing since the
characteristics' effects on portfolio returns are not independent, as noted by Kozak, Nagel, and Santosh (2020).\footnote{For example the $\theta$ coefficient on
a characteristic may be centered close to 0 but takes on large positive values when the coefficients on other characteristics are small and
takes on sizeable negative values when the coefficients on those other characteristics are large.  This is especially important in light of the large sampling variation
on all characteristics' $\theta$ estimates.}

Table 1 shows that this power utility investor's mean (95\% confidence interval) out-of-sample certainty equivalent return without regularization (i.e., with $\lambda = 0$), 
over the 180 months ending in 1989 of -41\% $(-100\% \, , \, 52\%)$ per month under the updating protocol and -4\% $(-100\% \, , \, 31\%)$ per month
under the rolling protocol.
The power utility function is  indeterminate at returns less than or equal to -100\%.
Therefore we define the certainty equivalent return for a pseudosample to be -100\% (-10,000 basis points) per month if the return in any month in the relevant period does not exceed -1.
The mean (95\% confidence interval) of the $\theta$ coefficients in this case are: momentum, 5.8 $(-30.5 \, , \, 10.8)$; book-to-market ratio, 5.3  $(-2.7 \, , \, 9.5)$;
log size, -16.5  $(-37.5 \, , \, -10.9)$; residual volatility, -7.1  $(-13.8 \, , \, 41.1)$; and average same-month return, 10.5  $(-30.5 \, , \, 10.8)$.
The enormous sampling variation in the weight coefficients shows the reason for the poor performance of this unregularized estimation.  The asymmetry in the sampling distributions
results from the correlatedness amongst the characteristics and cross-effects.
This sampling variation in itself is enough to warn off investors from this configuration. 

This type of estimation risk arises because of large absolute sampling correlations between the $\theta$ coefficients.  In this case, the sampling correlations between the $\theta$ coefficient
on residual volatility and the $\theta$ coefficients on momentum, average same-month return, and log size are -93\%, -88\%, and -86\%, respectively.  The sampling correlations between
the $\theta$ coefficient on momentum and the $\theta$ coefficients on average same-month return and log size are 88\% and 86\%, respectively.
As such, this type of overfitting
cannot be solved by restricting some of the $\theta$ coefficients to zero (e.g., using an L1-norm penalty on $\theta$).  
None of the 63 characteristic sets generates a portfolio which dominates the benchmark with $\lambda = 0$.  By increasing the implicit penalty on the conditional
return variance we are able to take advantage of the cross-effects amongst the characteristics.

Since the utility (statistical loss) function depends on higher moments, we report sampling distributions of measures of skewness and kurtosis.
Our robust measures of skewness ($S_4$) and kurtosis ($K_3$) are recommended by Kim and White (2003):
\begin{eqnarray}
S_4 = \frac{\mu - r_{.5}}{\sigma}
\end{eqnarray}
and
\begin{eqnarray}
K_3 = \left\{\frac{\overline{r}^{+}_{.95} - \overline{r}^{-}_{.05}}{\overline{r}^{+}_{.5} - \overline{r}^{-}_{.5}} - 2.63\right\} \cdot 100
\end{eqnarray}
where:
$\mu$ is the mean portfolio return over the out-of-sample period, $\sigma$ is the return standard deviation,
$\overline{r}^{+}_{.95}$ is the mean of the highest 5\% of returns, $\overline{r}^{-}_{.05}$ is the mean of the smallest 5\% of returns,
$\overline{r}^{+}_{.5}$ is the mean of the top half of returns, $\overline{r}^{-}_{.5}$ is the mean of the bottom half of returns, and
$r_{.5}$ is the median out-of-sample return.\footnote{As per Kim and White (2003), in our pseudosamples $S_4$, the Pearson skewness coefficient is
similar to, and more reliable than $S_3$, the Bowley skewness coefficient-integrated
over the tail size.  Similarly $K_3$, the Hogg coefficient, is more reliable than $K_4$, the Crow and Siddiqui parameter.}

\vskip .2in

\leftline{\em 3.1 Out-of-sample regularization: Ex-post optimal configurations}

The optimal configurations are very similar between the two protocols through the year 2000.  Both specify $\lambda = 1$, and for the most part use all six 
characteristics.
This similarity is somewhat surprising in light of the large
sampling variation in the optimal portfolios' certainty equivalent returns.  Many of the 882 alternative configurations have similar certainty equivalent 
distributions.  For example, the optimal configuration under the updating portfolio for 1992 uses only the four characteristics: momentum, log size, average same-month
return, and residual volatility, along with $\lambda = 1$.  This configuration's 1\%ile out-of-sample certainty equivalent return over the preceding 204 months is
512 basis points per month.  The 1\%ile of the configuration with $\lambda = 1$ and all six characteristics (i.e., add the book-to-market ratio and beta to the optimal set) 
is 509 basis points per month.  The configuration
with $\lambda = 2$ and all six characteristics has an analogous 1\%ile of 452 basis points per month.  Both characteristic sets with $\lambda = 0$ have 1\%ile out-of-sample
certainty equivalents of -100\% per month.

Figure 1 shows the sampling distributions of the $\theta$ coefficients from the ex-post optimal portfolios under the rolling protocol preceding each year in the out-of-sample period.
The figure illustrates the apparent structural change in the conditional return generating process around the year 2000.  Prior to that point, the $\theta$ coefficients on all six
characteristics tend to be large in absolute value, and their 95\% confidence intervals are far from zero as well.

These optimal portfolios tend to tilt the portfolio most aggressively to low residual volatility stocks.  They also tilt toward high beta stocks.\footnote{The joint 
results on beta and residual volatility during this era are consistent with the findings in Liu, Stambaugh, and Yuan (2018).
It is clear that prior to 1999, there is a relationship between variance and
future expected returns.  When the characteristic set contains both beta and residual volatility beta becomes desirable, and high residual volatility stocks appear undesirable, ceteris paribus  
When the characteristic set includes beta, but not residual volatility, then the sign on the beta $\theta$ coefficient becomes negative.}
Prior to 2000, these ex-post optimal portfolios also tilt toward stocks whose average same-month return is high (in the next month) and away from those with relatively 
low average same-month return over the previous five years.  The ex-post optimal portfolios also tilt toward stocks that have relatively high returns over months $[-13, \; -2]$,
small stocks, and value stocks (those with relatively high book-to-market values).

The rolling protocol results in Table 1 show that the PPP's performance starts to deteriorate around 1999.  Around this time the variance of the ex-post optimal portfolio's
certainty equivalent return, with $\lambda = 1$, increases four-fold, and then $\lambda$ increases to 5 and 6.  Furthermore, the table shows that the optimal portfolio for the
years 2010, 2011, 2013, 2014, 2015, and 2018, is the equally-weighted index of all sample stocks.  This means that the 1\%ile value of the certainty equivalent return 
over the preceding 15 years on all 882 configurations is less than this index's 1\%ile certainty equivalent in that period, which is reported in the table.
There is no evident breakdown under the updating protocol, as its ex-post optimal configurations are remarkably stable by comparison.

\vskip .2in

\leftline{\em 3.2 Optimal regularized portfolios' out-of-sample performance}

\leftline{\em 3.2.1 Subperiod 1 (1990 - 1998)}

Table 2 shows that both regularized updating and rolling protocols generate out-of-sample certainty equivalent returns that are statistically and 
economically significantly larger than the benchmarks in the first subperiod.  
The updating protocol produces a mean (95\%ile confidence interval) certainty equivalent of 428 $(329 \, , \, 529)$ basis 
points per month compared to the value-weighted index
of the stocks in the universe of 129 $(121 \, , \, 137)$.  The optimal portfolios' Sharpe ratio means (95\%ile confidence interval)
under both protocols are  1.5 $(1.3 \, , \, 1.8)$, significantly higher than the benchmark's
0.93 $(0.86 \, , \, 1.00)$.  The mean (95\%ile confidence interval) certainty equivalent from the rolling protocol is: 497 $(291 \, , \, 688)$ basis points per month.
The optimal portfolios under both protocols have higher means and
scales than both benchmark portfolios.  Table 2 shows that both benchmarks are significantly negatively skewed: the 95\%ile confidence interval on the value-weighted benchmark's $S_4$ (SKEW)
is $(-11.6 \, , \, -0.4)$.
Both benchmarks also have significantly fatter tails than a Gaussian distribution. The 95\%ile confidence interval  on the value-weighted benchmark's $K_3$ (KURT) is $(17 \, , \, 35)$.  
The dynamic optimal portfolios
from both rolling and updating, in contrast, are symmetric and not significantly more leptokurtic than a Gaussian distribution.  
The rolling protocol generates a portfolio with higher scale and sampling variation, and lower minimum returns than the updating protocol.  

The left-hand panel of Figure 2 contrasts the optimal dynamic portfolio from the updating protocol with the value-weighted benchmark.  
The difference in scale is apparent, as is the fact that the
distribution of the optimal portfolio seems shifted to the right of the benchmark (higher mean and median returns).  

Comparing characteristic-tilted portfolios and benchmarks in the metric of certainty equivalent returns makes no assumptions about the sources of systematic risk, or the 
factor structure of returns.  
Since the dynamic PPP dominates the benchmarks we next examine the relationship between these portfolios' returns and the Fama-French 6-factor model of returns.
Table 3, Panel A provides information on {\em how} characteristics achieved higher utility in the pre-1999 era.  
This shows the projection of optimal portfolio returns minus the monthly riskless rate on the six Fama-French factors:
the excess return on the value-weighted US stock market (Mkt),
the value factor (HML), the (small) size factor (SMB), the momentum factor (MOM), the profitability factor (RMW), and the investment factor (CMA), obtained
from Professor Kenneth French's website at Dartmouth.
The loading on the market factor is significantly negative under both protocols.  That is characteristics tilt the portfolio weights so that it is short
the overall stock market factor.  The power utility investor with $\gamma = 2$ instead seeks positive exposure to the value factor (HML), small stock factor (SMB), momentum factor (MOM), 
and the profitability factor (RMW).  This portfolio has a significant negative loading on the conservative investment factor (CMA).   The mean (95\% confidence interval) portfolio alpha 
from the updating protocol is 264 $(157 \, , \, 371)$ basis points per month.  The analogous sampling statistics on alpha from the rolling protocol's optimal 
portfolio are 387 $(210 \, , \, 578)$ basis points per month.

In this subperiod the characteristics shift the
portfolio to lie outside the span of these six factors.  Panel B of Table 3 decomposes the mean and variance of the updating protocol dynamic 
PPP's portfolio returns in excess of the risk free rate within the space of the 6-factor model.  The orthogonal variance for each sample is the ratio of the residual variance from this 
regression to the portfolio variance.  Variances attributed to the factors are the squared factor beta in the sample times the factor's variance.
The variance values in the table do not sum to 100 in any bootstrap sample because the factors are not orthogonal.  The largest pairwise correlations amongst the six
factors in this first subperiod are: 77\% between HML and CMA, -63\% between the market and CMA, and -42\% between SMB and RMW.
On average (95\% confidence interval), 53\% $(45\% \, , \, 62\%)$ of the 
variance in excess returns is not spanned by the 6-factor model.  And 48\% $(33\% \, , \, 61\%)$ of the portfolio mean excess return comes in the form of alpha 
from this six-factor model.  Within the factor span, momentum accounts for an average (95\% confidence interval) of 48\% $(38\% \, , \, 60\%)$ of the optimal portfolio's 
expected excess return and 34\% $(25\% \, , \, 43\%)$ of its variance.
HML is the third largest source of this portfolio's returns.  
HML accounts for 14\% $(11\% \, , \, 19\%)$ of the optimal portfolio's expected excess return and 38\% $(23\% \, , \, 54\%)$ of its
variance.  Table 3 Panel B shows that 
on average (95\% confidence interval), 2\% $(0\% \, , \, 4\%)$ of the optimal portfolio's variance comes from exposure to the market factor.

\vskip .2in

\leftline{\em 3.2.2 Subperiod 2 (1999 - 2021)}

Table 4 reports the properties of the dynamically optimized portfolios in the second-stage out-of-sample subperiod, 1999 - 2021.  The right-hand panel in Figure 2
shows the return densities of the optimal portfolio from the updating protocol and the equally-weighted portfolio of sample stocks in this period.  
After 1998, both benchmark portfolios dominate the regularized dynamic
parametric portfolio.  The mean  (95\%ile confidence interval) certainty equivalent from the updating dynamic PPP is -2 $(-110 \, , \, 83)$ basis points per month, whereas 
these statistics are 77 $(73 \, , \, 80)$ basis points per month for the equally-weighted 
benchmark and 60 $(53 \, , \, 67)$ basis points per month for the value-weighted benchmark.  
The dynamic parametric portfolios under both protocols  have significantly higher mean returns than the benchmarks. 
The Sharpe ratio from the updating protocol is not statistically different from the benchmarks' Sharpe ratios.  However the equally weighted benchmark's Sharpe ratio is
significantly larger than that generated by the rolling protocol's dynamic PPP.

As in the first subperiod,
characteristic-tilts generate portfolios that are not negatively skewed--unlike the benchmarks.  However, in this subperiod the dynamic PPP is dominated by the 
benchmarks for several reasons.
First, characteristics are used to increase the scale of the distribution (measured by the interquartile range) by some 3.4 fold in both subperiods.
The portfolio median return is only 1.9 times higher than the benchmark in this subperiod, whereas this ratio is 3.5 in the first subperiod.  Another
feature of the characteristic-based portfolios that changes from the first to the second subperiod is heightened kurtosis--manifest in the  large negative returns.  
This is especially 
true for the rolling protocol under which at least one monthly return is less than -100\% in more than 25\% of the bootstrap samples, and less than -87\% in more than 75\% of these
samples.  Even though the power utility investor with $\gamma = 2$ is relatively risk tolerant, losing 100\% results in a certainty equivalent return of that same magnitude.  

Table 3, Panel C shows that the optimal characteristic-based portfolio from the updating protocol has a mean (95\% confidence interval) alpha of 81 $(14 \, , \, 150)$ 
basis points per month, which is significantly positive.  This highlights the possibility
for discrepancies between the performance metrics, as this portfolio's Sharpe ratio is not significantly different from the benchmark. Because it results in a negative certainty
equivalent return in more than 25\% of the bootstrap samples, we consider it dominated in the metric of expected utility.
This discrepancy is even more pronounced by the optimal dynamic parametric portfolio from
the rolling protocol.  This portfolio's  alpha is larger (although not significantly so) than that from the updating protocol.  
Yet this rule produces a portfolio whose certainty
equivalent is less than -21\% (per month) in more than half of the bootstrap samples.  Comparing the two periods, the biggest effect on the portfolio mean is the drop in the
loading on momentum, whose mean drops from 2.6 to 1.0. The mean return
on this factor dropped from 100 to 26 basis points per month.\footnote{This phenomenon likely related to the momentum crash of 2009, documented by Barroso and Santa-Clara (2015a) and
Daniel and Moskowitz (2016).}
The second largest effect is the drop in alpha--the mean alpha drops by 183 basis points per month.

These effects are consistent with the properties of the $\theta$ coefficients under the rolling protocol, shown in Figure 1.  The optimal portfolios shift toward the
market factor in this period.  The mean (95\% confidence interval) loading on the market portfolio in this period is 0.9 $(0.7 \, , \, 1.1)$ under the updating protocol 
and 1.3 $(1.1 \, , \, 1.5)$ under the rolling protocol.

\vskip .2in

\leftline{\em 3.3 Updating versus rolling}

In our empirical design we do not allow the investor to choose between rolling and updating, since the out-of-sample periods used for identifying the ex-post 
optimal configurations are
not the same across the two protocols.  Instead, as econometricians we compare results across protocols to make inferences about the nature of the data generating process.

Table 1 shows that the updating protocol does not provide any warning that characteristics' predictive value has vanished.  Instead, the ex-post optimal model
appears temporally stable.  The large gains over the first 15 years of the configuration selection period disguise the fact that the PPPs underperform benchmarks
in the more recent years.   For example, were we to evaluate the model using the ex-post optimal portfolio ending in 2018 (Table 1),
we would infer that the optimal value for $\lambda$ is 1, and the optimal characteristic set comprises: 
momentum, log size, beta, average same-month return, and residual volatility.  Indeed, over the out-of-sample period from 1975 - 2018 this portfolio 
dominates the benchmarks.  However this masks the fact that this portfolio's certainty equivalent mean (95\% confidence interval) over the preceding 180 months
is -23 $(-113 \, , \, 45)$ basis points per month.
The salutary historical performance over the 44-year out-of-sample period trades off the positive effects of characteristic-based tilting in the years 1975 - 1998
against the negative effects in the latter years.
This underscores the issues raised in Martin and Nagel (2022).  The flexibility to choose the characteristic 
set and the hyperparameter, $\lambda$ expands the dimensionality of the model space.  The results of using an ex-post optimal model configuration (on out-of-sample returns)
must be analyzed in a subsequent (truly) out-of-sample period to evaluate the model.

Table 1 suggests that some of the usefulness of $\lambda$ as a regularization parameter may derive from its capacity to detect a structural break.  However this is primarily
true under the rolling protocol.  Once there is a year in which the out-of-sample results deteriorate there is a jump in $\lambda$, as the algorithm scales back its reliance on
characteristics.  When 2000 replaces 1985 in the out-of-sample validation period $\lambda$ jumps from 1 to 6.  
This increase in $\lambda$ has a large impact on the
out-of-sample certainty equivalent returns.  For contrast, the 1\%ile (95\% confidence interval) on
the certainty equivalent return from the configuration that was optimal in the preceding year, with $\lambda = 1$, is -10,000 $(-10,000 \, , \, 398.4)$.

Since Table 2 describes a noisy,
structurally stable environment, the updating protocol produces portfolios with higher 2.5\%ile certainty equivalents than rolling for all three investors.  However
the rolling protocol produces higher 97.5\%ile certainty equivalents than updating.  In a structurally stable environment more data is
strictly preferred.  Table 1 shows that the two protocols have very similar ex-post optimal configurations ($\lambda$ and characteristic sets) through
1999.

\vskip .2in
        
\leftline{\em 3.4 Increasing risk aversion}

Figures IA-1 - IA-6 show the confidence intervals of the ex-post optimal $\theta$ values from the rolling protocol preceding each year in the second-stage out-of-sample period, for each of the 
six characteristics for three power utility investors, with coefficients of relative risk-aversion: 2, 5, and 8.  The values for the least risk-averse investor
correspond to those reported in Figure 1.  The figures show that as risk-aversion increases the optimal effect of characteristics on portfolio weights decreases in
absolute value.  Further, the apparent appeal of characteristic tilt appears to vanish more quickly for the more risk-averse investors.  For example, the ex-post optimal
out-of-sample portfolio put a significant positive weight on average same month return for the first 20 years in the second-stage out-of-sample period, for the most risk-tolerant
investor--although there is a
significant drop in the coefficient after 2000 (for the 2001 out-of-sample portfolio).  However, this coefficient is zero for the most risk-averse investor in years 12 - 26 and 
28 - 32.

Tables IA-2 and IA-3 provide the ex-post optimal configurations prior to each year in the second-stage out-of-sample period, from both protocols (analogous to Table 1) for the power utility investors
with $\gamma = 5$ and $\gamma = 8$, respectively.
Both power utility investors optimally set $\lambda = 0$ for most of the first subsample, confirming Brandt, Santa-Clara, and Valkanov's conjecture that 
their algorithm has much less estimation risk than mean-variance optimization.  Contrasting this with the importance of $\lambda > 0$ for the most risk-tolerant of these
three investors provides additional evidence that estimation risk (the tendency to overfit) is linked to the return variance.  The power utility
investor with $\gamma = 2$ is tolerant enough of variance (and kurtosis) so that maximizing this utility function directly on the data accepts positions with attractive
in-sample expected return and skew
that result from noise in the conditional return generating process.  In this case using $\lambda = 1$ helps to separate the predictability from the noise.  

These tables also demonstrate similar evidence from the rolling protocol, as in Table 1 that the fit of the model starts to deteriorate around 1999.  The equally-weighted
benchmark is the ex-post optimal portfolio (i.e., the 1\%ile of its certainty equivalent sampling distribution is greater than all 882 configurations) in 2009 - 2015
for the power utility investor with $\gamma = 5$.  Similarly the power utility investor chooses either the value-weighted benchmark or the equally-weighted
benchmark above all 882 optimized portfolios prior to the years: 2002, 2005, and 2009 - 2015.

Table IA-4 is analogous to Table 2.  It shows that in the first true out-of-sample period (1990 - 1998), the dynamic PPP portfolios under the updating protocol
dominate the benchmark portfolios for both of these more risk-averse power utility investors.  However, unlike the case for our most risk-tolerant investor, the optimal
dynamic PPP portfolios from the rolling protocol do not dominate the benchmarks in this period for these investors.  These results jointly lend credence to the hypothesis
that the conditional return generating process was largely stable over the 1955 - 1998 period, as using more information reduces estimation risk.
This table also shows that the optimal dynamic portfolios from the
updating protocol are neither negatively skewed nor more leptokurtic than the Gaussian distribution.  

The left-hand panels in Figures IA-7 and IA-8 contrast the 
return distributions of the optimal portfolio from the updating protocol with the preferred benchmark, which is the value-weighted benchmark in both cases in the first subperiod.
Figure IA-8
is especially revealing as our most risk-averse investor's optimal portfolio has a similar scale as the benchmark.  The mean (95\% confidence interval) 
interquartile range of the optimal dynamic portfolio is 702 $(576 \, , \, 840)$ basis points per month compared with the value-weighted benchmark's 
470 $(427 \, , \, 515)$ basis points per month.  This figure's left panel shows that the optimal portfolio has more significant mass above a 5\% monthly return than the benchmark,
while having similar left-tail properties in the first subperiod.  Both dynamic optimal portfolios have similar Sharpe ratios to that in Table 2 for our most risk-tolerant investor.
Unlike the certainty equivalent, the Sharpe ratios from both protocols are very similar--both are significantly larger than the benchmarks'.  

Table IA-5 reports the out-of-sample properties of the dynamic optimal portfolios for these two investors in our second-stage out-of-sample subperiod, 1999 - 2021.  
The mean certainty equivalent is negative for both investors'
dynamic optimal portfolios under both protocols in this subperiod.  The updating portfolios' Sharpe ratios are not significantly different from the benchmarks', however
both investors' optimal dynamic portfolios' Sharpe ratios under the rolling protocol are significantly lower than the equally-weighted benchmarks'.

Table IA-6 contrasts the dynamic PPP with the benchmarks for the entire 384 month out-of-sample period (1990 - 2021).  The optimal portfolios for for our focal investor
(whose $\gamma = 2$) is the equally weighted benchmark, and the optimal portfolio for the two more risk-averse investors is the value weighted benchmark.  As in the second subperiod,
the dynamic PPP is less attractive in the expected utility metric because of its additional kurtosis.  These portfolios from the updating protocol,
for the two more risk-averse investors have a significantly higher Sharpe ratio than the benchmarks.  Unlike the utility function, the Sharpe ratio does not 
over-weight the smaller minimum returns.  The mean (95\% confidence interval) of the value weighted benchmark's minimum monthly return over this period is -17\%
$(-18\% \, , \, -15\%)$ per month.  These statistics for our most risk-averse investor's dynamic PPP are -33\% $(-44\% \, , \, -25\%)$ per month, from the updating protocol.
These distributions are compared graphically in the right-hand panels of Figures IA-7 and IA-8.  Both of these graphs show that the left tail of the dynamic PPP's return distribution
dominates that of the preferred benchmark.

Tables IA-7 and IA-8 report the relationships between these two more risk-averse investors' optimal portfolios and the six Fama-French factors for all three out-of-sample
periods, the first and second subperiods and the full 32-year period.  Consistent with the analysis of these portfolios above, alpha decreases in $\gamma$.  In the first
subperiod our most risk-averse investor's optimal portfolio has mean (95\% confidence interval) alpha of 104  $(60 \, , \, 149)$ basis points per month, (Table IA-8).
Table IA-9 provides the sampling distributions of the power utility investor with $\gamma = 2$ dynamic optimal portfolio return projection on the Fama-French six factors
for the full 32-year out-of-sample period.  Table IA-10 reports the mean and variance decompositions for the two more risk-averse investors' optimal portfolios in the first
subperiod (analogous to Table 3, Panel B).  During this subperiod, when the characteristics provide useful information about future returns the variance decompositions are
constant across risk aversion.  For all three investors the six factors account for 40 to 55\% (95\% confidence intervals) of the optimal portfolio's return variance.
Similarly, HML and MOM are the largest sources of return variance for all three portfolios amongst the six factors for all three investors.

In the period when characteristics were efficacious, the optimal portfolios' scales decrease in risk aversion.  The interquartile range of returns for our
most risk-tolerant investor has a mean (95\% confidence interval) of 1,495 $(1,201 \, , \, 1,822)$ basis points per month; the analogous statistics for our most risk-averse investor's 
dynamic optimal portfolio are 702 $(576 \, , \, 840)$ basis points per month.  However the sources of variance in the span of the six Fama-French factors are flat in $\gamma$.
This result is fully consistent with the notion that sources of predictable excess return require exposure to non-diversifiable variance, as discussed by Kozak, Nagel,
and Santosh (2020).  Since all three investors tilt their optimal portfolios toward stocks that have done relatively well in the same month over the past five years, 
this result is consistent
with Keloharju, Linnainmaa, and Nyberg's (2016) hypothesis that there are many small seasonal (month-of-the-year) factors, and return in the month serves as an instrument for exposure
to these factors.

\vskip .2 in

\leftline{\bf 4. Conclusions}

We explore the nature of estimation risk in conjunction with Brandt, Santa-Clara, and Valkanov's parametric portfolio policy.  We use a bootstrapped out-of-sample maxmin criterion 
to select the optimal portfolio configuration at the beginning of each year in a second-stage (optimized) out-of-sample period.
To gain insight into the interactions between portfolio optimization and estimation risk,
we introduce a new form of regularization.  Rather than penalizing the parameter space we introduce a hyperparameter that can increase the curvature 
of the loss function used to estimate portfolio weights.
For power utility investors with moderate to high risk aversion (coefficients of relative
risk aversion of 5 and 8), Brandt, Santa-Clara, and Valkanov's (2009) PPP algorithm afforded a way to exploit this predictability without significant estimation
error.  The most risk-tolerant
power utility investor we consider, with coefficient of relative risk aversion of two, experiences much higher estimation risk with the parametric portfolio policy.  
This investor reduces overfitting by optimizing a power utility function  with a higher coefficient of relative risk aversion than her own in sample.  
This loss function increases the shadow cost of variance in terms of expected return.  It works because estimation error increases in portfolio variance.

Our results suggest that measurable characteristics did have economically and statistically significant predictive content for portfolio construction prior to the
year 1999.  That result considers all moments of the predictive distribution and is not
specific to expected return, alpha, or Sharpe ratio.
Since during this period (the 20th Century), optimal portfolios' characteristic-tilts diminished in risk aversion, we do not infer that any of the opportunities
afforded by conditioning on characteristics represented a free lunch.  Rather there were dimensions wherein the tradeoffs between mean (and skewness) and variance (and kurtosis) were more attractive than 
those afforded by the market portfolio.  Two of those dimensions appear to be the well-known momentum and value factors.
Optimal portfolio returns were virtually orthogonal to the market factor and one-half of their return variance 
came from outside the span of the Fama-French 6-factor model.

Characteristics'
predictive efficacy for portfolio optimization has vanished starting in 1999.  The market portfolio dominates all of the dynamically optimal parametric portfolios over the
1999 - 2021 period.  This finding is consistent with the literature, which contains several non-mutually exclusive hypotheses.  Martin and Nagel (2022) show that a
complex economy in which agents learn about predictive relationships econometricians should expect to find in-sample predictability that vanishes out of sample.
McLean and Pontiff (2016) suggest that investors adapt to academic research.  Green, Hand, and Zhang (2017) note that the 21st Century has seen new regulations and
technological advances that serve to reduce trading frictions.  This, in turn, allows investors to more fully exploit predictive relationships.

\newpage

\leftline{\bf Appendix}

\renewcommand*{\thefootnote}{\fnsymbol{footnote}}

\vspace{.2cm}

\noindent
This appendix provides a pseudo code for our bootstrap dynamically regularized out-of-sample empirical design, with both updating and rolling sample protocols.
Each year 
we optimize expected utility in sample to construct the next year's out-of-sample returns.  Once we have at least 180 months of out-of-sample returns,  we select the 
optimal configuration with numerical minmax of out-of-sample certainty equivalent returns.  The optimal dynamically regularized policy--or second out-of-sample
stage--has this configuration's out-of-sample returns in the next year.  As in the text, we use the following notation:  $y = 1, \, \ldots, \, 62$ references the number of each
year in our sample. Uppercase $Y$ is the year: $Y_1 = 1960$, $Y_{15} = 1974$, $Y_{30} = 1989$, and $Y_{62} = 2021$.


\newpage

\noindent
\vspace{-.2cm}

\hrule
\hrule

\begin{description}

\item [FOR] each year, $y = 15,\,  16,  \, \ldots, \, 29:$ \hfill \\

\vspace{-.9cm}

\begin{description}
\item [FOR] each configuration, $c = 1, \, 2, \,  \ldots, \,  882:$  \hfill \\

\vspace{-.7cm}

  \begin{description}
    \item [FOR] each bootstrap sample, $b = 1, \, 2, \, \ldots, \,  10,000:$  \hfill \\

\vspace{-.7cm}

  \begin{description}
    \item[Form out-of-sample returns:]  Maximize (1) over $\theta$ -- using  $\left[Y_s, \; Y_{y}\right]$.\footnote[1]{Under the updating protocol, $Y_s \equiv Y_1 = 1960$, and $Y_s = Y_{y-14}$ 
          under the rolling protocol.}
           These $\hat{\theta}_{y,c,b}$ are used to construct the out-of-sample portfolio returns in the year $Y_{y+1}$, for configuration $c$ and bootstrap sample $b$, respectively.
  \end{description}

\vspace{-.1cm}

    \item[END FOR]
  \end{description}

\vspace{-.2cm}

\item[END FOR]

\end{description}

\vspace{-.43cm}

\item[END FOR]

\end{description}


\noindent

\vspace{-.4cm}


\begin{description}

\item [FOR] each year, $y = 30,\,  31,  \, \ldots, \, 61:$  \hfill \\

\vspace{-1.03cm}

  \begin{description}
    \item[Dynamic Regularization:] Identify which of the 882 configurations' {\em out-of-sample portfolio returns} over years $\left[Y_\upsilon, \, Y_y\right]$ has the maximum 1\%ile value
            certainty equivalent return.\footnote[7]{Under the updating protocol, $Y_\upsilon \equiv Y_{16} = 1975$, and $Y_\upsilon = Y_{y-14}$ under the rolling protocol.}
            This optimal configuration, $c^*_y$,  is reported by year ($Y_{y+1}$) in Table 1 for the power utility investor with coefficient of relative risk aversion of 2.  Tables IA-2 and IA-3 report this optimal
configuration by year for power utility investors with coefficients of relative risk aversion of 5 and 8, respectively.
  \end{description}

\vspace{-.2cm}

\begin{description}
\item [FOR] each configuration, $c = 1, \, 2, \, \ldots$, 882:  \hfill \\

\vspace{-.7cm}

  \begin{description}
    \item [FOR] each bootstrap sample, $b = 1, \, 2, \, \ldots$, 10,000:  \hfill \\

\vspace{-.7cm}

  \begin{description}
    \item[Form out-of-sample returns:]  Maximize (1) over $\theta$ -- using  $\left[Y_s, \; Y_y\right]$.$^*$
          These $\hat{\theta}_{y,c,b}$ are used to construct the out-of-sample portfolio returns in the year $Y_{y+1}$, for configuration $c$ and bootstrap sample $b$, respectively.
  \end{description}

\vspace{-.1cm}

    \item[END FOR]
  \end{description}

\vspace{-.2cm}

\item[END FOR]

\item[] The bootstrap set of second stage,
{\em out-of-sample} dynamically regularized optimal portfolio returns for the 12 months in year $y+1$ is: $\{r_{y+1,c^*_y}\}$.

\end{description} 

\vspace{-.65cm}

\item[END FOR]

\end{description}

\hrule
\hrule

\newpage

\noindent
The replication code is available at https://doi.org/10.7910/DVN/LK4DCN.

\hoffset -.4in

\begin{center}
{\bf References}
\end{center}

\parindent -22pt
\parskip .1in


A\"{i}t-Sahalia, Yacine and Michael W. Brandt, 2001, Variable selection for portfolio choice, {\em Journal of
    Finance} 56, 1297--1351.



Ao, Mengment, Yingying Li, and Xinghua Zheng, 2019, Approaching mean-variance efficiency for large portfolios,
       {\em Review of Financial Studies} 32, 2890--2919.



Barroso, Pedro, Jurij-Andrei Reichenecker, Michael Reichenecker, and Florent Rouxelin, 2023, An international equity and
        currency optimisation with frictions, Working paper, Catloica-Lisbon School of Business and Economics.

Barroso, Pedro and Pedro Santa-Clara, 2015a, Momentum has its moments, {\em Journal of Financial Economics}
    116, 111--120.

Barroso, Pedro and Pedro Santa-Clara, 2015b, Beyond the carry trade: Optimal currency portfolios, {\em Journal of Financial 
         and Quantitative Analysis}, 50, 1037--1056.

Barroso, Pedro and Konark Saxena, 2022, Lest we forget: Learn from out-of-sample forecast errors when optimizing portfolios,
     {\em Review of Financial Studies} 35, 1222-1278.


Bertsimas, Dimitris, Vishal Gupta, and Nathan Kallus, 2018, Data-driven robust optimization, {\em Mathematical Programming,
    Series A} 167, 235--292.

Best, Michael J. and Robert R. Grauer, 1991, On the sensitivity of mean-variance-efficient portfolios to changes in 
     asset means: Some analytical and computational results, {\em Review of Financial Studies} 4, 315--342.

Boguth, Oliver, Murray Carlson, Adlai Fisher, and Mikhail Simutin, 2011, Conditional risk and
    performance evaluation: Volatility timing, overconditioning, and new estimates of momentum
    alphas, {\em Journal of Financial Economics} 102, 363--389.

Brandt, Michael W., Pedro Santa-Clara, and Rossen Valkanov, 2009,  Parametric portfolio policies:
    Exploiting characteristics in the cross-section of equity returns,
    {\em Review of Financial Studies} 22, 3411--3447.






Daniel, Kent and Tobias J. Moskowitz, 2016, Momentum crashes, {\em Journal of Financial Economics} 122, 221--247.

DeMiguel, Victor, Lorenzo Garlappi, and Raman Uppal, 2009, Optimal versus naive diversification: How inefficient is the
      1/N portfolio strategy?, {\em Review of Financial Studies} 22, 1915--1953.

\newpage

DeMiguel, Victor, Alberto Mart\'{i}n-Utrera, Francisco J. Nogales, and Raman Uppal, 2020, A transaction-cost perspective on the
      multitude of firm characteristics, {\em Review of Financial Studies} 33, 2180--2222.


DeMiguel, Victor, Yuliya Plyakha, Raman Uppal, and Grigory Vilkov, 2013, Improving portfolio selection using option-implied
     volatility and skewness, {\em Journal of Financial and Quantitative Analysis} 48, 1813--1845.


Faias, José and Pedro Santa-Clara, 2017, Optimal option portfolio strategies: Deepening the puzzle of index option mispricing,
     {\em Journal of Financial and Quantitative Analysis} 52, 277--303.



    


Freyberger, Joachim, Andreas Neuhierl, and Michael Weber, 2020, Dissecting characteristics nonparametrically,
    {\em Review of Financial Studies} 33, 2326--2377.


Gay, David M., 1983, Subroutines for unconstrained minimization using a model/trust-region approach,
    {\em ACM Transactions on Mathematical Software} 9, 503--524.




Giglio, Stefano, Yuan Liao, and Dacheng Xiu, 2021, Thousands of alpha tests, {\em The Review of Financial 
       Studies} 34, 3456--3496.


Gilboa, Itzhak, Andrew W. Postlewaite, and David Schmeidler, 2008, Probability and uncertainty in economic 
    modeling, {\em Journal of Economic Perspectives} 22, 173--188.

Gilboa, Itzhak and David Schmeidler, 1989, Maxmin expected utility with non-unique prior, 
    {\em Journal of Mathematical Economics} 18, 141--153.



Green, Jeremiah, John R.M. Hand, and X. Frank Zhang, 2017, The characteristics that provide independent
    information about average U.S. monthly stock returns, {{\em Review of Financial Studies} 30, 
    4389--4436.




Hansen, Lars P. and Thomas J. Sargent, 2008,  {\em Robustness}, Princeton University Press, Princeton, NJ.







Heston, Steve L. and Ronnie Sadka, 2008,  Seasonality in the cross-section of stock returns,
    {\em Journal of Financial Economics} 87, 418--445.





Jagannathan, Ravi and Tongshu Ma, 2003, Risk reduction in large portfolios: Why imposing the
    wrong constraint helps, {\em Journal of Finance} 58, 1651--1683.






Kadan, Ohad and Fang Liu, 2014, Performance evaluation with high moments and disaster risk,
    {\em Journal of Financial Economics} 113, 131--155.


Kan, Raymond and Guofu Zhou, 2007, Optimal portfolio choice with parameter uncertainty, 
      {\em Journal of Financial and Quantitative Analysis} 42, 621--656.

Kapetanios, George, 2008, A bootstrap procedure for panel data sets with many cross-sectional units,
    {\em The Econometrics Journal} 11, 377--395.

Keloharju, Matti, Juhani T. Linnainmaa, and Peter Nyberg, 2016,  Return
    seasonalities, {\em Journal of Finance} 71, 1557--1589.


Kim, Tae-Hwan and Halbert White,  2003, On more robust estimation of skewness and kurtosis: Simulation and application to the S\&P 500 Index,
    Working Paper, University of California, San Diego.





Kozak, Serhiy, Stefan Nagel, and Shrihari Santosh, 2020, Shrinking the cross section, {\em Journal of Financial Economics}
    135, 271--292.

Kroencke, Tim, Felix Schindler, and Andreas Schrimpf, 2014, International diversification benefits with foreign investment
    styles, {\em Review of Finance} 18, 1847--1883.







Lewellen, Jonathan, 2015, The cross section of expected stock returns, {\em Critical Finance
    Review} 4, 1--44.


Lewellen, Jonathan, Stefan Nagel, and Jay Shanken, 2010, A skeptical appraisal of asset pricing
    tests, {\em Journal of Financial Economics} 96, 175--194.




Liu, Jianan, Robert F. Stambaugh, and Yu Yuan, 2018, Absolving beta of volatility's effects, {\em Journal
     of Financial Economics} 128, 1--15.


MacKinlay, A. Craig and Lubo\v{s} P\'{a}stor, 2000, Asset pricing models: Implications for expected
      returns and portfolio selection, {\em Review of Financial Studies}  13, 883--916.



Martin, Ian and Stefan Nagel, 2022, Market efficiency in the age of big data, {\em Journal of Financial Economics} 
       145, 154--177.

McLean, R. David and Jeffrey Pontiff, 2016, Does academic research destroy stock return
    predictability?  {\em Journal of Finance} 71, 5--31.


Nagel, Stefan, 2021, {\em Machine Learning in Asset Pricing}, Princeton University Press.







P\'{a}stor, Lubo\v{s}, 2000, Portfolio selection and asset pricing models, {\em Journal of Finance}
        60, 179--223.

P\'{a}stor, Lubo\v{s} and Robert F. Stambaugh, 2000, Comparing asset pricing models: An investment 
    perspective, {\em Journal of Financial Economics}  56, 335--381.



Rahimian, Hamed and Sanjay Mehrotra, 2019, Distributionally robust optimization: A review, Working paper, 
      Northwestern University.









\newpage

\setlength{\intextsep}{0pt plus 2pt}
\setlength{\abovecaptionskip}{8pt}

\flushbottom

\captionsetup{width=\textwidth}

\restylefloat{figure}

\hoffset=-1.5in

\long\def\symbolfootnote[#1]#2{\begingroup%
\def\thefootnote{\fnsymbol{footnote}}\footnote[#1]{#2}\endgroup}

\begin{figure}[H]
\includegraphics*[scale=1.18]{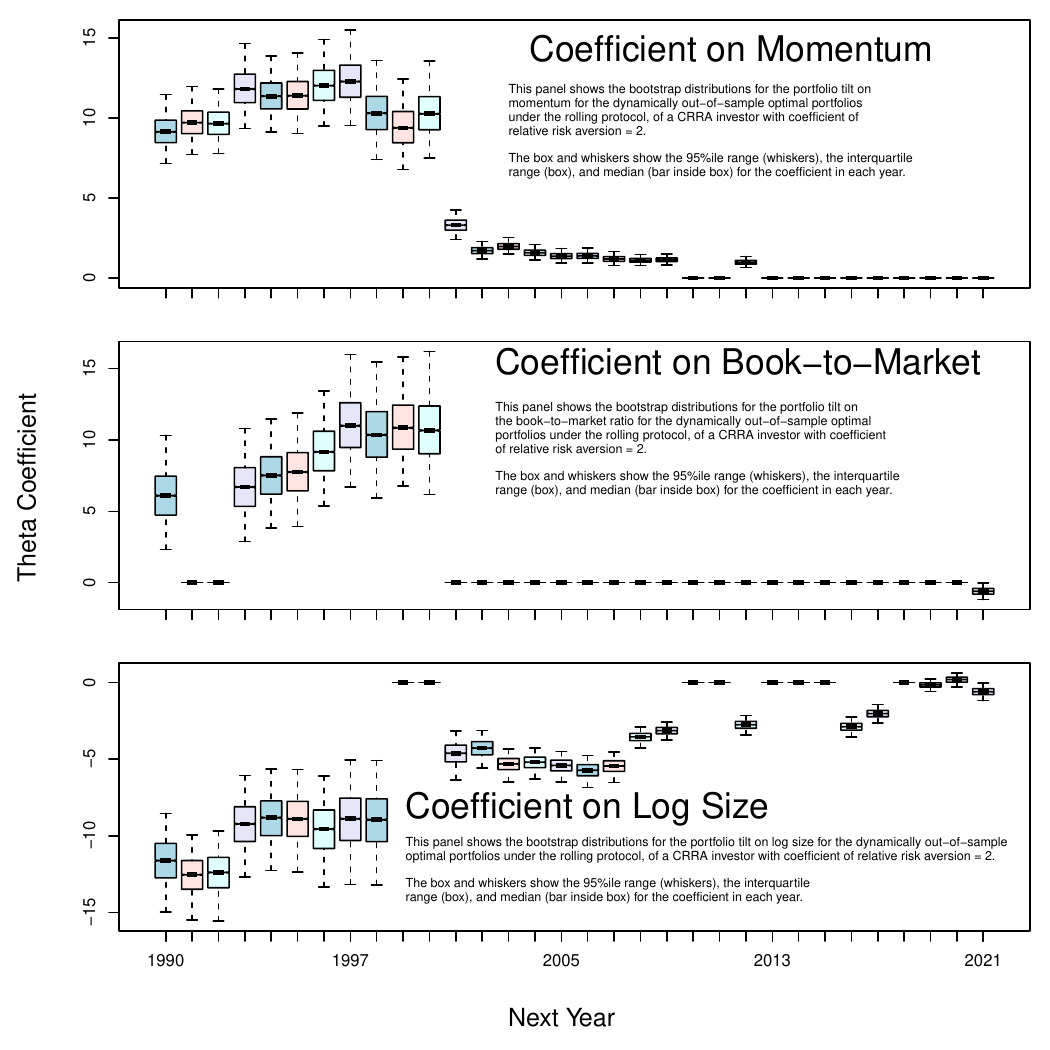}

{\bf Figure 1.  Characteristic tilts (\boldmath{$\theta$}) over time.} 
Sampling distributions of the $\theta$ coefficient on (standardized) characteristics from the optimal model over the preceding 180 out-of-sample months, used to
construct the optimal portfolio in the indicated year.

\end{figure}

\newpage

\begin{figure}[H]
\includegraphics*[scale=1.18]{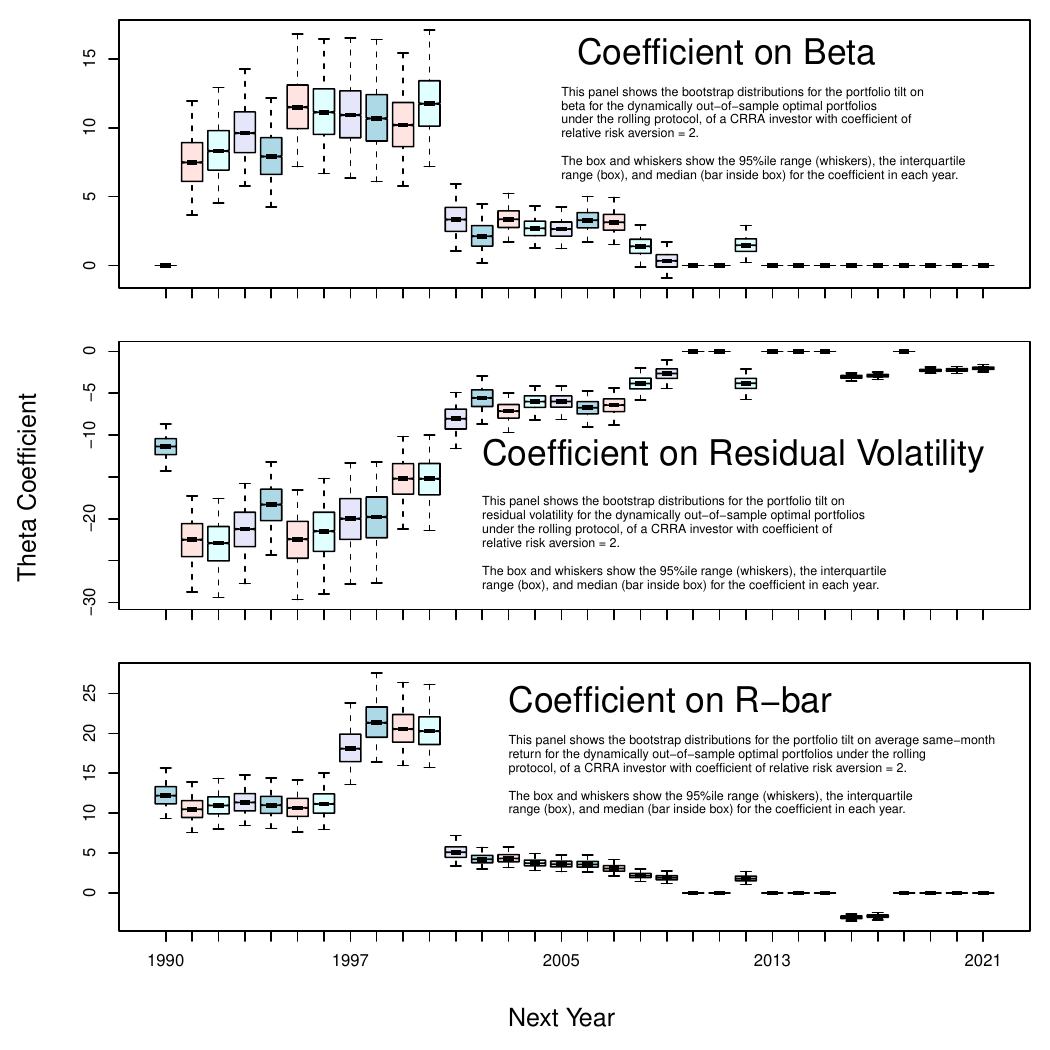}

{\bf Figure 1 (Cont'd.).  Characteristic tilts (\boldmath{$\theta$}) over time.} 
Sampling distributions of the $\theta$ coefficient on (standardized) characteristics from the optimal model over the preceding 180 out-of-sample months, used to
construct the optimal portfolio in the indicated year.

\end{figure}

\newpage

\renewcommand{\textwidth}{7.5 in}
\textheight 10.8in

\begin{landscape}

\voffset -1.9in

\begin{figure}[H]
\includegraphics*[scale=0.905]{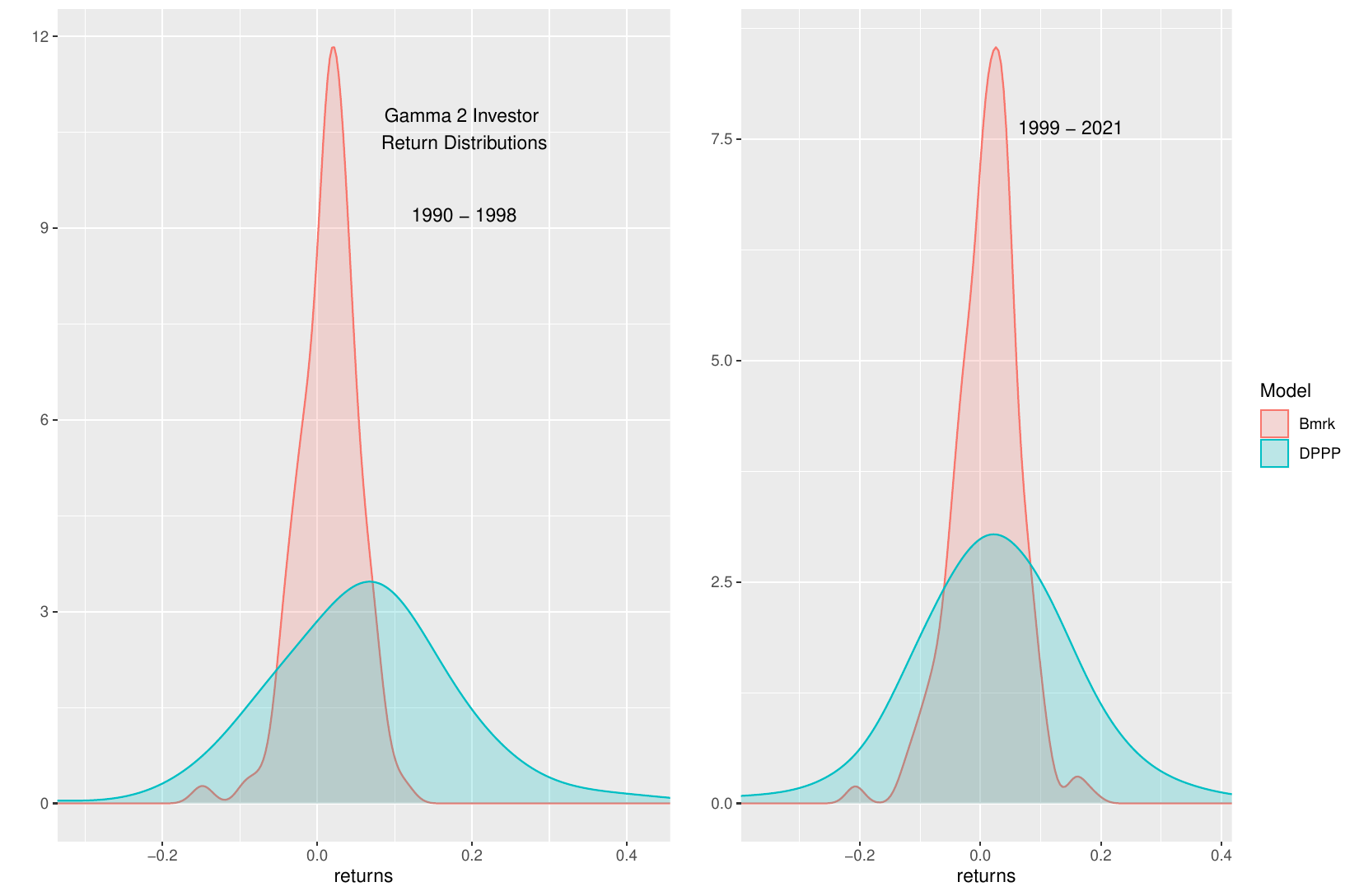}

\captionsetup{width=19.5cm}
{\bf Figure 2  Portfolio return densities in the 2 subperiods.}
``DPPP'' is the optimal dynamic parametric portfolio under the\\ updating protocol--selected at the beginning of
each year.  ``Bmrk'' is the preferred benchmark in the subperiod.  For this power\\ utility investor with coefficient of relative risk aversion,
$\gamma$, $= \; 2$ that is the value-weighted portfolio of all stocks in
the first subperiod\\ and the equally-weighted portfolio of all stocks in the second subperiod.

\end{figure}

\end{landscape}

\newpage

\pagestyle{empty}

\voffset=-1.6in
\hoffset=-1.2in

\renewcommand{\textwidth}{7.35 in}

\captionsetup{width=\textwidth}

\interfootnotelinepenalty=100

\setlength{\intextsep}{0pt plus 2pt}
\setlength{\abovecaptionskip}{-28pt}

\makeatletter
\def\normaljustify{%
  \let\\\@centercr\rightskip\z@skip \leftskip\z@skip%
  \parfillskip=0pt plus 1fil}
\makeatother

\newcolumntype{C}[1]{>{\Centering}m{#1}}
\renewcommand\tabularxcolumn[1]{C{#1}}

\flushbottom

\captionsetup[figure]{labelformat=empty, labelsep=period}

\restylefloat{table}
\restylefloat{figure}

\long\def\symbolfootnote[#1]#2{\begingroup%
\def\thefootnote{\fnsymbol{footnote}}\footnote[#1]{#2}\endgroup}

\voffset=-1.2in
\hoffset=-0.6in
\renewcommand{\textwidth}{7.35 in}
\textheight 10.5in

\setlength{\intextsep}{0pt plus 2pt}

\setlength{\abovecaptionskip}{-200pt}

\restylefloat{table}
\restylefloat{figure}

\textheight 10.5in


\hfuzz 50pt

\flushbottom


\voffset=-1.15in
\hoffset=-1.2in

\baselineskip 14pt

\begin{center}

{\Large {\bf Table 1}}\\
{\bf Optimal $\gamma^*$ and portfolio (characteristic sets):}\\
{\bf Sampling properties of certainty equivalent returns}\\
{\bf For investor with power utility and coefficient of relative risk aversion,} \boldmath{$\gamma = 2.$}\\
\end{center}

\begin{tabular}{lccrrrcccrrr}

\multicolumn{12}{l}{Weight tilts ($\theta$) are estimated for 63 characteristic sets under each of 14 values of the loss function curvature 
($\gamma^*$),}\\
\multicolumn{12}{l}{using both rolling and updating protocols.  Of these 882 cases that with the highest 1\%ile value of the out of sample}\\ 
\multicolumn{12}{l}{certainty equivalent is reported for each of three investors in basis points per month.  The characteristic symbols}\\
\multicolumn{12}{l}{are: $\zeta$: momentum, V: book-to-market ratio, S: log size, $\beta$: from lagged 60-month market model, $\overline{r}$: average}\\
\multicolumn{12}{l}{same-month return over the previous 5 years, $\sigma_\epsilon$: standard deviation of lagged 60-month market model residual.}\\

   &    &   &   &  & &\hspace{.1in}  &  & & & & \\

     & \multicolumn{5}{c}{Updating Protocol} & & \multicolumn{5}{c}{Rolling Protocol}\\
\cline{2-6} \cline{8-12}
Next & \multicolumn{2}{c}{Optimal} & \multicolumn{3}{c}{Certainty Equivalent} & &  \multicolumn{2}{c}{Optimal} &  \multicolumn{3}{c}{Certainty Equivalent}\\
\cline{2-3} \cline{8-9}
Year & Chars & $\gamma^*$ & 1\%ile & Mean & Std Dev & & Chars & $\gamma^*$ & 1\%ile & Mean & Std Dev\\
\hline
1990 &                   {\bf VWI} &          &   113.3 & 120.7 & 3.2  & & {\bf VWI}                              &   & 113.3 & 120.7 & 3.2\\
1990 &                   {\bf EWI} &          &   145.0 & 149.1 & 1.8  & & {\bf EWI}                              &   & 145.0 & 149.1 & 1.8\\
1990 & $\zeta$,V,S,$\overline{r}$,$\sigma_\epsilon$ & 2 & -10,000.0 & -4,146.6 & 5,245.0 & & $\zeta$,V,S,$\overline{r}$,$\sigma_\epsilon$ & 2 & -10,000.0 & -437.6 & 3,148.4\\
\hline
1990 & $\zeta$,V,S,$\overline{r}$,$\sigma_\epsilon$ & 3 & 527.8 & 625.5 & 46.1 & & $\zeta$,V,S,$\overline{r}$,$\sigma_\epsilon$ & 3 & 523.1 & 622.2 & 46.0\\
1991 & $\zeta$,S,$\overline{r}$,$\sigma_\epsilon$ & 3 & 497.7 & 589.7 & 43.5 &   & $\zeta$,S,$\beta$,$\overline{r}$,$\sigma_\epsilon$ & 3 & 498.4 & 597.8 & 46.1\\
1992 & $\zeta$,S,$\overline{r}$,$\sigma_\epsilon$ & 3 & 512.1 & 603.4 & 43.4  & & $\zeta$,S,$\beta$,$\overline{r}$,$\sigma_\epsilon$ & 3 & 497.4 & 601.6 & 48.1\\
1993 & $\zeta$,V,S,$\beta$,$\overline{r}$,$\sigma_\epsilon$ & 3 & 512.3 & 615.5 & 48.1  & & $\zeta$,V,S,$\beta$,$\overline{r}$,$\sigma_\epsilon$ & 3 & 486.8 & 597.4 & 51.3\\
1994 & $\zeta$,V,S,$\beta$,$\overline{r}$,$\sigma_\epsilon$ & 3 & 518.2 & 619.6 & 47.3  & & $\zeta$,V,S,$\beta$,$\overline{r}$,$\sigma_\epsilon$ & 3 & 493.3 & 606.4 & 52.3\\
1995 & $\zeta$,V,S,$\beta$,$\overline{r}$,$\sigma_\epsilon$ & 3 & 500.8 & 601.9 & 46.2  & & $\zeta$,V,S,$\beta$,$\overline{r}$,$\sigma_\epsilon$ & 3 & 471.1 & 593.3 & 54.2\\
1996 & $\zeta$,V,S,$\beta$,$\overline{r}$,$\sigma_\epsilon$ & 3 & 489.7 & 584.7 & 44.2  & & $\zeta$,V,S,$\beta$,$\overline{r}$,$\sigma_\epsilon$ & 3 & 496.1 & 615.5 & 54.3\\
1997 & $\zeta$,V,S,$\beta$,$\overline{r}$,$\sigma_\epsilon$ & 3 & 477.8 & 568.1 & 42.6  & & $\zeta$,V,S,$\beta$,$\overline{r}$,$\sigma_\epsilon$ & 3 & 503.1 & 620.6 & 54.5\\
1998 & $\zeta$,V,S,$\beta$,$\overline{r}$,$\sigma_\epsilon$ & 3 & 503.9 & 598.6 & 44.3  & & $\zeta$,V,S,$\beta$,$\overline{r}$,$\sigma_\epsilon$ & 3 & 540.8 & 670.4 & 60.4\\
1999 & $\zeta$,V,S,$\beta$,$\overline{r}$,$\sigma_\epsilon$ & 3 & 477.0 & 565.3 & 42.2  & & $\zeta$,V,$\beta$,$\overline{r}$,$\sigma_\epsilon$ & 3 & 488.7 & 624.1 & 61.9\\
2000 & $\zeta$,V,S,$\beta$,$\overline{r}$,$\sigma_\epsilon$ & 3 & 472.2 & 562.5 & 42.9  & & $\zeta$,V,$\beta$,$\overline{r}$,$\sigma_\epsilon$ & 3 & 485.7 & 643.8 & 240.0\\
2001 & $\zeta$,S,$\beta$,$\overline{r}$,$\sigma_\epsilon$ & 3 & 443.8 & 530.1 & 45.3  & & $\zeta$,S,$\beta$,$\overline{r}$,$\sigma_\epsilon$ & 7 & 229.4 & 322.9 & 39.9\\
2002 & $\zeta$,S,$\beta$,$\overline{r}$,$\sigma_\epsilon$ & 3 & 415.5 & 503.0 & 44.4  & & $\zeta$,S,$\beta$,$\overline{r}$,$\sigma_\epsilon$ & 8 & 199.7 & 278.5 & 35.1\\
2003 & $\zeta$,S,$\beta$,$\overline{r}$,$\sigma_\epsilon$ & 3 & 432.4 & 521.3 & 44.9  & & $\zeta$,S,$\beta$,$\overline{r}$,$\sigma_\epsilon$ & 8 & 219.1 & 303.6 & 38.1\\
2004 & $\zeta$,S,$\beta$,$\overline{r}$,$\sigma_\epsilon$ & 3 & 401.5 & 483.8 & 41.7  & & $\zeta$,S,$\beta$,$\overline{r}$,$\sigma_\epsilon$ & 8 & 200.9 & 283.1 & 36.6\\
2005 & $\zeta$,S,$\beta$,$\overline{r}$,$\sigma_\epsilon$ & 3 & 395.8 & 476.4 & 40.6  & & $\zeta$,S,$\beta$,$\overline{r}$,$\sigma_\epsilon$ & 8 & 181.7 & 265.6 & 36.8\\
2006 & $\zeta$,S,$\beta$,$\overline{r}$,$\sigma_\epsilon$ & 3 & 388.9 & 468.3 & 39.5  & & $\zeta$,S,$\beta$,$\overline{r}$,$\sigma_\epsilon$ & 8 & 177.9 & 260.1 & 36.3\\
2007 & $\zeta$,S,$\beta$,$\overline{r}$,$\sigma_\epsilon$ & 3 & 374.5 & 450.9 & 37.9  & & $\zeta$,S,$\beta$,$\overline{r}$,$\sigma_\epsilon$ & 8 & 154.8 & 235.2 & 35.0\\
2008 & $\zeta$,S,$\beta$,$\overline{r}$,$\sigma_\epsilon$ & 3 & 345.9 & 415.7 & 35.4  & & $\zeta$,S,$\beta$,$\overline{r}$,$\sigma_\epsilon$ & 9 & 115.0 & 181.5 & 28.2\\
2009 & $\zeta$,S,$\beta$,$\overline{r}$,$\sigma_\epsilon$ & 3 & 339.8 & 410.5 & 35.1  & & $\zeta$,S,$\beta$,$\overline{r}$,$\sigma_\epsilon$ & 10 & 87.7 & 146.7 & 26.0\\
2010 & $\zeta$,S,$\beta$,$\overline{r}$,$\sigma_\epsilon$ & 3 & 305.7 & 376.1 & 35.1  & & {\bf EWI} &   & 72.8 & 77.6 & 2.1\\
2011 & $\zeta$,S,$\beta$,$\overline{r}$,$\sigma_\epsilon$ & 3 & 306.6 & 376.0 & 34.4  & & {\bf EWI} &   & 69.2 & 74.1 & 2.1\\
2012 & $\zeta$,S,$\beta$,$\overline{r}$,$\sigma_\epsilon$ & 3 & 306.2 & 375.4 & 34.1  & & $\zeta$,S,$\beta$,$\overline{r}$,$\sigma_\epsilon$ & 9 & 58.6 & 128.2 & 30.0\\
2013 & $\zeta$,S,$\beta$,$\overline{r}$,$\sigma_\epsilon$ & 3 & 299.6 & 366.8 & 33.0  & & {\bf EWI} &   & 50.3 & 55.4 & 2.1\\
2014 & $\zeta$,S,$\beta$,$\overline{r}$,$\sigma_\epsilon$ & 3 & 301.4 & 367.4 & 32.4  & & {\bf EWI} &  & 69.7 & 74.6 & 2.1\\
2015 & $\zeta$,S,$\beta$,$\overline{r}$,$\sigma_\epsilon$ & 3 & 284.5 & 348.2 & 31.0  & & {\bf EWI} &  & 65.9 & 70.5 & 2.0\\
2016 & $\zeta$,S,$\beta$,$\overline{r}$,$\sigma_\epsilon$ & 3 & 279.0 & 339.6 & 29.9  & & S,$\overline{r}$,$\sigma_\epsilon$ & 6 & 84.8 & 114.0 & 13.8\\
2017 & $\zeta$,S,$\beta$,$\overline{r}$,$\sigma_\epsilon$ & 3 & 273.7 & 333.0 & 29.1  & & S,$\overline{r}$,$\sigma_\epsilon$ & 7 & 77.2 & 96.2 & 9.1\\
2018 & $\zeta$,S,$\beta$,$\overline{r}$,$\sigma_\epsilon$ & 3 & 260.7 & 317.3 & 27.8  & & {\bf EWI} &  & 81.0 & 85.1 & 1.8\\
2019 & $\zeta$,S,$\beta$,$\overline{r}$,$\sigma_\epsilon$ & 4 & 249.3 & 295.1 & 20.9  & & S,$\sigma_\epsilon$ & 16 & 63.0 & 73.3 & 4.5\\
2020 & $\zeta$,S,$\beta$,$\overline{r}$,$\sigma_\epsilon$ & 4 & 243.1 & 287.3 & 20.3  & & S,$\sigma_\epsilon$ & 22 & 68.7 & 80.0 & 4.9\\
2021 & $\zeta$,S,$\beta$,$\overline{r}$,$\sigma_\epsilon$ & 4 & 236.3 & 278.6 & 19.3  & & V,S,$\sigma_\epsilon$ & 22 & 69.5 & 84.5 & 6.5\\
\hline

\end{tabular}

\newpage

\begin{landscape}

\voffset=-1.7in
\hoffset=-0.94in

\begin{center}

\baselineskip 14pt
{\Large {\bf Table 2}}\\
{\bf Sampling properties of out-of-sample Portfolio Performance Statistics}\\
{\bf 108-month out-of-sample period, 1990 -- 1998}\\
\end{center}

\vskip .1in

\begin{tabular}{lrrrrrrrcrrrrrrr}

\multicolumn{16}{l}{Sampling properties of dynamic optimal PPPs.  Portfolio characteristic tilts from the best out-of-sampling performer over the relevant preceding period}\\
\multicolumn{16}{l}{(shown in Table 1) each year.  $\cal{CE}$$_2$ is the certainty equivalent return in basis points per month for a power utility investor with coefficient of relative}\\
\multicolumn{16}{l}{risk aversion $(\gamma) = 2$.  $E(r)$, $\sigma$, Median, IQR, and MIN are the mean monthly return, the
standard deviation of monthly returns, the median}\\
\multicolumn{16}{l}{monthly return, the interquartile range of monthly returns, and the minimum monthly return--all expressed in basis points per month.}\\
\multicolumn{16}{l}{SKEW and KURT are the return skewness and kurtosis measures, and SR is the Sharpe ratio.}\\
\multicolumn{16}{l}{Results are for the first 9-year out-of-sample subperiod (1990 -- 1998).}\\

    &    &   &   &  &  & & & \hspace{.1in}  &  & & & &  &  &  \\

\multicolumn{16}{l}{{\bf Panel A: Benchmark portfolios}}\\
     & \multicolumn{7}{c}{VWI} & & \multicolumn{7}{c}{EWI}\\
\cline{2-8} \cline{10-16}
Statistic  & Mean & Std Dev &  2.5\%ile & 25\%ile & Median & 75\%ile & 97.5\%ile & &  Mean & Std Dev &  2.5\%ile & 25\%ile & Median & 75\%ile & 97.5\%ile\\
$\cal{CE}$$_2$ & 128.8 & 4.0 & 121.0 & 126.1 & 128.8 & 131.5 & 136.6 & & 109.4 & 2.3 & 104.8 & 107.9 & 109.4 & 110.9 & 113.9\\
$E(r)$ & 144.2 & 4.0 & 136.2 & 141.5 & 144.2 & 146.9 & 152.0 & & 128.6 & 2.3 &  124.1 &  127.1 &   128.6 & 130.1 & 133.1\\
$\sigma$ & 388.6 & 4.6 &  379.6 & 385.5 & 388.6 & 391.7 & 397.7 & & 430.0 &  2.3 &   425.1 &  428.3 &   430.0 &  431.6 &  434.9\\
Median & 167.4 & 11.6 & 144.9 & 159.3 & 167.6 & 175.3 & 189.7 & & 187.5 & 9.8 &  168.1 & 180.0 & 187.6 & 194.1 & 206.5\\
IQR & 469.9 &  22.4 & 426.6 & 454.5 & 469.8 & 484.9 & 514.7 & & 508.7 & 17.5 &     474.6 & 496.7 & 508.7 &  520.5 & 543.3\\
MIN & -1,482.9 & 59.1 & -1,602.5 & -1,521.9 & -1,481.9 & -1,442.2 & -1,368.6 & & -1,747.3 & 28.2 & -1,803.2 & -1,766.3 & -1,747.3 & -1,728.2 & -1,691.9\\
SKEW & -6.0 & 2.9 & -11.6 & -8.0 & -6.0 & -4.0 & -0.4 & & -13.7 & 2.2 & -18.0 & -15.2 & -13.7 &  -12.2 &  -9.3\\
KURT & 26.4 & 4.6 &  17.4 & 23.3 & 26.4 & 29.4 & 35.6 & & 35.2 & 2.5 & 30.3 &  33.5 &  35.2 & 36.9 &  40.2\\
SR & 0.9337 &  0.0364 &  0.8627 &  0.9095 &  0.9337 &  0.9581 &  1.0051 & & 0.7158 &  0.0187 &  0.6787 &  0.7034 &  0.7159 &  0.7284 &  0.7520\\
\hline

   &    &   &   &  &  & & & \hspace{.1in}  &  & & & &  &  &  \\

\multicolumn{16}{l}{{\bf Panel B: Dynamic PPP}}\\
     & \multicolumn{7}{c}{Dyn. Opt. Updating} & & \multicolumn{7}{c}{Dyn. Opt. Rolling}\\
\cline{2-8} \cline{10-16}
Statistic  & Mean & Std Dev &  2.5\%ile & 25\%ile & Median & 75\%ile & 97.5\%ile & &  Mean & Std Dev &  2.5\%ile & 25\%ile & Median & 75\%ile & 97.5\%ile\\
$\cal{CE}$$_2$ & 427.7 & 50.9 & 328.6 & 393.6 & 428.1 & 461.1 & 529.0 &  &  497.2 & 246.8 & 291.1 & 444.1 & 507.7 & 570.1 & 688.4\\
$E(r)$ & 587.9 & 54.4 & 486.4 & 550.1 & 586.1 & 623.1 & 700.9 & & 900.1 &  93.5 & 725.2 & 835.7 & 897.0 &  960.5 & 1,094.8\\
$\sigma$ & 1,254.8 & 88.6 & 1,090.0 & 1,193.7 & 1,251.6 & 1,311.1 & 1,441.2 & & 1,952.8 & 129.8 & 1,714.8 & 1,862.7 & 1,947.4 & 2,035.4 & 2,221.6\\
Median & 593.3 & 85.2 & 431.2 & 535.3 & 591.7 & 649.8 & 767.1 & & 878.5 & 136.0 & 617.7 &  786.6 & 877.4 &  966.9 & 1,153.3\\
IQR & 1,495.0 & 157.8 & 1,200.8 & 1,385.5 & 1,490.0 & 1,598.4 & 1,822.1 & & 2,463.4 & 247.0 & 1,995.5 & 2,294.0 & 2,456.5 & 2,627.1 & 2,961.3\\
MIN & -3,887.9 & 705.9 & -5,336.6 & -4,344.9 & -3,871.2 & -3,399.8 & -2,567.2 & & -5,145.8 & 1,160.3 & -7,575.9 & -5,901.6 & -5,063.5 &  -4,286.7 & -3,177.4\\
SKEW & -0.4 & 5.3 & -10.9 & -4.1 & -0.4 & 3.2 & 10.2 & & 1.1 & 5.7 &  -10.0 &  -2.7 &  1.0 & 4.9 & 12.5\\
KURT & 24.1 & 13.4 & -1.2 & 14.9 & 23.9 & 32.9 & 50.7 & & 11.8 & 14.2 & -15.2 & 1.9 & 11.4 & 21.2 & 40.8\\
SR & 1.5281 &  0.1383 &  1.2626 &  1.4353 &  1.5266 &  1.6191 &  1.8050 & & 1.5400 &  0.1391 &  1.2741 &  1.4451 &  1.5369 &  1.6333 &  1.8157\\
\hline

\end{tabular}

\end{landscape}

\newpage

\begin{landscape}

\voffset=-1.7in
\hoffset=-1.04in

\begin{center}  
     
\baselineskip 14pt
{\Large {\bf Table 3}}\\
{\bf Out-of-Sample 6-factor Fama-French regressions}\\
$
r_{i,t} - r_f = \alpha + \beta_1 \cdot (R_{m,t} - r_f) + \beta_2 \cdot {\rm HML} + \ \beta_3 \cdot {\rm SMB}  + \beta_4 \cdot {\rm MOM} + \beta_5 \cdot {\rm RMW} + \beta_6 \cdot {\rm CMA} + \epsilon_{i,t}
$\\
For power utility investor with coefficient of relative risk aversion, $\gamma = 2.$  Monthly returns; $\alpha$ in basis points per month.\\
\end{center}

\baselineskip 12pt

\vskip .1in

\hoffset=-1in

\begin{tabular}{lrrrrrrrcrrrrrrr}

\multicolumn{16}{l}{{\bf Panel A. Subperiod 1: 1990 - 1998}}\\

     & \multicolumn{7}{c}{Updating protocol} & & \multicolumn{7}{c}{Rolling protocol}\\
\cline{2-8} \cline{10-16}
Coefficient  & Mean & Std Dev &  2.5\%ile & 25\%ile & Median & 75\%ile & 97.5\%ile & &  Mean & Std Dev &  2.5\%ile & 25\%ile & Median & 75\%ile & 97.5\%ile\\
$\alpha$ / Orthog. & 263.75 &  54.97 & 156.53 & 226.66 & 263.04 & 299.71 & 371.25 &  & 387.20 &  94.63 & 209.65 & 323.21 & 385.11 & 450.49 & 577.66\\
Mkt &  -0.36 &   0.17 &  -0.69 &  -0.47 &  -0.36 &  -0.25 &  -0.04 &   & -0.51 & 0.27 &  -1.05 &  -0.69 &  -0.51 &  -0.33 &  -0.02\\
HML &  3.16 &   0.43 &  2.35 &  2.86 &  3.15 &   3.44 &   4.02 &    & 5.83 &   0.66 &  4.59 &  5.39 &  5.82 &  6.27 &  7.14\\
SMB &   1.68 &   0.28 &   1.12 &   1.48 &   1.68 &   1.87 &   2.23 &  & 0.78 &   0.43 &   -0.06 &   0.50 &   0.78 &   1.06 &   1.64\\
MOM &   2.62 &   0.25 &   2.14 &   2.44 &   2.61 &   2.78 &   3.13 &  & 3.80 &   0.39 &   3.04 &   3.53 &   3.80 &   4.06 &   4.58\\
RMW &   0.93 &   0.41 &  0.14 &   0.65 &   0.92 &   1.20 &   1.75 &  &  1.42 &   0.74 &  -0.03 &  0.93 &  1.41 &  1.92 &  2.89\\
CMA &  -1.67 &   0.53 &  -2.73 &  -2.01 &  -1.66 &  -1.30 &  -0.66 &  &  -4.08 &   0.86 &  -5.81 &  -4.65 &  -4.07 &  -3.50 &  -2.42\\
\hline

   &    &   &   &  &  & & & \hspace{.1in}  &  & & & &  &  &  \\

\multicolumn{16}{l}{{\bf Panel B. Subperiod 1: 1990 - 1998 Updating Protocol: Decompositions}}\\

     & \multicolumn{7}{c}{\% of Portfolio Mean due to:} & & \multicolumn{7}{c}{\% of Portfolio Variance due to:}\\
\cline{2-8} \cline{10-16}
Coefficient  & Mean & Std Dev &  2.5\%ile & 25\%ile & Median & 75\%ile & 97.5\%ile & &  Mean & Std Dev &  2.5\%ile & 25\%ile & Median & 75\%ile & 97.5\%ile\\
$\alpha$ / Orthog. & 47.86 &   7.07 &  32.94 &  43.41 &  48.22 &  52.74 &  60.76 & & 53.16 &   4.26 &  45.09 &  50.24 &  53.12 &  56.00 &  61.72\\
Mkt &  -6.64 &   3.15 & -13.04 &  -8.71 &  -6.56 &  -4.50 &  -0.65 & &  1.52 &   1.17 &   0.03 &   0.63 &   1.27 &   2.17 &   4.42\\
HML & 14.27 &   2.02 &  10.63 &  12.90 &  14.18 &  15.54 &  18.52 & & 37.75 &   7.77 &  23.31 &  32.39 &  37.46 &  42.86 &  53.65\\
SMB &   -7.90 &   1.62 & -11.23 &  -8.94 &  -7.85 &  -6.79 &  -4.90 & & 13.45 &   4.34 &   5.74 &  10.36 &  13.16 &  16.25 &  22.65\\
MOM &   48.11 &   5.57 &  38.05 &  44.32 &  47.83 &  51.57 &  59.98 & &  34.07 &   4.65 &  25.09 &  30.97 &  34.05 &  37.17 &  43.41\\
RMW &   7.35 &   3.29 &   1.10 &   5.12 &   7.26 &   9.49 &  14.02 & & 1.22 &   0.92 &   0.04 &   0.52 &   1.03 &   1.72 &   3.51\\
CMA &  -3.06 &   0.97 &  -5.01 &  -3.68 &  -3.05 &  -2.40 &  -1.23 & & 6.61 &   3.73 &   1.03 &   3.84 &   6.11 &   8.75 &  15.32\\
\hline

   &    &   &   &  &  & & & \hspace{.1in}  &  & & & &  &  &  \\

\multicolumn{16}{l}{{\bf Panel C. Subperiod 2: 1999 - 2021}}\\

     & \multicolumn{7}{c}{Updating protocol} & & \multicolumn{7}{c}{Rolling protocol}\\
\cline{2-8} \cline{10-16}
Coefficient  & Mean & Std Dev &  2.5\%ile & 25\%ile & Median & 75\%ile & 97.5\%ile & &  Mean & Std Dev &  2.5\%ile & 25\%ile & Median & 75\%ile & 97.5\%ile\\
$\alpha$ / Orthog. &  80.86 &  34.66 & 14.47 & 56.94 &  80.18 &  104.21 &  150.37 &  & 99.24 &  30.89 & 40.38 & 78.48 & 98.04 & 119.61 & 161.97\\
Mkt &  0.91 &   0.11 &   0.70 &   0.84 &   0.91 &   0.98 &   1.12 &   & 1.27 &   0.09 &   1.10 &   1.20 &   1.27 &   1.33 &   1.47\\
HML &   1.38 &   0.16 &   1.07 &   1.27 &   1.38 &   1.49 &   1.70 &    &  0.74 &   0.14 &   0.47 &   0.65 &   0.73 &   0.83 &   1.02\\
SMB &   0.62 &   0.22 &   0.18 &   0.47 &   0.62 &   0.77 &   1.04 &  & -1.14 &   0.26 &  -1.69 &  -1.31 &  -1.13 &  -0.95 &  -0.67\\
MOM &   1.01 &   0.15 &   0.71 &   0.91 &   1.01 &   1.12 &   1.32 &  & 0.50 &   0.14 &   0.23 &   0.41 &   0.50 &   0.59 &   0.78\\
RMW &   1.26 &   0.22 &   0.83 &   1.11 &   1.26 &   1.41 &   1.70 &  &   0.01 &   0.31 &  -0.59 &  -0.19 &   0.02 &   0.22 &   0.61\\
CMA &  -0.22 &   0.25 &  -0.70 &  -0.39 &  -0.22 &  -0.05 &  0.20 &  &  -0.13 &   0.25 &  -0.62 &  -0.29 &  -0.12 &  0.04 &  0.35\\
\hline

\end{tabular}

\newpage

\begin{center}

\baselineskip 14pt
{\Large {\bf Table 4}}\\
{\bf Sampling properties of out-of-sample Portfolio Performance Statistics}\\
{\bf 276-month out-of-sample period, 1999 -- 2021}\\
\end{center}

\vskip .1in

\begin{tabular}{lrrrrrrrcrrrrrrr}

\multicolumn{16}{l}{Sampling properties of dynamic optimal PPPs.  Portfolio characteristic tilts from the best out-of-sampling performer over the relevant preceding period}\\
\multicolumn{16}{l}{(shown in Table 1) each year.  $\cal{CE}$$_2$ is the certainty equivalent return in basis points per month for a power utility investor with coefficient of relative}\\
\multicolumn{16}{l}{risk aversion $(\gamma) = 2$.  $E(r)$, $\sigma$, Median, IQR, and MIN are the mean monthly return, the
standard deviation of monthly returns, the median}\\
\multicolumn{16}{l}{monthly return, the interquartile range of monthly returns, and the minimum monthly return--all expressed in basis points per month.}\\
\multicolumn{16}{l}{SKEW and KURT are the return skewness and kurtosis measures, and SR is the Sharpe ratio.}\\
\multicolumn{16}{l}{Results are for the second 23-year out-of-sample subperiod (1999 -- 2021).}\\

    &    &   &   &  &  & & & \hspace{.1in}  &  & & & &  &  &  \\

\multicolumn{16}{l}{{\bf Panel A: Benchmark portfolios}}\\
     & \multicolumn{7}{c}{VWI} & & \multicolumn{7}{c}{EWI}\\
\cline{2-8} \cline{10-16}
Statistic  & Mean & Std Dev &  2.5\%ile & 25\%ile & Median & 75\%ile & 97.5\%ile & &  Mean & Std Dev &  2.5\%ile & 25\%ile & Median & 75\%ile & 97.5\%ile\\
$\cal{CE}$$_2$ & 60.0 & 3.5 & 53.1 & 57.6 & 60.0 & 62.4 & 66.9 & & 76.6 & 1.7 & 73.4 & 75.4 & 76.6 & 77.7 & 79.8\\
$E(r)$ & 79.2 & 3.5 & 72.4 & 76.8 & 79.2 & 81.6 & 85.9 & & 106.8 & 1.7 & 103.6 & 105.6 & 106.8 & 107.9 & 110.0\\
$\sigma$ & 433.3 & 4.2 & 425.2 & 430.4 & 433.3 & 436.3 & 441.6 & & 542.7 & 2.1 & 538.6 & 541.3 & 542.7 & 544.1 & 546.8\\
Median & 124.0 & 8.0 & 107.9 & 118.6 & 124.1 & 129.4 & 139.4 & & 141.9 & 6.7 & 128.8 & 137.4 & 141.9 & 146.4 & 155.3\\
IQR & 511.2 & 13.9 & 484.6 & 501.8 & 510.9 & 520.6 & 538.8 & & 655.0 & 12.1 & 612.0 & 626.8 &  635.0 &     642.9 &     659.3\\
MIN & -1,667.4 & 78.6 & -1,819.4 & -1,720.6 & -1,667.6 & -1,613.6 & -1,514.3 & & -2119.7 & 42.4 & -2,206.5 & -2,147.9 & -2,118.6 & -2,090.3 & -2,039.7\\
SKEW & -10.3 & 1.8 & -13.9 & -11.5 & -10.4 & -9.1 &  -6.8 & & -6.5 & 1.2 & -8.9 &  -7.3 & -6.5 & -5.6 & -4.1\\
KURT & 27.9 & 3.8 & 20.6 & 25.3 &  27.9 & 30.4 & 35.4 & & 32.5 & 1.5 & 29.5 & 31.5 & 32.5 & 33.5 & 35.5\\
SR & 0.5245 &  0.0286 &  0.4687 &  0.5052 &  0.5246 &  0.5436 &  0.5807 & & 0.5959 &  0.0105 &  0.5756 &  0.5887 &  0.5957 &  0.6029 &  0.6165\\
\hline

   &    &   &   &  &  & & & \hspace{.1in}  &  & & & &  &  &  \\

\multicolumn{16}{l}{{\bf Panel B: Dynamic PPP}}\\
     & \multicolumn{7}{c}{Dyn. Opt. Updating} & & \multicolumn{7}{c}{Dyn. Opt. Rolling}\\
\cline{2-8} \cline{10-16}
Statistic  & Mean & Std Dev &  2.5\%ile & 25\%ile & Median & 75\%ile & 97.5\%ile & &  Mean & Std Dev &  2.5\%ile & 25\%ile & Median & 75\%ile & 97.5\%ile\\
$\cal{CE}$$_2$ & -1.5 & 172.3 & -110.2 & -24.9 & 5.9 & 33.5 & 82.9 &  & -5,017.2 & 4,7619.8 & -10,000 & -10,000 & -2,099.6 & -222.1 & 4.1\\
$E(r)$ & 242.7 & 32.8 & 180.1 & 220.1 & 242.2 & 264.3 & 309.4 & & 175.2 & 28.1 & 121.4 & 156.3 & 174.7 & 193.0 & 232.9\\
$\sigma$ & 1,472.3 & 72.6 & 1,339.3 & 1,421.4 & 1,469.2 & 1,519.9 & 1,622.7 & & 1,370.2 & 144.2 & 1,120.7 & 1,269.2 & 1,356.7 & 1,459.4 & 1,685.5\\
Median & 233.0 & 52.2 & 133.8 & 196.7 & 232.7 & 267.9 & 337.9 & & 149.2 & 22.8 & 107.0 & 132.9 & 148.5 & 164.9 & 194.5\\
IQR & 1,718.9 & 113.3 & 1,510.8 & 1,639.5 & 1,714.3 & 1,792.7 & 1,954.6 & & 748.9 & 37.6 & 680.4 & 722.2 & 747.6 & 773.7 & 826.2\\
MIN & -5,490.7 & 933.4 & -7,756.8 & -5,958.2 & -5,352.3 & -4,856.6 & -4,072.5 & & -10,082.7 & 1,862.8 & -14,309.7 & -11,127.4 & -9,861.2 & -8,787.2 & -7,071.3\\
SKEW & 0.6 & 3.2 &  -5.7 & -1.5 & 0.7 & 2.8 & 6.8 & & 1.9 & 2.1 & -2.3 & 0.5 & 1.9 & 3.3 & 6.0\\
KURT & 40.2 & 11.5 & 18.6 & 32.3 & 39.7 & 47.7 & 63.2 & & 209.8 &   23.5 & 164.1 & 193.6 & 209.6 &  225.4 & 256.9\\
SR &  0.5415 &  0.0750 &  0.3971 &  0.4895 &  0.5414 &  0.5931 &  0.6902 & & 0.4130 &  0.0726 &  0.2717 &  0.3641 &  0.4127 &  0.4617 &  0.5566\\
\hline

\end{tabular}

\end{landscape}


\voffset=-.9in
\hoffset=-0.4in
\renewcommand{\textwidth}{6.25 in}

\interfootnotelinepenalty=10000

\setlength{\intextsep}{0pt plus 2pt}
\setlength{\abovecaptionskip}{-28pt}

\captionsetup{justification=raggedright,singlelinecheck=false}

\makeatletter
\def\normaljustify{%
  \let\\\@centercr\rightskip\z@skip \leftskip\z@skip%
  \parfillskip=0pt plus 1fil}
\makeatother

\textheight 9.2in
\flushbottom

\captionsetup[figure]{labelformat=empty, labelsep=period}

\restylefloat{table}
\restylefloat{figure}

\pagestyle{empty}

\long\def\symbolfootnote[#1]#2{\begingroup%
\def\thefootnote{\fnsymbol{footnote}}\footnote[#1]{#2}\endgroup}

\hfuzz 50pt

\begin{center}

\vskip .4cm

{\bf Internet Appendix} {\em for}\\
{\bf An Empirical Assessment of Characteristics and Optimal Portfolios}
\end{center}
\baselineskip 15pt

\parindent 18pt

\vskip .1in

\normalsize

\noindent
This appendix provides supporting results for the paper An Empirical Assessment of Characteristics and Optimal Portfolios.\\
Contents:
\begin{itemize}
   \item {\bf Figure IA-1}: Year-by year $\theta$ coefficent on momentum for all 3 investors.
   \item {\bf Figure IA-2}: Year-by year $\theta$ coefficent on the book-to-market ratio for all 3 investors.
   \item {\bf Figure IA-3}: Year-by year $\theta$ coefficent on log size for all 3 investors.
   \item {\bf Figure IA-4}: Year-by year $\theta$ coefficent on beta for all 3 investors.
   \item {\bf Figure IA-5}: Year-by year $\theta$ coefficent on residual volatility for all 3 investors.
   \item {\bf Figure IA-6}: Year-by year $\theta$ coefficent on 5-year average same-month return for all 3 investors.
   \item {\bf Figure IA-7}: Benchmark and PPP Return densities for the power utility investor with $\gamma = 5$ in both subperiods.
   \item {\bf Figure IA-8}: Benchmark and PPP Return densities for the power utility investor with $\gamma = 8$ in both subperiods.
   \item {\bf Table IA-1}: Month-by-month sample construction and profile.
   \item {\bf Table IA-2}: Optimal configuration prior to each year in the out-of-sample period for the power utility investor with $\gamma = 5$.
   \item {\bf Table IA-3}: Optimal configuration prior to each year in the out-of-sample period for the power utility investor with $\gamma = 8$.
   \item {\bf Table IA-4}: Sampling properties of out-of-sample benchmark and PPP returns in the first subperiod for both power utility investors with $\gamma = 5$ 
         and $\gamma = 8$.
   \item {\bf Table IA-5}: Sampling properties of out-of-sample benchmark and PPP returns in the second subperiod for both power utility investors with $\gamma = 5$ 
         and $\gamma = 8$.
   \item {\bf Table IA-6}: Sampling properties of out-of-sample benchmark and PPP returns in the entire out-of-sample period for all three power utility investors with $\gamma = 2$,
         $\gamma = 5$, and $\gamma = 8$.
   \item {\bf Table IA-7}: Six-factor Fama-French regressions for the power utility investor with $\gamma = 5$, in the first and second subperiods as well as the
         entire out-of-sample period.
   \item {\bf Table IA-8}: Six-factor Fama-French regressions for the power utility investor with $\gamma = 8$, in the first and second subperiods as well as the
         entire out-of-sample period.
   \item {\bf Table IA-9}: Six-factor Fama-French regression for the power utility investor with $\gamma = 2$, in the full 32-year out-of-sample period.
   \item  {\bf Table IA-10}: Six-factor Fama-French regression decompositions for the power utility investors with $\gamma = 5$ and $\gamma = 8$, in the first subperiod.
\end{itemize}

\newpage

\setcounter{footnote}{0}

\baselineskip 19pt

\parindent 18pt

\setcounter{page}{1} 

\hoffset -.2in
\parindent -22pt
\parskip .1in

\setlength{\intextsep}{0pt plus 2pt}
\setlength{\abovecaptionskip}{8pt}

\flushbottom

\captionsetup{width=\textwidth}

\restylefloat{figure}

\hoffset=-1.5in

\long\def\symbolfootnote[#1]#2{\begingroup%
\def\thefootnote{\fnsymbol{footnote}}\footnote[#1]{#2}\endgroup}

\begin{figure}[H]
\includegraphics*[scale=1.15]{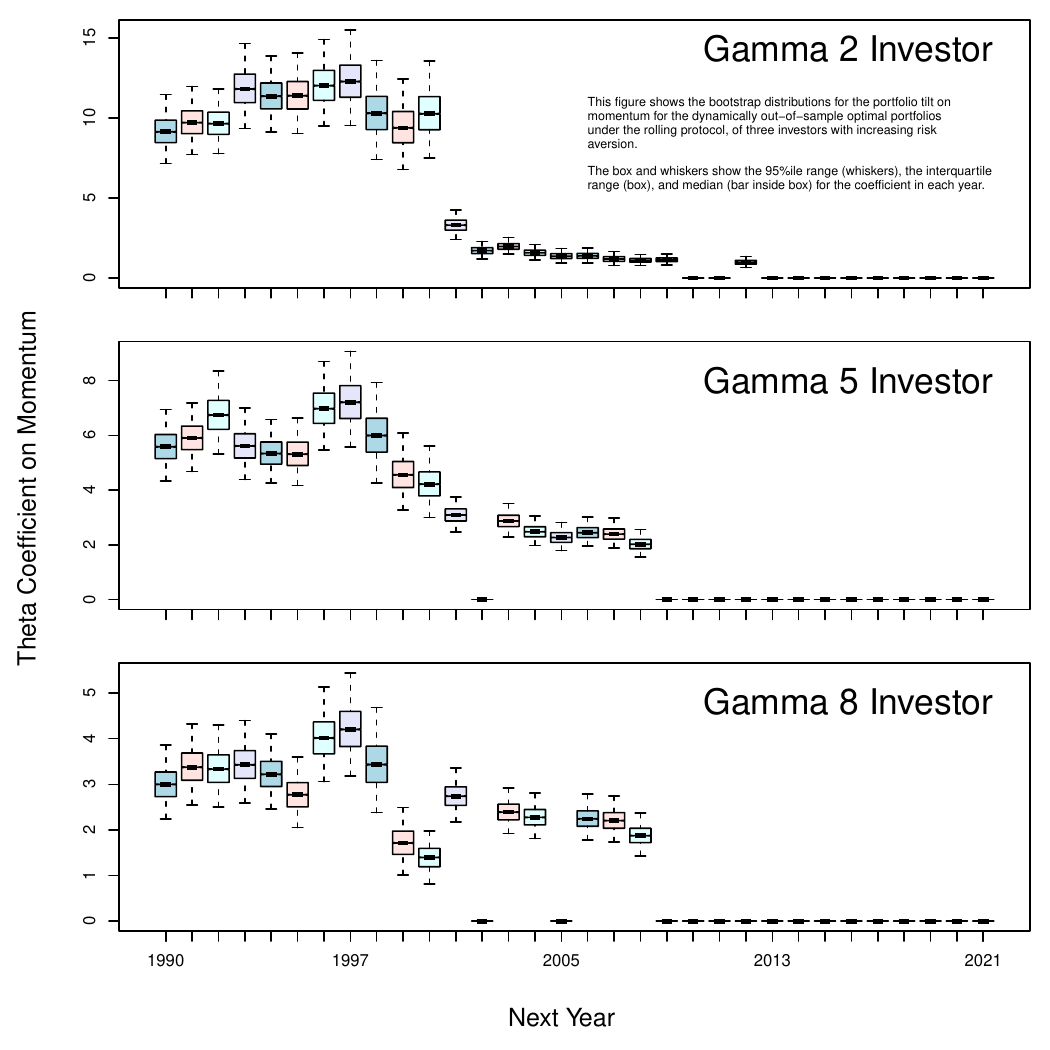}

\caption{{\bf Figure IA-1.  Momentum tilt.}
Sampling distributions of the $\theta$ coefficient on (standardized) momentum from the optimal model over the preceding 180 months--out-of-sample, used to
construct the optimal portfolio in the indicated year.}

\end{figure}

\newpage

\begin{figure}[H]
\includegraphics*[scale=1.15]{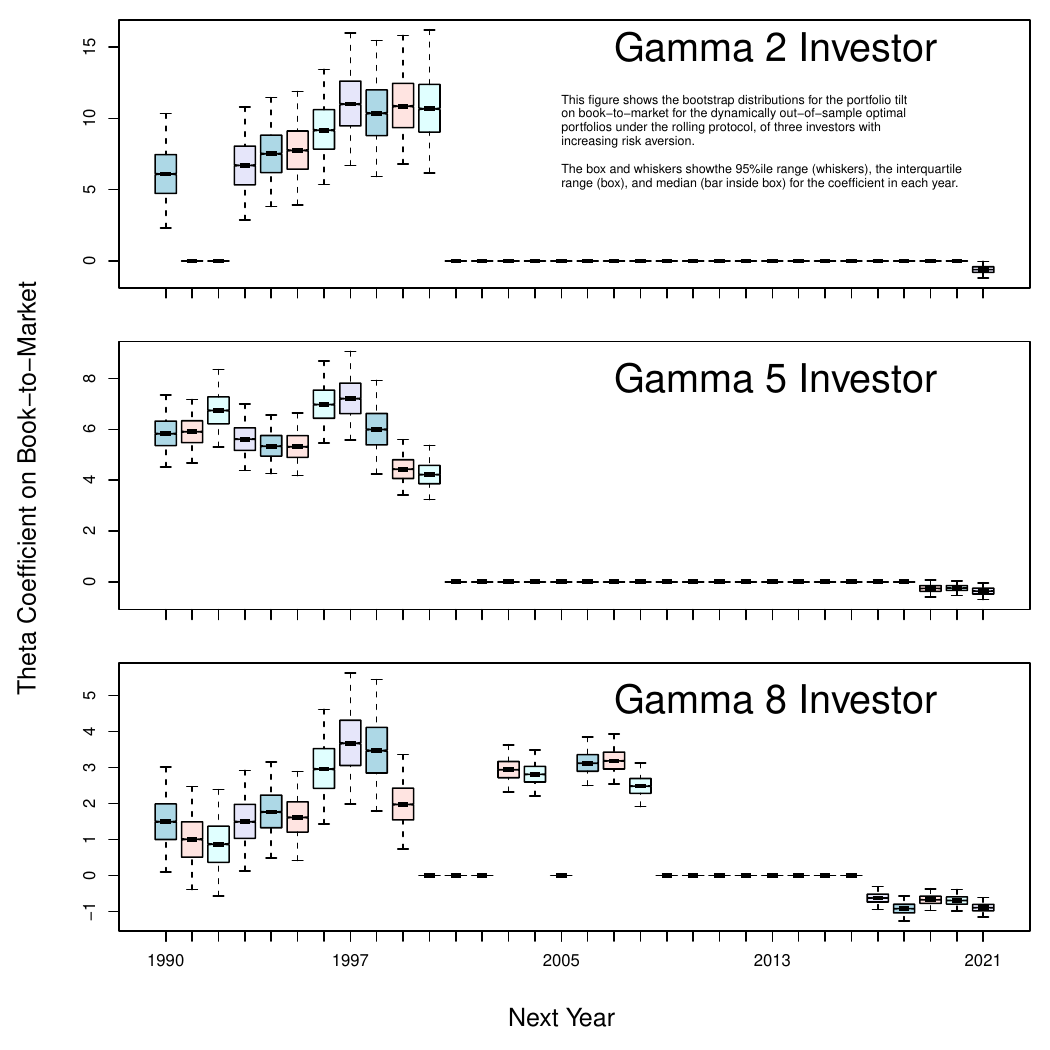}

\caption{{\bf Figure IA-2.  Value tilt.}
Sampling distributions of the $\theta$ coefficient on the (standardized) book-to-market ratio from the optimal model over the preceding 180 months--out-of-sample, used to
construct the optimal portfolio in the indicated year.}

\end{figure}

\newpage

\begin{figure}[H]
\includegraphics*[scale=1.15]{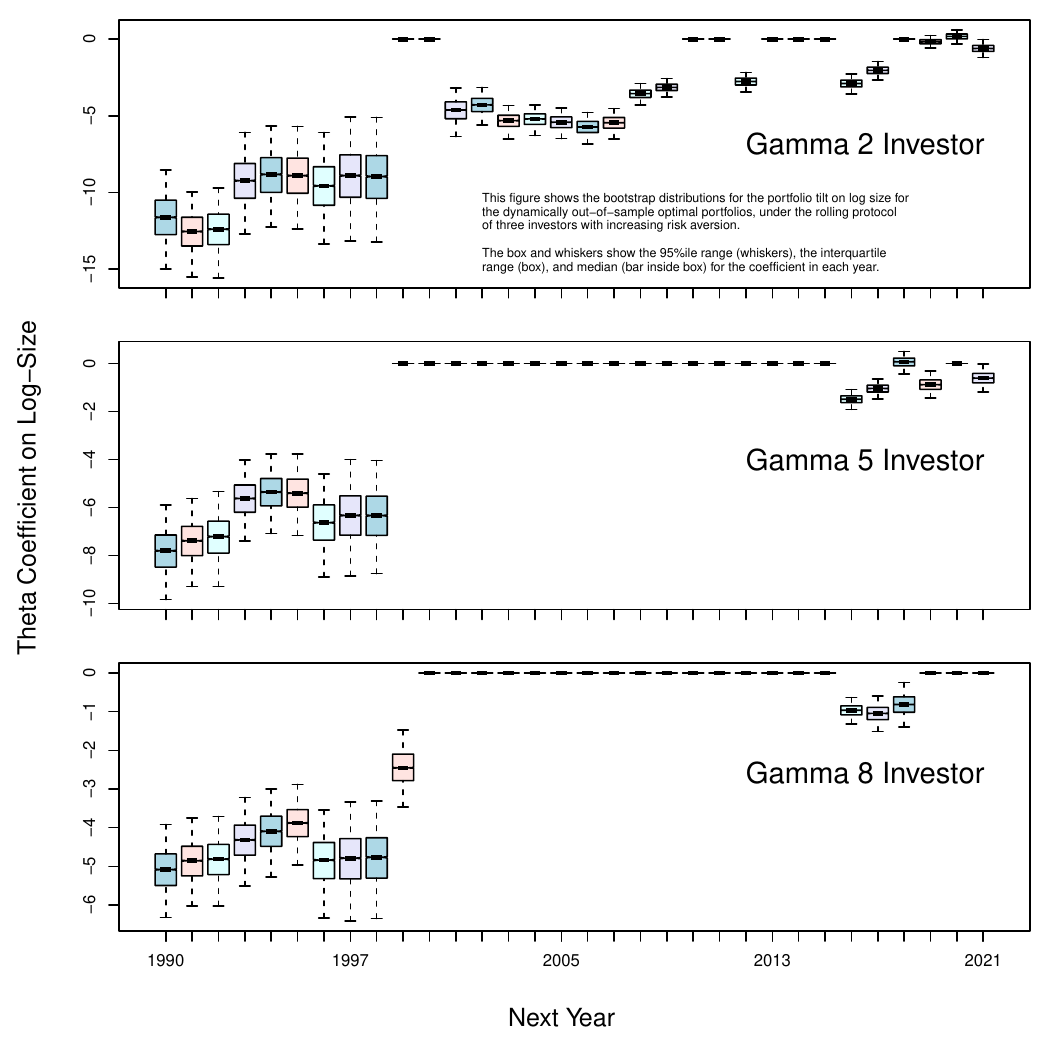}

\caption{{\bf Figure IA-3.  Size tilt.}
Sampling distributions of the $\theta$ coefficient on (standardized) log market capitalization from the optimal model over the preceding 180 months--out-of-sample, used to
construct the optimal portfolio in the indicated year.}

\end{figure}

\newpage

\begin{figure}[H]
\includegraphics*[scale=1.15]{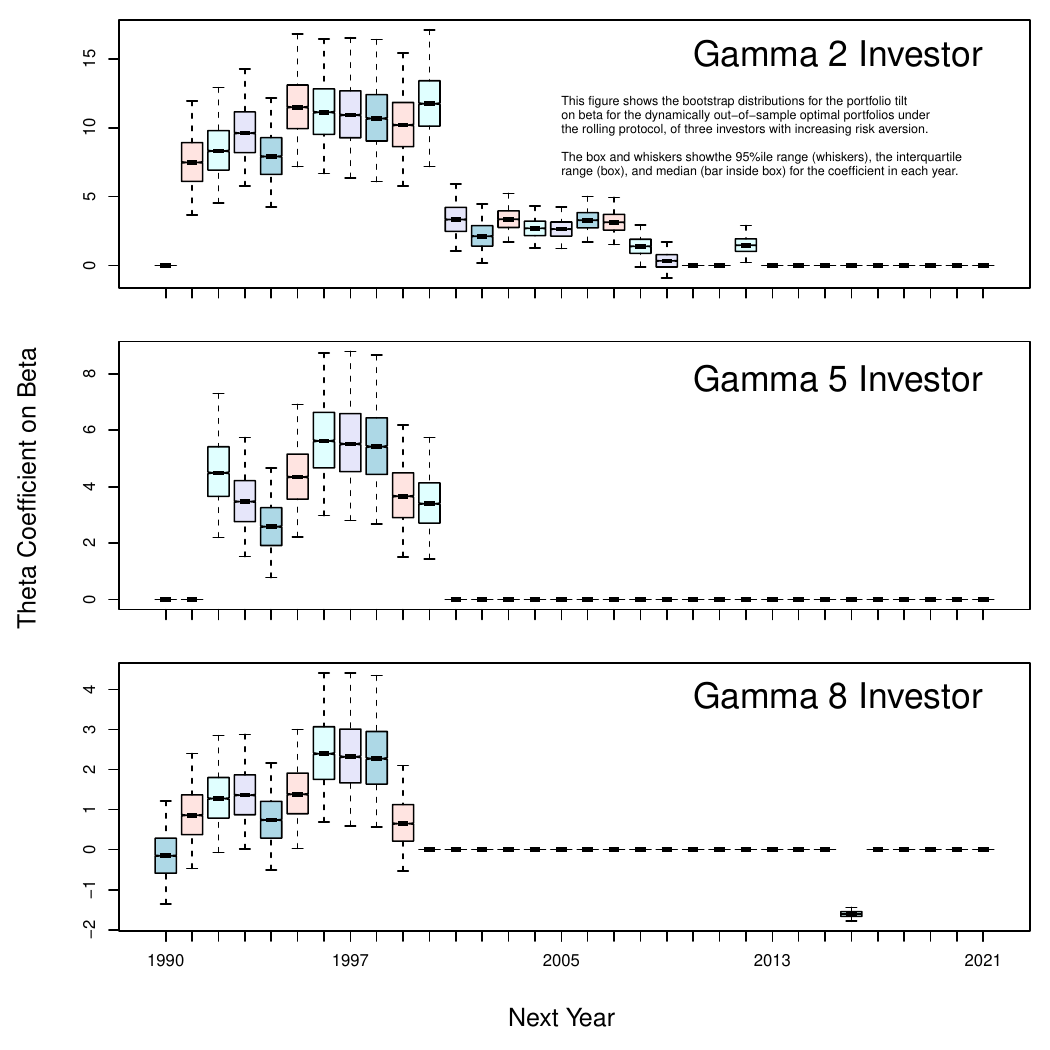}

\caption{{\bf Figure IA-4.  Beta tilt.}
Sampling distributions of the $\theta$ coefficient on (standardized) beta from the optimal model over the preceding 180 months--out-of-sample, used to
construct the optimal portfolio in the indicated year.}

\end{figure}

\newpage

\begin{figure}[H]
\includegraphics*[scale=1.15]{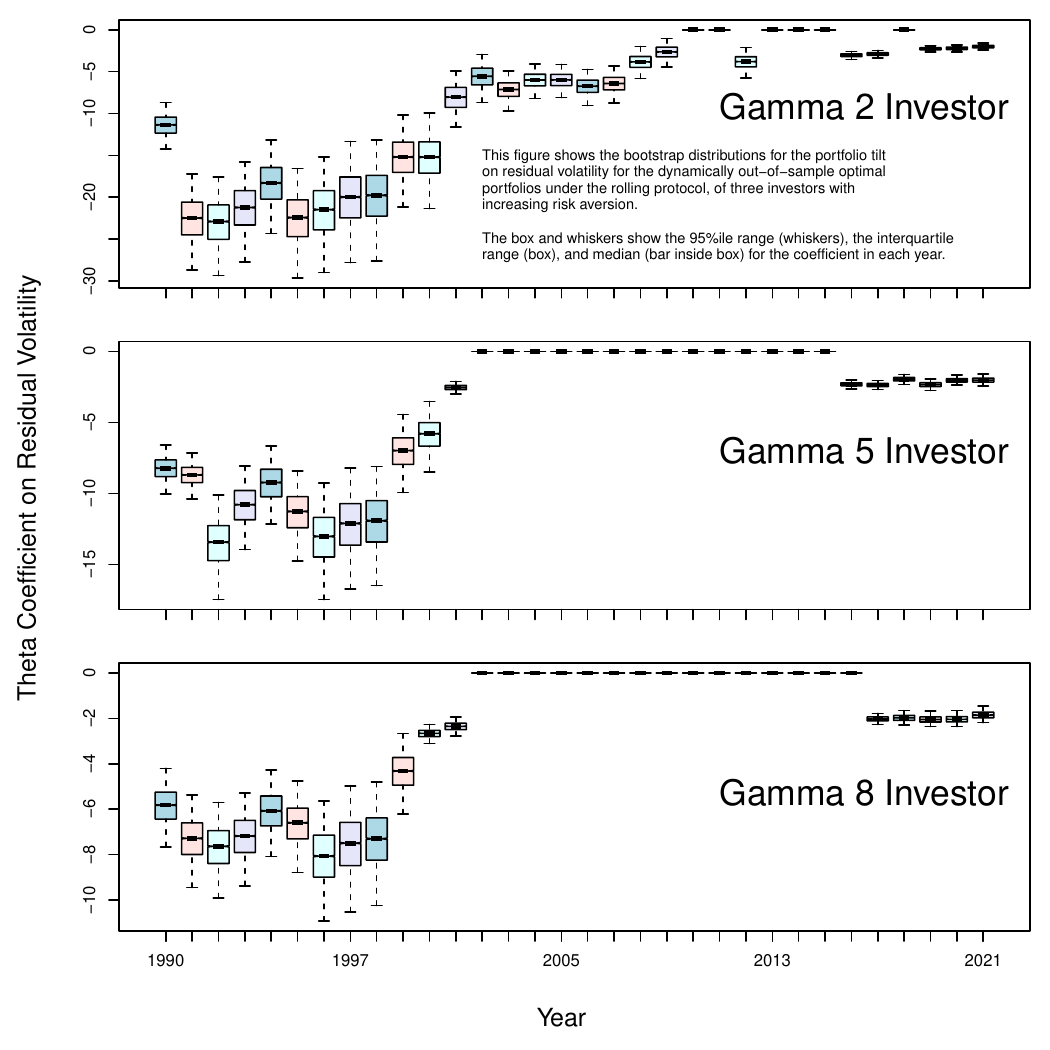}

\caption{{\bf Figure IA-5.  Residual volatility tilt.}
Sampling distributions of the $\theta$ coefficient on (standardized) residual standardized deviation from the optimal model over the preceding 180 months--out-of-sample, used to
construct the optimal portfolio in the indicated year.}

\end{figure}

\newpage

\begin{figure}[H]
\includegraphics*[scale=1.15]{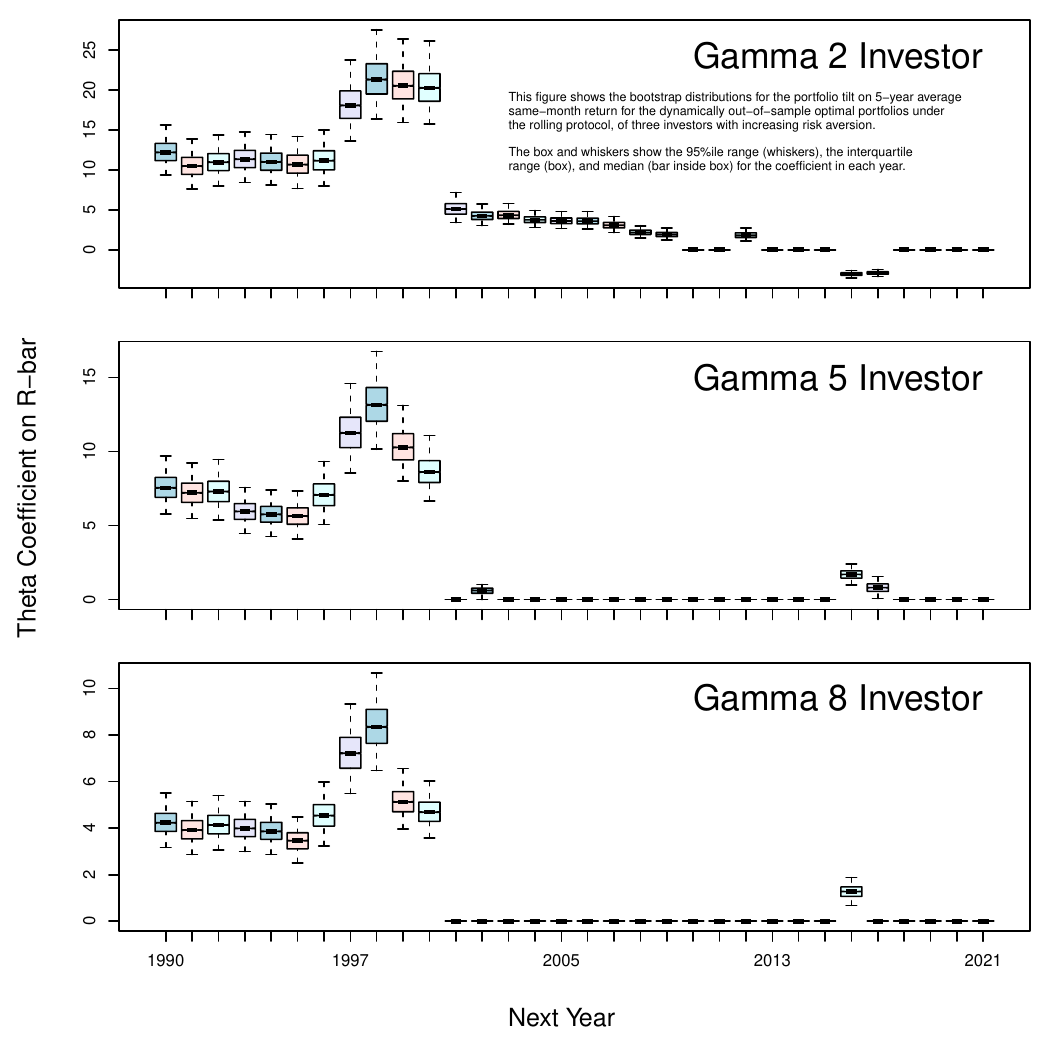}

\caption{{\bf Figure IA-6.  Average same-month return tilt.}
Sampling distributions of the $\theta$ coefficient on the (standardized) average same-month return from the optimal model over the preceding 180 months--out-of-sample, used to
construct the optimal portfolio in the indicated year.}

\end{figure}


\renewcommand{\textwidth}{7.75 in}
\textheight 10.8in

\begin{landscape}

\voffset -1.9in

\newpage

\begin{figure}[H]
\includegraphics*[scale=0.915]{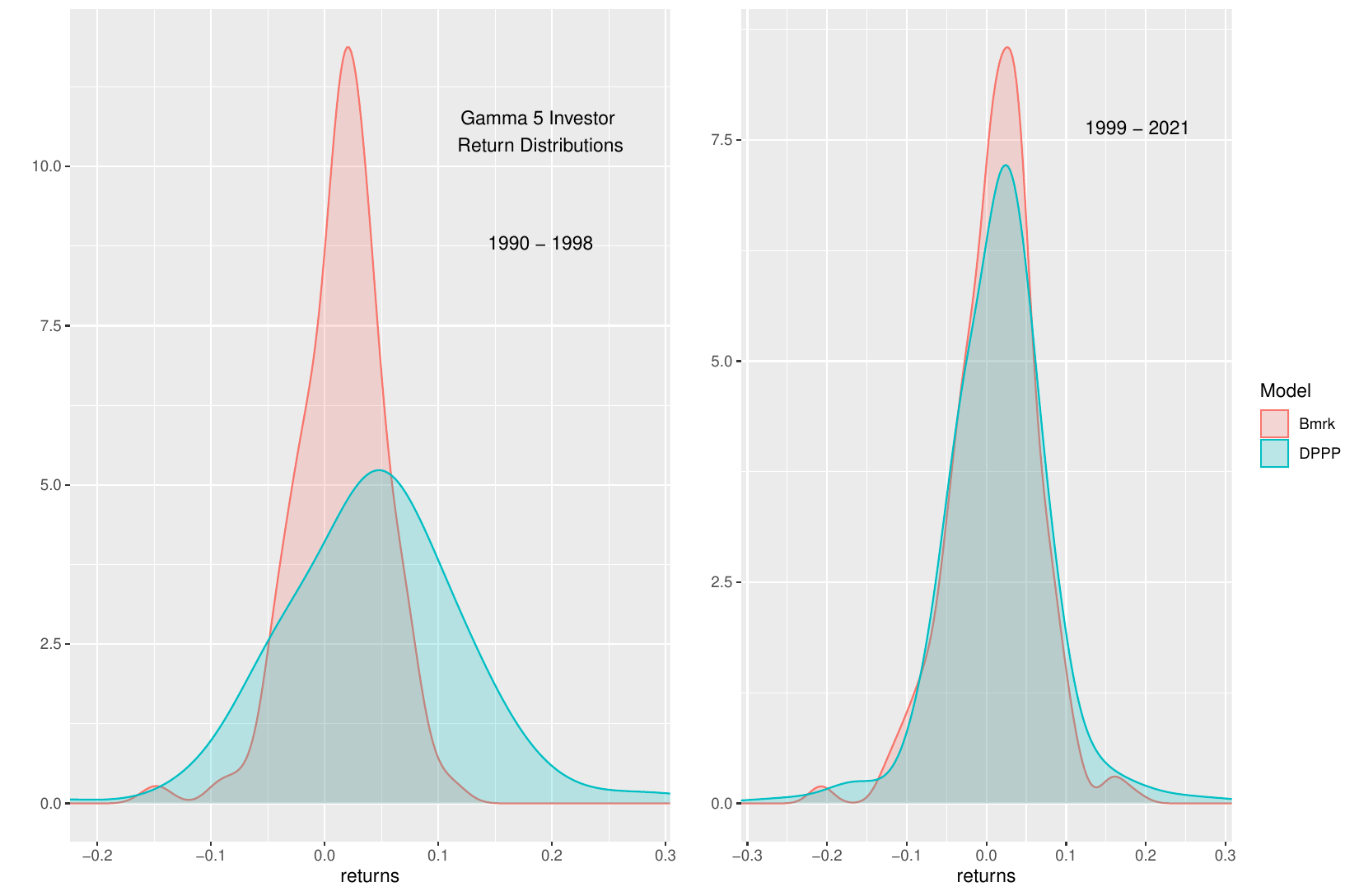}

\captionsetup{width=19.5cm}
{\bf Figure IA-7.  Portfolio return densities in the 2 subperiods.}
``DPPP'' is the optimal dynamic parametric portfolio under the\\ updating protocol--selected at the beginning of
each year.  ``Bmrk'' is the preferred benchmark in the subperiod.  For this power\\ utility investor with coefficient of relative risk aversion,
$\gamma$ $= \; 5$: the value-weighted portfolio of all stocks in
the first subperiod\\ and the equally-weighted portfolio of all stocks in the second subperiod.

\end{figure}

\newpage

\begin{figure}[H]
\includegraphics*[scale=0.915]{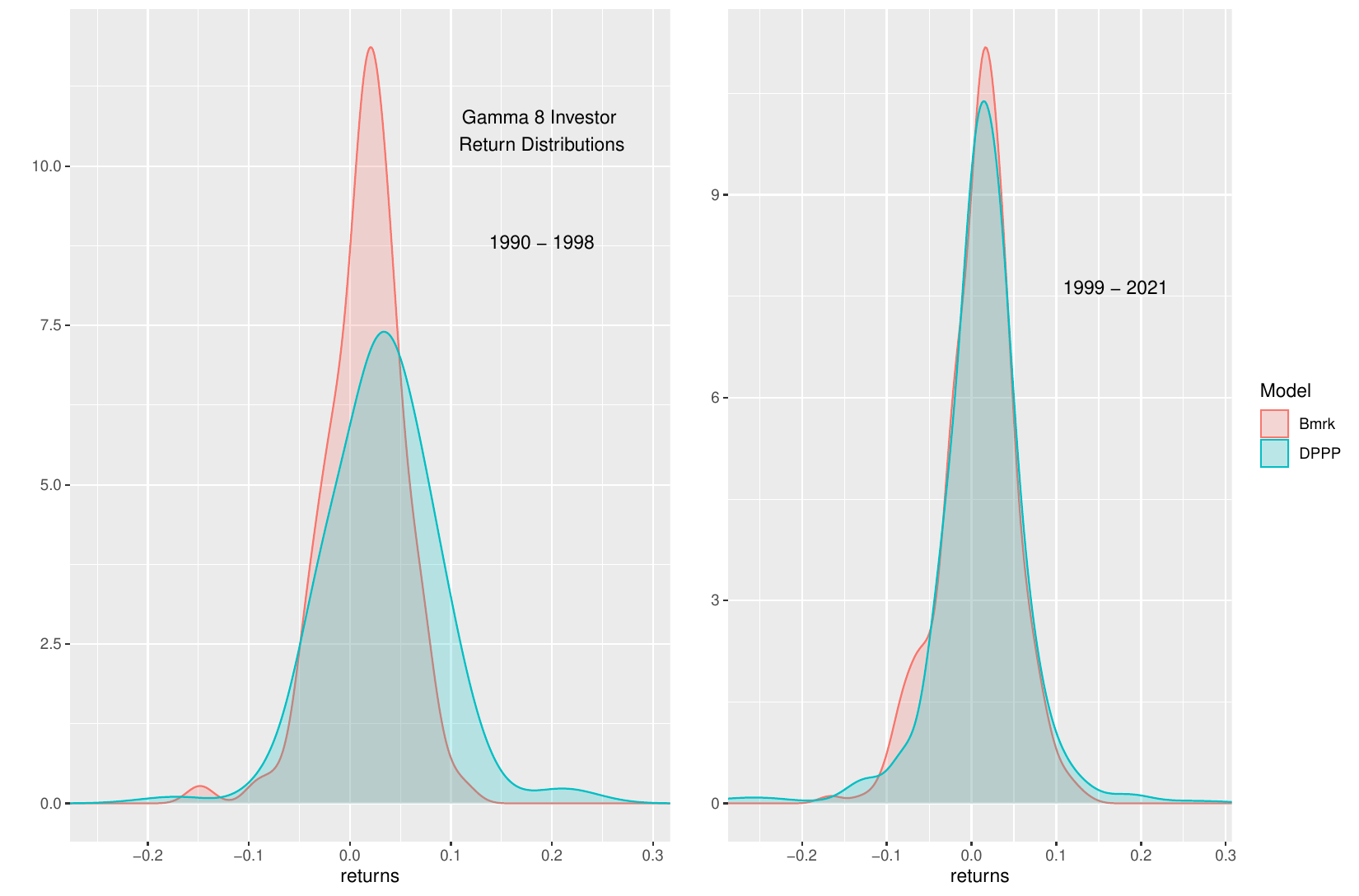}

\captionsetup{width=19.5cm}
{\bf Figure IA-8.  Portfolio return densities in the 2 subperiods.}
``DPPP'' is the optimal dynamic parametric portfolio under the\\ updating protocol--selected at the beginning of
each year.  ``Bmrk'' is the preferred benchmark in the subperiod.  For this power\\ utility investor with coefficient of relative risk aversion,
$\gamma$ $= \; 8$: the value-weighted portfolio of all stocks in both subperiods.

\end{figure}

\end{landscape}

\voffset=-1.6in
\hoffset=-1.2in

\renewcommand{\textwidth}{7.35 in}

\captionsetup{width=\textwidth}

\interfootnotelinepenalty=100

\setlength{\intextsep}{0pt plus 2pt}
\setlength{\abovecaptionskip}{-28pt}

\makeatletter
\def\normaljustify{%
  \let\\\@centercr\rightskip\z@skip \leftskip\z@skip%
  \parfillskip=0pt plus 1fil}
\makeatother

\newcolumntype{C}[1]{>{\Centering}m{#1}}
\renewcommand\tabularxcolumn[1]{C{#1}}

\flushbottom

\captionsetup[figure]{labelformat=empty, labelsep=period}

\restylefloat{table}
\restylefloat{figure}

\long\def\symbolfootnote[#1]#2{\begingroup%
\def\thefootnote{\fnsymbol{footnote}}\footnote[#1]{#2}\endgroup}

\voffset=-1.6in
\hoffset=-0.6in
\renewcommand{\textwidth}{7.35 in}
\textheight 10.5in

\setlength{\intextsep}{0pt plus 2pt}

\setlength{\intextsep}{0pt plus 2pt}
\setlength{\abovecaptionskip}{-200pt}

\restylefloat{table}
\restylefloat{figure}

\textheight 10.5in

\newpage

\voffset=-1in
\hoffset=-0.5in

\baselineskip 14pt

\begin{center}           
{\Large {\bf Table IA-1}}\\
{\bf Sample Construction}\\
\end{center}

\baselineskip 12pt
\noindent
Stocks must have 60 months non-missing data on CRSP and a book value in CRSP/Compustat in months $[-18, \; -6]$ to be eligible.  Next a real dollar minimum is applied
(\$110 million in December 2021).  Finally, the smallest 10\% of stocks are removed prior to Nasdaq eligibility, and 20\% after Nasdaq stocks enter the sample (after December,
1977).  This table shows the effect of each filter on the sample size, as well as the smallest and median sized stocks eligible for the ensuing month in \$ millions.

\vskip .1in

\hoffset=-0.85in

\begin{center}

\begin{tabular}{lrrrrr}

      & Eligible & \multicolumn{1}{c}{After} & \multicolumn{1}{c}{Final} & \multicolumn{1}{c}{Minimum} & \multicolumn{1}{c}{Median}\\
 \multicolumn{1}{c}{Month End}      & \multicolumn{1}{c}{Stocks}   & \$ criterion & Sample & Mkt. Cap. & Mkt. Cap. \\
\hline
December-1959	&	 469 	&	 456 	&	 411 	&	26,277	&	198,482\\
January-1960	&	 472 	&	 457 	&	 412 	&	26,862	&	181,843\\
February-1960	&	 473 	&	 456 	&	 411 	&	26,738	&	186,863\\
March-1960	&	 473 	&	 454 	&	 409 	&	26,623	&	179,322\\
April-1960	&	 473 	&	 455 	&	 410 	&	25,752	&	177,732\\
May-1960	&	 476 	&	 460 	&	 414 	&	25,538	&	178,108\\
June-1960	&	 476 	&	 460 	&	 414 	&	26,988	&	180,770\\
July-1960	&	 478 	&	 459 	&	 414 	&	25,820	&	177,441\\
August-1960	&	 478 	&	 462 	&	 416 	&	25,698	&	178,343\\
September-1960	&	 480 	&	 462 	&	 416 	&	24,024	&	168,739\\
October-1960	&	 481 	&	 459 	&	 414 	&	23,952	&	168,416\\
November-1960	&	 484 	&	 463 	&	 417 	&	23,250	&	175,208\\
December-1960	&	 487 	&	 464 	&	 418 	&	25,082	&	187,619\\
January-1961	&	 491 	&	 471 	&	 424 	&	25,798	&	195,168\\
February-1961	&	 495 	&	 474 	&	 427 	&	26,802	&	209,228\\
March-1961	&	 497 	&	 481 	&	 433 	&	27,547	&	207,533\\
April-1961	&	 499 	&	 481 	&	 433 	&	28,017	&	203,797\\
May-1961	&	 534 	&	 503 	&	 453 	&	25,183	&	200,238\\
June-1961	&	 535 	&	 503 	&	 453 	&	24,541	&	194,714\\
July-1961	&	 536 	&	 504 	&	 454 	&	24,544	&	198,551\\
August-1961	&	 536 	&	 502 	&	 452 	&	25,523	&	201,934\\
September-1961	&	 537 	&	 499 	&	 450 	&	25,420	&	200,271\\
October-1961	&	 538 	&	 499 	&	 450 	&	24,567	&	208,202\\
November-1961	&	 539 	&	 505 	&	 455 	&	25,394	&	203,906\\
December-1961	&	 542 	&	 509 	&	 459 	&	25,628	&	209,549\\
January-1962	&	 542 	&	 510 	&	 459 	&	25,846	&	202,489\\
February-1962	&	 542 	&	 513 	&	 462 	&	24,956	&	198,553\\
March-1962	&	 542 	&	 510 	&	 459 	&	25,344	&	204,148\\
April-1962	&	 542 	&	 504 	&	 454 	&	24,854	&	194,457\\
May-1962	&	 547 	&	 501 	&	 451 	&	23,209	&	179,199\\
June-1962	&	 544 	&	 496 	&	 447 	&	22,813	&	166,498\\
July-1962	&	 546 	&	 504 	&	 454 	&	23,391	&	168,964\\
August-1962	&	 550 	&	 506 	&	 456 	&	24,750	&	174,250\\
September-1962	&	 550 	&	 500 	&	 450 	&	25,253	&	166,191\\
October-1962	&	 552 	&	 497 	&	 448 	&	25,424	&	168,204\\
November-1962	&	 558 	&	 513 	&	 462 	&	25,970	&	179,241\\
December-1962	&	 562 	&	 515 	&	 464 	&	24,917	&	179,326\\
\hline

\end{tabular}

\end{center}

\newpage

\begin{center}
{\Large {\bf Table IA-1}}\\
{\bf Sample Construction}\\
\end{center}

\baselineskip 12pt       
\noindent
Stocks must have 60 months non-missing data on CRSP and a book value in CRSP/Compustat in months $[-18, \; -6]$ to be eligible.  Next a real dollar minimum is applied
(\$110 million in December 2021).  Finally, the smallest 10\% of stocks are removed prior to Nasdaq eligibility, and 20\% after Nasdaq stocks enter the sample (after December,
1977).  This table shows the effect of each filter on the sample size as well as the smallest and median sized stocks eligible for the ensuing month in \$ millions.

\vskip .1in     

\hoffset=-0.85in

\begin{center}

\begin{tabular}{lrrrrr}  

      & Eligible & \multicolumn{1}{c}{After} & \multicolumn{1}{c}{Final} & \multicolumn{1}{c}{Minimum} & \multicolumn{1}{c}{Median}\\
 \multicolumn{1}{c}{Month End}      & \multicolumn{1}{c}{Stocks}   & \$ criterion & Sample & Mkt. Cap. & Mkt. Cap. \\
\hline
January-1963	&	 565 	&	 525 	&	 473 	&	24,011	&	180,719\\
February-1963	&	 571 	&	 531 	&	 478 	&	25,234	&	174,471\\
March-1963	&	 575 	&	 535 	&	 482 	&	25,051	&	186,411\\
April-1963	&	 575 	&	 536 	&	 483 	&	25,278	&	191,703\\
May-1963	&	 615 	&	 578 	&	 521 	&	27,393	&	185,043\\
June-1963	&	 617 	&	 578 	&	 521 	&	26,202	&	184,613\\
July-1963	&	 617 	&	 576 	&	 519 	&	26,838	&	186,339\\
August-1963	&	 619 	&	 583 	&	 525 	&	25,667	&	187,841\\
September-1963	&	 620 	&	 579 	&	 522 	&	27,246	&	190,534\\
October-1963	&	 620 	&	 582 	&	 524 	&	27,743	&	184,228\\
November-1963	&	 620 	&	 580 	&	 522 	&	28,196	&	186,675\\
December-1963	&	 622 	&	 580 	&	 522 	&	28,130	&	195,023\\
January-1964	&	 624 	&	 584 	&	 526 	&	28,123	&	192,954\\
February-1964	&	 624 	&	 585 	&	 527 	&	28,013	&	194,858\\
March-1964	&	 627 	&	 593 	&	 534 	&	26,877	&	193,158\\
April-1964	&	 630 	&	 593 	&	 534 	&	28,518	&	193,054\\
May-1964	&	 640 	&	 600 	&	 540 	&	28,928	&	196,121\\
June-1964	&	 643 	&	 604 	&	 544 	&	28,013	&	195,869\\
July-1964	&	 645 	&	 608 	&	 548 	&	28,187	&	195,636\\
August-1964	&	 643 	&	 604 	&	 544 	&	29,715	&	192,946\\
September-1964	&	 645 	&	 608 	&	 548 	&	29,417	&	204,829\\
October-1964	&	 647 	&	 610 	&	 549 	&	31,185	&	207,468\\
November-1964	&	 652 	&	 616 	&	 555 	&	29,464	&	208,542\\
December-1964	&	 653 	&	 618 	&	 557 	&	29,775	&	199,523\\
January-1965	&	 653 	&	 623 	&	 561 	&	29,633	&	208,249\\
February-1965	&	 655 	&	 628 	&	 566 	&	29,416	&	210,972\\
March-1965	&	 658 	&	 633 	&	 570 	&	29,584	&	209,008\\
April-1965	&	 657 	&	 632 	&	 569 	&	31,899	&	219,375\\
May-1965	&	 689 	&	 663 	&	 597 	&	31,862	&	206,585\\
June-1965	&	 696 	&	 666 	&	 600 	&	27,030	&	186,960\\
July-1965	&	 700 	&	 672 	&	 605 	&	27,958	&	185,763\\
August-1965	&	 704 	&	 677 	&	 610 	&	28,860	&	191,818\\
September-1965	&	 708 	&	 684 	&	 616 	&	28,760	&	197,295\\
October-1965	&	 709 	&	 686 	&	 618 	&	29,458	&	204,931\\
November-1965	&	 709 	&	 685 	&	 617 	&	30,475	&	207,690\\
December-1965	&	 713 	&	 692 	&	 623 	&	31,363	&	211,730\\
\hline

\end{tabular}

\end{center}

\newpage

\begin{center}
{\Large {\bf Table IA-1 -Cont'd.}}\\
{\bf Sample Construction}\\
\end{center}

\baselineskip 12pt       
\noindent
Stocks must have 60 months non-missing data on CRSP and a book value in CRSP/Compustat in months $[-18, \; -6]$ to be eligible.  Next a real dollar minimum is applied
(\$110 million in December 2021).  Finally, the smallest 10\% of stocks are removed prior to Nasdaq eligibility, and 20\% after Nasdaq stocks enter the sample (after December,
1977).  This table shows the effect of each filter on the sample size, as well as the smallest and median sized stocks eligible for the ensuing month in \$ millions.

\vskip .1in     

\hoffset=-0.85in

\begin{center}

\begin{tabular}{lrrrrr}  

      & Eligible & \multicolumn{1}{c}{After} & \multicolumn{1}{c}{Final} & \multicolumn{1}{c}{Minimum} & \multicolumn{1}{c}{Median}\\
 \multicolumn{1}{c}{Month End}      & \multicolumn{1}{c}{Stocks}   & \$ criterion & Sample & Mkt. Cap. & Mkt. Cap. \\
\hline
January-1966    &        716    &        698    &        629    &       32,702  &       206,904\\
February-1966   &        716    &        700    &        630    &       33,722  &       205,932\\
March-1966      &        716    &        699    &        630    &       32,872  &       206,818\\
April-1966      &        720    &        704    &        634    &       32,780  &       210,266\\
May-1966        &        727    &        703    &        633    &       31,020  &       196,721\\
June-1966       &        731    &        707    &        637    &       29,799  &       197,799\\
July-1966       &        734    &        707    &        637    &       30,509  &       195,896\\
August-1966     &        736    &        703    &        633    &       28,909  &       182,819\\
September-1966  &        742    &        708    &        638    &       27,808  &       179,487\\
October-1966    &        749    &        710    &        639    &       28,899  &       176,234\\
November-1966   &        753    &        715    &        644    &       28,886  &       181,985\\
December-1966   &        753    &        716    &        645    &       29,000  &       187,575\\
January-1967	&	 756 	&	 733 	&	 660 	&	31,000	&	197,309\\
February-1967	&	 759 	&	 736 	&	 663 	&	31,125	&	201,510\\
March-1967	&	 763 	&	 743 	&	 669 	&	32,895	&	210,937\\
April-1967	&	 764 	&	 748 	&	 674 	&	31,734	&	219,412\\
May-1967	&	 767 	&	 748 	&	 674 	&	33,070	&	216,716\\
June-1967	&	 768 	&	 753 	&	 678 	&	34,750	&	215,572\\
July-1967	&	1,145	&	 978 	&	 881 	&	22,155	&	152,183\\
August-1967	&	1,152	&	 988 	&	 890 	&	21,672	&	150,631\\
September-1967	&	1,164	&	1,006	&	 906 	&	22,514	&	148,847\\
October-1967	&	1,166	&	 998 	&	 899 	&	22,897	&	147,687\\
November-1967	&	1,172	&	 995 	&	 896 	&	23,500	&	147,199\\
December-1967	&	1,179	&	1,022	&	 920 	&	23,030	&	154,618\\
January-1968    &       1,180   &       1,030   &        927    &       23,385  &       148,004\\
February-1968   &       1,182   &       1,012   &        911    &       23,621  &       149,284\\
March-1968      &       1,190   &       1,017   &        916    &       23,925  &       145,821\\
April-1968      &       1,198   &       1,039   &        936    &       25,665  &       154,682\\
May-1968        &       1,210   &       1,079   &        972    &       23,599  &       152,028\\
June-1968       &       1,214   &       1,082   &        974    &       24,413  &       155,456\\
July-1968       &       1,213   &       1,074   &        967    &       24,750  &       152,260\\
August-1968     &       1,218   &       1,091   &        982    &       25,067  &       152,600\\
September-1968  &       1,222   &       1,109   &        999    &       25,398  &       158,815\\
October-1968    &       1,222   &       1,108   &        998    &       24,408  &       161,185\\
November-1968   &       1,229   &       1,118   &       1,007   &       25,642  &       167,999\\
December-1968   &       1,238   &       1,130   &       1,017   &       26,898  &       170,701\\
\hline

\end{tabular}

\end{center}

\newpage

\begin{center}
{\Large {\bf Table IA-1 -Cont'd.}}\\
{\bf Sample Construction}\\
\end{center}

\baselineskip 12pt
\noindent
Stocks must have 60 months non-missing data on CRSP and a book value in CRSP/Compustat in months $[-18, \; -6]$ to be eligible.  Next a real dollar minimum is applied
(\$110 million in December 2021).  Finally, the smallest 10\% of stocks are removed prior to Nasdaq eligibility, and 20\% after Nasdaq stocks enter the sample (after December,
1977).  This table shows the effect of each filter on the sample size, as well as the smallest and median sized stocks eligible for the ensuing month in \$ millions.

\vskip .1in

\hoffset=-0.85in

\begin{center}

\begin{tabular}{lrrrrr}

      & Eligible & \multicolumn{1}{c}{After} & \multicolumn{1}{c}{Final} & \multicolumn{1}{c}{Minimum} & \multicolumn{1}{c}{Median}\\
 \multicolumn{1}{c}{Month End}      & \multicolumn{1}{c}{Stocks}   & \$ criterion & Sample & Mkt. Cap. & Mkt. Cap. \\
\hline
January-1969	&	1,246	&	1,137	&	1,024	&	26,928	&	169,884\\
February-1969	&	1,262	&	1,131	&	1,018	&	26,786	&	161,004\\
March-1969	&	1,261	&	1,130	&	1,017	&	26,856	&	164,496\\
April-1969	&	1,271	&	1,130	&	1,017	&	27,114	&	163,191\\
May-1969	&	1,288	&	1,150	&	1,035	&	26,483	&	164,405\\
June-1969	&	1,299	&	1,126	&	1,014	&	26,082	&	152,692\\
July-1969	&	1,306	&	1,108	&	 998 	&	25,326	&	144,156\\
August-1969	&	1,304	&	1,109	&	 999 	&	25,400	&	150,713\\
September-1969	&	1,317	&	1,106	&	 996 	&	24,640	&	152,133\\
October-1969	&	1,322	&	1,129	&	1,017	&	25,568	&	156,599\\
November-1969	&	1,327	&	1,117	&	1,006	&	24,648	&	151,620\\
December-1969	&	1,333	&	1,107	&	 997 	&	23,962	&	155,108\\
January-1970	&	1,339	&	1,101	&	 991 	&	24,317	&	148,749\\
February-1970	&	1,347	&	1,112	&	1,001	&	24,340	&	147,026\\
March-1970	&	1,351	&	1,110	&	 999 	&	23,798	&	144,923\\
April-1970	&	1,360	&	1,063	&	 957 	&	22,715	&	131,445\\
May-1970	&	1,365	&	1,032	&	 929 	&	23,077	&	124,230\\
June-1970	&	1,372	&	1,018	&	 917 	&	21,954	&	120,225\\
July-1970	&	1,374	&	1,028	&	 926 	&	22,544	&	127,966\\
August-1970	&	1,378	&	1,046	&	 942 	&	22,515	&	130,708\\
September-1970	&	1,378	&	1,073	&	 966 	&	23,184	&	133,555\\
October-1970	&	1,385	&	1,057	&	 952 	&	22,788	&	132,833\\
November-1970	&	1,386	&	1,056	&	 951 	&	23,180	&	138,671\\
December-1970	&	1,399	&	1,076	&	 969 	&	24,929	&	148,933\\
January-1971    &       1,405   &       1,118   &       1,007   &       24,711  &       152,665\\
February-1971   &       1,403   &       1,134   &       1,021   &       24,393  &       153,459\\
March-1971      &       1,404   &       1,143   &       1,029   &       24,885  &       158,011\\
April-1971      &       1,405   &       1,143   &       1,029   &       26,004  &       165,055\\
May-1971        &       1,406   &       1,129   &       1,017   &       25,305  &       158,353\\
June-1971       &       1,412   &       1,124   &       1,012   &       25,476  &       161,626\\
July-1971       &       1,425   &       1,112   &       1,001   &       25,185  &       155,701\\
August-1971     &       1,432   &       1,127   &       1,015   &       25,146  &       164,238\\
September-1971  &       1,440   &       1,126   &       1,014   &       25,200  &       161,701\\
October-1971    &       1,444   &       1,111   &       1,000   &       24,645  &       155,283\\
November-1971   &       1,446   &       1,103   &        993    &       24,262  &       153,243\\
December-1971   &       1,445   &       1,129   &       1,017   &       24,960  &       168,139\\
\hline

\end{tabular}

\end{center}

\newpage

\begin{center}
{\Large {\bf Table IA-1 -Cont'd.}}\\
{\bf Sample Construction}\\
\end{center}

\baselineskip 12pt
\noindent
Stocks must have 60 months non-missing data on CRSP and a book value in CRSP/Compustat in months $[-18, \; -6]$ to be eligible.  Next a real dollar minimum is applied
(\$110 million in December 2021).  Finally, the smallest 10\% of stocks are removed prior to Nasdaq eligibility, and 20\% after Nasdaq stocks enter the sample (after December,
1977).  This table shows the effect of each filter on the sample size, as well as the smallest and median sized stocks eligible for the ensuing month in \$ millions.

\vskip .1in

\hoffset=-0.85in

\begin{center}

\begin{tabular}{lrrrrr}

      & Eligible & \multicolumn{1}{c}{After} & \multicolumn{1}{c}{Final} & \multicolumn{1}{c}{Minimum} & \multicolumn{1}{c}{Median}\\
 \multicolumn{1}{c}{Month End}      & \multicolumn{1}{c}{Stocks}   & \$ criterion & Sample & Mkt. Cap. & Mkt. Cap. \\
\hline
January-1972	&	1,451	&	1,156	&	1,041	&	25,397	&	170,630\\
February-1972	&	1,451	&	1,162	&	1,046	&	26,035	&	178,619\\
March-1972	&	1,453	&	1,166	&	1,050	&	25,239	&	178,408\\
April-1972	&	1,457	&	1,165	&	1,049	&	26,130	&	173,812\\
May-1972	&	1,465	&	1,161	&	1,045	&	27,016	&	174,563\\
June-1972	&	1,474	&	1,149	&	1,035	&	27,543	&	173,099\\
July-1972	&	1,476	&	1,151	&	1,036	&	27,312	&	172,140\\
August-1972	&	1,484	&	1,152	&	1,037	&	27,263	&	174,641\\
September-1972	&	1,494	&	1,145	&	1,031	&	26,729	&	178,604\\
October-1972	&	1,501	&	1,148	&	1,034	&	26,104	&	173,773\\
November-1972	&	1,510	&	1,156	&	1,041	&	28,031	&	182,792\\
December-1972	&	1,518	&	1,164	&	1,048	&	28,633	&	185,024\\
January-1973	&	1,528	&	1,159	&	1,044	&	27,869	&	168,607\\
February-1973	&	1,533	&	1,143	&	1,029	&	26,760	&	158,445\\
March-1973	&	1,543	&	1,142	&	1,028	&	26,338	&	157,846\\
April-1973	&	1,553	&	1,125	&	1,013	&	25,730	&	152,246\\
May-1973	&	1,554	&	1,101	&	 991 	&	25,436	&	147,990\\
June-1973	&	1,561	&	1,089	&	 981 	&	25,500	&	146,954\\
July-1973	&	1,567	&	1,125	&	1,013	&	25,530	&	154,840\\
August-1973	&	1,578	&	1,111	&	1,000	&	25,525	&	154,707\\
September-1973	&	1,582	&	1,134	&	1,021	&	27,683	&	166,756\\
October-1973	&	1,591	&	1,133	&	1,020	&	27,200	&	168,873\\
November-1973	&	1,600	&	1,071	&	 964 	&	26,118	&	151,040\\
December-1973	&	1,610	&	1,057	&	 952 	&	26,418	&	157,706\\
January-1974    &       1,621   &       1,107   &        997    &       27,054  &       153,915\\
February-1974   &       1,636   &       1,112   &       1,001   &       26,955  &       152,928\\
March-1974      &       1,648   &       1,113   &       1,002   &       27,224  &       153,504\\
April-1974      &       1,659   &       1,100   &        990    &       26,985  &       147,341\\
May-1974        &       1,673   &       1,078   &        971    &       26,537  &       140,207\\
June-1974       &       1,683   &       1,073   &        966    &       25,880  &       139,228\\
July-1974       &       1,692   &       1,053   &        948    &       27,050  &       138,117\\
August-1974     &       1,707   &       1,021   &        919    &       27,738  &       135,984\\
September-1974  &       1,714   &        993    &        894    &       27,176  &       129,658\\
October-1974    &       1,710   &       1,023   &        921    &       27,666  &       135,506\\
November-1974   &       1,716   &       1,002   &        902    &       27,793  &       138,147\\
December-1974   &       1,721   &        977    &        880    &       27,319  &       142,810\\
\hline

\end{tabular}

\end{center}

\newpage

\begin{center}
{\Large {\bf Table IA-1 -Cont'd.}}\\
{\bf Sample Construction}\\
\end{center}

\baselineskip 12pt
\noindent
Stocks must have 60 months non-missing data on CRSP and a book value in CRSP/Compustat in months $[-18, \; -6]$ to be eligible.  Next a real dollar minimum is applied
(\$110 million in December 2021).  Finally, the smallest 10\% of stocks are removed prior to Nasdaq eligibility, and 20\% after Nasdaq stocks enter the sample (after December,
1977).  This table shows the effect of each filter on the sample size, as well as the smallest and median sized stocks eligible for the ensuing month in \$ millions.

\vskip .1in

\hoffset=-0.85in

\begin{center}

\begin{tabular}{lrrrrr}

      & Eligible & \multicolumn{1}{c}{After} & \multicolumn{1}{c}{Final} & \multicolumn{1}{c}{Minimum} & \multicolumn{1}{c}{Median}\\
 \multicolumn{1}{c}{Month End}      & \multicolumn{1}{c}{Stocks}   & \$ criterion & Sample & Mkt. Cap. & Mkt. Cap. \\
\hline
January-1975	&	1,730	&	1,067	&	 961 	&	28,247	&	145,871\\
February-1975	&	1,742	&	1,074	&	 967 	&	29,013	&	154,957\\
March-1975	&	1,754	&	1,110	&	 999 	&	28,951	&	153,146\\
April-1975	&	1,757	&	1,119	&	1,008	&	28,627	&	157,762\\
May-1975	&	1,771	&	1,131	&	1,018	&	30,254	&	162,335\\
June-1975	&	1,771	&	1,161	&	1,045	&	29,722	&	163,768\\
July-1975	&	1,777	&	1,153	&	1,038	&	29,631	&	163,863\\
August-1975	&	1,779	&	1,117	&	1,006	&	29,780	&	169,567\\
September-1975	&	1,783	&	1,104	&	 994 	&	29,997	&	168,518\\
October-1975	&	1,790	&	1,108	&	 998 	&	30,760	&	176,060\\
November-1975	&	1,795	&	1,115	&	1,004	&	31,155	&	176,361\\
December-1975	&	1,753	&	1,106	&	 996 	&	31,388	&	177,837\\
January-1976	&	1,764	&	1,188	&	1,070	&	30,199	&	178,962\\
February-1976	&	1,749	&	1,217	&	1,096	&	30,195	&	180,812\\
March-1976	&	1,749	&	1,221	&	1,099	&	29,984	&	179,046\\
April-1976	&	1,751	&	1,215	&	1,094	&	31,017	&	177,453\\
May-1976	&	1,757	&	1,212	&	1,091	&	31,167	&	176,868\\
June-1976	&	1,762	&	1,214	&	1,093	&	31,839	&	187,299\\
July-1976	&	1,641	&	1,200	&	1,080	&	32,863	&	190,857\\
August-1976	&	1,645	&	1,194	&	1,075	&	33,569	&	187,032\\
September-1976	&	1,636	&	1,193	&	1,074	&	33,616	&	192,962\\
October-1976	&	1,640	&	1,182	&	1,064	&	34,004	&	190,791\\
November-1976	&	1,647	&	1,193	&	1,074	&	33,860	&	198,162\\
December-1976	&	1,641	&	1,221	&	1,099	&	33,942	&	205,271\\
January-1977    &       1,644   &       1,231   &       1,108   &       33,561  &       199,309\\
February-1977   &       1,642   &       1,222   &       1,100   &       33,363  &       195,582\\
March-1977      &       1,648   &       1,224   &       1,102   &       34,164  &       195,052\\
April-1977      &       1,656   &       1,227   &       1,105   &       34,688  &       197,363\\
May-1977        &       1,666   &       1,230   &       1,107   &       34,941  &       193,923\\
June-1977       &       1,668   &       1,254   &       1,129   &       35,488  &       197,351\\
July-1977       &       1,672   &       1,248   &       1,124   &       35,211  &       197,328\\
August-1977     &       1,677   &       1,234   &       1,111   &       36,013  &       197,855\\
September-1977  &       1,669   &       1,224   &       1,102   &       37,635  &       202,003\\
October-1977    &       1,672   &       1,224   &       1,102   &       36,772  &       197,232\\
November-1977   &       1,669   &       1,251   &       1,001   &       56,084  &       249,400\\
December-1977   &       2,829   &       1,775   &       1,420   &       44,712  &       171,134\\
\hline

\end{tabular}

\end{center}

\newpage

\begin{center}
{\Large {\bf Table IA-1 -Cont'd.}}\\
{\bf Sample Construction}\\
\end{center}

\baselineskip 12pt
\noindent
Stocks must have 60 months non-missing data on CRSP and a book value in CRSP/Compustat in months $[-18, \; -6]$ to be eligible.  Next a real dollar minimum is applied
(\$110 million in December 2021).  Finally, the smallest 10\% of stocks are removed prior to Nasdaq eligibility, and 20\% after Nasdaq stocks enter the sample (after December,
1977).  This table shows the effect of each filter on the sample size, as well as the smallest and median sized stocks eligible for the ensuing month in \$ millions.

\vskip .1in

\hoffset=-0.85in

\begin{center}

\begin{tabular}{lrrrrr}

      & Eligible & \multicolumn{1}{c}{After} & \multicolumn{1}{c}{Final} & \multicolumn{1}{c}{Minimum} & \multicolumn{1}{c}{Median}\\
 \multicolumn{1}{c}{Month End}      & \multicolumn{1}{c}{Stocks}   & \$ criterion & Sample & Mkt. Cap. & Mkt. Cap. \\
\hline
January-1978	&	2,837	&	1,748	&	1,399	&	45,627	&	167,855\\
February-1978	&	2,842	&	1,771	&	1,417	&	44,594	&	165,400\\
March-1978	&	2,842	&	1,819	&	1,456	&	45,117	&	169,908\\
April-1978	&	2,838	&	1,858	&	1,487	&	45,639	&	176,291\\
May-1978	&	2,832	&	1,878	&	1,503	&	45,201	&	178,126\\
June-1978	&	2,815	&	1,856	&	1,485	&	46,023	&	178,354\\
July-1978	&	2,803	&	1,880	&	1,504	&	46,795	&	185,598\\
August-1978	&	2,787	&	1,907	&	1,526	&	47,403	&	189,195\\
September-1978	&	2,780	&	1,895	&	1,516	&	47,685	&	187,478\\
October-1978	&	2,777	&	1,724	&	1,380	&	47,282	&	181,387\\
November-1978	&	2,777	&	1,754	&	1,404	&	47,821	&	186,627\\
December-1978	&	2,767	&	1,762	&	1,411	&	47,385	&	189,628\\
January-1979	&	2,753	&	1,803	&	1,443	&	48,662	&	190,752\\
February-1979	&	2,752	&	1,768	&	1,415	&	49,128	&	186,524\\
March-1979	&	2,761	&	1,818	&	1,455	&	50,693	&	195,735\\
April-1979	&	2,758	&	1,810	&	1,448	&	51,026	&	197,443\\
May-1979	&	2,769	&	1,802	&	1,442	&	50,879	&	193,419\\
June-1979	&	2,764	&	1,821	&	1,457	&	51,471	&	200,612\\
July-1979	&	2,753	&	1,816	&	1,453	&	53,057	&	208,264\\
August-1979	&	2,745	&	1,839	&	1,472	&	54,450	&	215,254\\
September-1979	&	2,739	&	1,815	&	1,452	&	55,220	&	219,713\\
October-1979	&	2,740	&	1,753	&	1,403	&	52,604	&	210,183\\
November-1979	&	2,739	&	1,779	&	1,424	&	54,503	&	221,857\\
December-1979	&	2,739	&	1,789	&	1,432	&	56,856	&	228,882\\
January-1980    &       2,738   &       1,815   &       1,452   &       59,693  &       230,340\\
February-1980   &       2,737   &       1,781   &       1,425   &       60,074  &       232,028\\
March-1980      &       2,728   &       1,673   &       1,339   &       56,808  &       218,453\\
April-1980      &       2,727   &       1,694   &       1,356   &       58,158  &       226,890\\
May-1980        &       2,734   &       1,719   &       1,376   &       60,146  &       238,050\\
June-1980       &       2,749   &       1,742   &       1,394   &       60,581  &       241,717\\
July-1980       &       2,762   &       1,787   &       1,430   &       62,578  &       247,373\\
August-1980     &       2,761   &       1,809   &       1,448   &       65,029  &       255,070\\
September-1980  &       2,754   &       1,819   &       1,456   &       63,232  &       255,930\\
October-1980    &       2,753   &       1,823   &       1,459   &       65,771  &       255,569\\
November-1980   &       2,748   &       1,834   &       1,468   &       67,970  &       263,619\\
December-1980   &       2,759   &       1,806   &       1,445   &       66,981  &       265,648\\
\hline

\end{tabular}

\end{center}

\newpage

\begin{center}
{\Large {\bf Table IA-1 -Cont'd.}}\\
{\bf Sample Construction}\\
\end{center}

\baselineskip 12pt
\noindent
Stocks must have 60 months non-missing data on CRSP and a book value in CRSP/Compustat in months $[-18, \; -6]$ to be eligible.  Next a real dollar minimum is applied
(\$110 million in December 2021).  Finally, the smallest 10\% of stocks are removed prior to Nasdaq eligibility, and 20\% after Nasdaq stocks enter the sample (after December,
1977).  This table shows the effect of each filter on the sample size, as well as the smallest and median sized stocks eligible for the ensuing month in \$ millions.

\vskip .1in

\hoffset=-0.85in

\begin{center}

\begin{tabular}{lrrrrr}

      & Eligible & \multicolumn{1}{c}{After} & \multicolumn{1}{c}{Final} & \multicolumn{1}{c}{Minimum} & \multicolumn{1}{c}{Median}\\
 \multicolumn{1}{c}{Month End}      & \multicolumn{1}{c}{Stocks}   & \$ criterion & Sample & Mkt. Cap. & Mkt. Cap. \\
\hline
January-1981	&	2,778	&	1,814	&	1,452	&	66,811	&	261,618\\
February-1981	&	2,774	&	1,815	&	1,452	&	67,062	&	261,526\\
March-1981	&	2,788	&	1,840	&	1,472	&	68,531	&	276,266\\
April-1981	&	2,788	&	1,850	&	1,480	&	70,058	&	277,539\\
May-1981	&	2,793	&	1,856	&	1,485	&	70,021	&	283,476\\
June-1981	&	2,804	&	1,847	&	1,478	&	71,320	&	280,811\\
July-1981	&	2,809	&	1,829	&	1,464	&	69,877	&	277,357\\
August-1981	&	2,954	&	1,810	&	1,448	&	67,863	&	261,935\\
September-1981	&	2,949	&	1,759	&	1,408	&	65,402	&	255,905\\
October-1981	&	2,949	&	1,806	&	1,445	&	66,041	&	255,782\\
November-1981	&	2,936	&	1,799	&	1,440	&	68,677	&	263,351\\
December-1981	&	2,919	&	1,786	&	1,429	&	67,074	&	258,255\\
January-1982	&	2,934	&	1,772	&	1,418	&	64,826	&	256,637\\
February-1982	&	2,920	&	1,732	&	1,386	&	64,803	&	249,571\\
March-1982	&	2,905	&	1,709	&	1,368	&	65,537	&	252,515\\
April-1982	&	2,905	&	1,722	&	1,378	&	67,803	&	257,076\\
May-1982	&	2,893	&	1,695	&	1,356	&	67,605	&	256,595\\
June-1982	&	2,886	&	1,671	&	1,337	&	67,084	&	255,345\\
July-1982	&	2,887	&	1,664	&	1,332	&	66,416	&	249,186\\
August-1982	&	2,876	&	1,688	&	1,351	&	66,749	&	265,166\\
September-1982	&	2,872	&	1,692	&	1,354	&	68,888	&	270,708\\
October-1982	&	2,859	&	1,755	&	1,404	&	71,993	&	289,404\\
November-1982	&	2,858	&	1,805	&	1,444	&	71,316	&	297,725\\
December-1982	&	2,849	&	1,799	&	1,440	&	71,396	&	295,346\\
January-1983    &       2,835   &       1,831   &       1,465   &       73,305  &       290,114\\
February-1983   &       2,830   &       1,863   &       1,491   &       73,584  &       294,856\\
March-1983      &       2,829   &       1,881   &       1,505   &       75,618  &       298,545\\
April-1983      &       2,832   &       1,908   &       1,527   &       76,768  &       308,700\\
May-1983        &       2,838   &       1,966   &       1,573   &       77,088  &       309,354\\
June-1983       &       2,843   &       1,974   &       1,580   &       79,328  &       321,485\\
July-1983       &       2,839   &       1,957   &       1,566   &       81,135  &       323,362\\
August-1983     &       2,838   &       1,946   &       1,557   &       79,501  &       320,425\\
September-1983  &       2,854   &       1,943   &       1,555   &       80,886  &       336,958\\
October-1983    &       2,861   &       1,915   &       1,532   &       79,515  &       323,366\\
November-1983   &       2,861   &       1,929   &       1,544   &       82,669  &       331,708\\
December-1983   &       2,853   &       1,916   &       1,533   &       81,679  &       336,259\\
\hline

\end{tabular}

\end{center}

\newpage

\begin{center}
{\Large {\bf Table IA-1 -Cont'd.}}\\
{\bf Sample Construction}\\
\end{center}

\baselineskip 12pt
\noindent
Stocks must have 60 months non-missing data on CRSP and a book value in CRSP/Compustat in months $[-18, \; -6]$ to be eligible.  Next a real dollar minimum is applied
(\$110 million in December 2021).  Finally, the smallest 10\% of stocks are removed prior to Nasdaq eligibility, and 20\% after Nasdaq stocks enter the sample (after December,
1977).  This table shows the effect of each filter on the sample size, as well as the smallest and median sized stocks eligible for the ensuing month in \$ millions.

\vskip .1in

\hoffset=-0.85in

\begin{center}

\begin{tabular}{lrrrrr}

      & Eligible & \multicolumn{1}{c}{After} & \multicolumn{1}{c}{Final} & \multicolumn{1}{c}{Minimum} & \multicolumn{1}{c}{Median}\\
 \multicolumn{1}{c}{Month End}      & \multicolumn{1}{c}{Stocks}   & \$ criterion & Sample & Mkt. Cap. & Mkt. Cap. \\
\hline
January-1983	&	2,835	&	1,831	&	1,465	&	73,305	&	290,114\\
February-1983	&	2,830	&	1,863	&	1,491	&	73,584	&	294,856\\
March-1983	&	2,829	&	1,881	&	1,505	&	75,618	&	298,545\\
April-1983	&	2,832	&	1,908	&	1,527	&	76,768	&	308,700\\
May-1983	&	2,838	&	1,966	&	1,573	&	77,088	&	309,354\\
June-1983	&	2,843	&	1,974	&	1,580	&	79,328	&	321,485\\
July-1983	&	2,839	&	1,957	&	1,566	&	81,135	&	323,362\\
August-1983	&	2,838	&	1,946	&	1,557	&	79,501	&	320,425\\
September-1983	&	2,854	&	1,943	&	1,555	&	80,886	&	336,958\\
October-1983	&	2,861	&	1,915	&	1,532	&	79,515	&	323,366\\
November-1983	&	2,861	&	1,929	&	1,544	&	82,669	&	331,708\\
December-1983	&	2,853	&	1,916	&	1,533	&	81,679	&	336,259\\
January-1984	&	2,853	&	1,908	&	1,527	&	82,418	&	329,745\\
February-1984	&	2,841	&	1,879	&	1,504	&	79,131	&	314,761\\
March-1984	&	2,843	&	1,880	&	1,504	&	80,146	&	320,934\\
April-1984	&	2,840	&	1,862	&	1,490	&	81,276	&	320,667\\
May-1984	&	2,850	&	1,837	&	1,470	&	80,528	&	315,956\\
June-1984	&	2,843	&	1,829	&	1,464	&	81,811	&	322,322\\
July-1984	&	2,835	&	1,808	&	1,447	&	80,441	&	320,420\\
August-1984	&	2,826	&	1,836	&	1,469	&	84,131	&	347,865\\
September-1984	&	2,824	&	1,821	&	1,457	&	85,008	&	352,286\\
October-1984	&	2,810	&	1,793	&	1,435	&	85,722	&	352,087\\
November-1984	&	2,800	&	1,771	&	1,417	&	84,123	&	348,478\\
December-1984	&	2,806	&	1,773	&	1,419	&	85,984	&	354,411\\
January-1985	&	2,806	&	1,824	&	1,460	&	86,736	&	372,393\\
February-1985	&	2,798	&	1,823	&	1,459	&	87,297	&	377,028\\
March-1985	&	2,805	&	1,820	&	1,456	&	85,638	&	371,961\\
April-1985	&	2,797	&	1,807	&	1,446	&	85,850	&	382,071\\
May-1985	&	2,778	&	1,806	&	1,445	&	86,638	&	390,781\\
June-1985	&	2,775	&	1,800	&	1,440	&	88,129	&	407,849\\
July-1985	&	2,776	&	1,810	&	1,448	&	87,304	&	398,066\\
August-1985	&	2,771	&	1,805	&	1,444	&	84,981	&	399,759\\
September-1985	&	2,776	&	1,774	&	1,420	&	84,416	&	393,679\\
October-1985	&	2,790	&	1,775	&	1,420	&	85,547	&	406,800\\
November-1985	&	2,801	&	1,796	&	1,437	&	87,146	&	423,115\\
December-1985	&	2,811	&	1,808	&	1,447	&	90,760	&	424,310\\
\hline

\end{tabular}

\end{center}

\newpage

\begin{center}
{\Large {\bf Table IA-1 -Cont'd.}}\\
{\bf Sample Construction}\\
\end{center}

\baselineskip 12pt
\noindent
Stocks must have 60 months non-missing data on CRSP and a book value in CRSP/Compustat in months $[-18, \; -6]$ to be eligible.  Next a real dollar minimum is applied
(\$110 million in December 2021).  Finally, the smallest 10\% of stocks are removed prior to Nasdaq eligibility, and 20\% after Nasdaq stocks enter the sample (after December,
1977).  This table shows the effect of each filter on the sample size, as well as the smallest and median sized stocks eligible for the ensuing month in \$ millions.

\vskip .1in

\hoffset=-0.85in

\begin{center}

\begin{tabular}{lrrrrr}

      & Eligible & \multicolumn{1}{c}{After} & \multicolumn{1}{c}{Final} & \multicolumn{1}{c}{Minimum} & \multicolumn{1}{c}{Median}\\
 \multicolumn{1}{c}{Month End}      & \multicolumn{1}{c}{Stocks}   & \$ criterion & Sample & Mkt. Cap. & Mkt. Cap. \\
\hline
January-1986	&	2,814	&	1,811	&	1,449	&	89,155	&	432,331\\
February-1986	&	2,815	&	1,848	&	1,479	&	89,559	&	446,721\\
March-1986	&	2,827	&	1,875	&	1,500	&	89,193	&	462,258\\
April-1986	&	2,850	&	1,866	&	1,493	&	89,960	&	457,179\\
May-1986	&	2,852	&	1,873	&	1,499	&	91,908	&	485,149\\
June-1986	&	2,872	&	1,872	&	1,498	&	91,315	&	480,470\\
July-1986	&	2,891	&	1,845	&	1,476	&	87,601	&	458,038\\
August-1986	&	2,895	&	1,851	&	1,481	&	87,435	&	476,741\\
September-1986	&	2,892	&	1,819	&	1,456	&	87,450	&	461,657\\
October-1986	&	2,888	&	1,823	&	1,459	&	87,937	&	458,894\\
November-1986	&	2,888	&	1,816	&	1,453	&	89,658	&	455,387\\
December-1986	&	2,864	&	1,786	&	1,429	&	88,225	&	456,993\\
January-1987	&	2,859	&	1,817	&	1,454	&	92,972	&	480,443\\
February-1987	&	2,859	&	1,848	&	1,479	&	92,862	&	491,320\\
March-1987	&	2,846	&	1,848	&	1,479	&	94,787	&	501,792\\
April-1987	&	2,842	&	1,832	&	1,466	&	91,896	&	480,356\\
May-1987	&	2,834	&	1,843	&	1,475	&	91,740	&	480,764\\
June-1987	&	2,826	&	1,845	&	1,476	&	93,004	&	506,244\\
July-1987	&	2,825	&	1,851	&	1,481	&	96,004	&	514,715\\
August-1987	&	2,816	&	1,839	&	1,472	&	97,943	&	529,616\\
September-1987	&	2,811	&	1,842	&	1,474	&	96,263	&	526,935\\
October-1987	&	2,798	&	1,663	&	1,331	&	86,883	&	448,550\\
November-1987	&	2,807	&	1,621	&	1,297	&	89,550	&	448,411\\
December-1987	&	2,816	&	1,637	&	1,310	&	93,713	&	467,014\\
January-1988	&	2,813	&	1,661	&	1,329	&	93,162	&	475,180\\
February-1988	&	2,806	&	1,695	&	1,356	&	91,044	&	487,994\\
March-1988	&	2,831	&	1,735	&	1,388	&	91,263	&	482,851\\
April-1988	&	2,833	&	1,720	&	1,376	&	92,701	&	492,393\\
May-1988	&	2,830	&	1,700	&	1,360	&	92,523	&	485,133\\
June-1988	&	2,865	&	1,733	&	1,387	&	97,860	&	500,714\\
July-1988	&	2,886	&	1,726	&	1,381	&	98,294	&	479,554\\
August-1988	&	2,925	&	1,733	&	1,387	&	95,282	&	460,562\\
September-1988	&	2,948	&	1,738	&	1,391	&	98,840	&	484,219\\
October-1988	&	2,975	&	1,735	&	1,388	&	97,203	&	487,318\\
November-1988	&	2,988	&	1,707	&	1,366	&	96,073	&	487,092\\
December-1988	&	3,022	&	1,724	&	1,380	&	97,646	&	475,925\\
\hline

\end{tabular}

\end{center}

\newpage

\begin{center}
{\Large {\bf Table IA-1 -Cont'd.}}\\
{\bf Sample Construction}\\
\end{center}

\baselineskip 12pt
\noindent
Stocks must have 60 months non-missing data on CRSP and a book value in CRSP/Compustat in months $[-18, \; -6]$ to be eligible.  Next a real dollar minimum is applied
(\$110 million in December 2021).  Finally, the smallest 10\% of stocks are removed prior to Nasdaq eligibility, and 20\% after Nasdaq stocks enter the sample (after December,
1977).  This table shows the effect of each filter on the sample size, as well as the smallest and median sized stocks eligible for the ensuing month in \$ millions.

\vskip .1in

\hoffset=-0.85in

\begin{center}

\begin{tabular}{lrrrrr}

      & Eligible & \multicolumn{1}{c}{After} & \multicolumn{1}{c}{Final} & \multicolumn{1}{c}{Minimum} & \multicolumn{1}{c}{Median}\\
 \multicolumn{1}{c}{Month End}      & \multicolumn{1}{c}{Stocks}   & \$ criterion & Sample & Mkt. Cap. & Mkt. Cap. \\
\hline
January-1989	&	3,029	&	1,737	&	1,390	&	98,225	&	499,012\\
February-1989	&	3,037	&	1,741	&	1,393	&	99,289	&	501,429\\
March-1989	&	3,044	&	1,761	&	1,409	&	98,432	&	499,230\\
April-1989	&	3,044	&	1,751	&	1,401	&	101,873	&	521,222\\
May-1989	&	3,065	&	1,777	&	1,422	&	104,145	&	529,630\\
June-1989	&	3,068	&	1,772	&	1,418	&	104,040	&	521,948\\
July-1989	&	3,066	&	1,782	&	1,426	&	103,250	&	542,209\\
August-1989	&	3,065	&	1,779	&	1,424	&	103,776	&	555,294\\
September-1989	&	3,057	&	1,773	&	1,419	&	105,665	&	539,674\\
October-1989	&	3,042	&	1,741	&	1,393	&	105,842	&	523,322\\
November-1989	&	3,025	&	1,730	&	1,384	&	108,324	&	534,105\\
December-1989	&	3,025	&	1,725	&	1,380	&	109,762	&	530,173\\
January-1990	&	3,017	&	1,673	&	1,339	&	106,099	&	515,912\\
February-1990	&	3,006	&	1,675	&	1,340	&	109,221	&	509,470\\
March-1990	&	3,004	&	1,691	&	1,353	&	108,373	&	514,046\\
April-1990	&	3,006	&	1,677	&	1,342	&	107,403	&	501,258\\
May-1990	&	3,004	&	1,709	&	1,368	&	107,991	&	514,933\\
June-1990	&	3,011	&	1,715	&	1,372	&	107,398	&	522,208\\
July-1990	&	3,010	&	1,690	&	1,352	&	109,436	&	515,513\\
August-1990	&	3,012	&	1,618	&	1,296	&	104,207	&	486,391\\
September-1990	&	3,019	&	1,572	&	1,258	&	105,375	&	476,382\\
October-1990	&	3,022	&	1,536	&	1,229	&	102,000	&	469,415\\
November-1990	&	3,029	&	1,556	&	1,245	&	106,422	&	499,758\\
December-1990	&	3,042	&	1,562	&	1,250	&	110,196	&	526,850\\
January-1991	&	3,041	&	1,590	&	1,272	&	113,346	&	543,114\\
February-1991	&	3,046	&	1,650	&	1,320	&	112,945	&	548,497\\
March-1991	&	3,057	&	1,688	&	1,351	&	113,080	&	546,596\\
April-1991	&	3,071	&	1,698	&	1,359	&	110,738	&	546,402\\
May-1991	&	3,076	&	1,712	&	1,370	&	114,361	&	575,570\\
June-1991	&	3,121	&	1,713	&	1,371	&	111,298	&	558,845\\
July-1991	&	3,140	&	1,738	&	1,391	&	110,195	&	557,589\\
August-1991	&	3,159	&	1,757	&	1,406	&	116,109	&	575,865\\
September-1991	&	3,182	&	1,770	&	1,416	&	114,983	&	573,093\\
October-1991	&	3,201	&	1,793	&	1,435	&	114,487	&	570,483\\
November-1991	&	3,224	&	1,790	&	1,432	&	112,594	&	550,078\\
December-1991	&	3,236	&	1,834	&	1,468	&	115,132	&	575,424\\
\hline

\end{tabular}

\end{center}

\newpage

\begin{center}
{\Large {\bf Table IA-1 -Cont'd.}}\\
{\bf Sample Construction}\\
\end{center}

\baselineskip 12pt
\noindent
Stocks must have 60 months non-missing data on CRSP and a book value in CRSP/Compustat in months $[-18, \; -6]$ to be eligible.  Next a real dollar minimum is applied
(\$110 million in December 2021).  Finally, the smallest 10\% of stocks are removed prior to Nasdaq eligibility, and 20\% after Nasdaq stocks enter the sample (after December,
1977).  This table shows the effect of each filter on the sample size, as well as the smallest and median sized stocks eligible for the ensuing month in \$ millions.

\vskip .1in

\hoffset=-0.85in

\begin{center}

\begin{tabular}{lrrrrr}

      & Eligible & \multicolumn{1}{c}{After} & \multicolumn{1}{c}{Final} & \multicolumn{1}{c}{Minimum} & \multicolumn{1}{c}{Median}\\
 \multicolumn{1}{c}{Month End}      & \multicolumn{1}{c}{Stocks}   & \$ criterion & Sample & Mkt. Cap. & Mkt. Cap. \\
\hline
January-1992	&	3,251	&	1,903	&	1,523	&	113,966	&	567,624\\
February-1992	&	3,264	&	1,928	&	1,543	&	114,452	&	578,987\\
March-1992	&	3,289	&	1,932	&	1,546	&	112,271	&	558,910\\
April-1992	&	3,290	&	1,907	&	1,526	&	111,915	&	557,105\\
May-1992	&	3,298	&	1,921	&	1,537	&	113,120	&	553,860\\
June-1992	&	3,301	&	1,905	&	1,524	&	111,176	&	550,902\\
July-1992	&	3,311	&	1,929	&	1,544	&	113,119	&	559,624\\
August-1992	&	3,324	&	1,920	&	1,536	&	111,569	&	555,116\\
September-1992	&	3,346	&	1,933	&	1,547	&	111,230	&	573,693\\
October-1992	&	3,342	&	1,949	&	1,560	&	111,117	&	570,888\\
November-1992	&	3,342	&	1,984	&	1,588	&	113,396	&	593,931\\
December-1992	&	3,340	&	2,007	&	1,606	&	113,646	&	601,934\\
January-1993	&	3,333	&	2,037	&	1,630	&	112,978	&	577,949\\
February-1993	&	3,340	&	2,017	&	1,614	&	116,219	&	596,701\\
March-1993	&	3,326	&	2,034	&	1,628	&	115,848	&	601,188\\
April-1993	&	3,330	&	2,011	&	1,609	&	116,790	&	605,492\\
May-1993	&	3,329	&	2,031	&	1,625	&	117,528	&	601,322\\
June-1993	&	3,338	&	2,045	&	1,636	&	119,469	&	632,551\\
July-1993	&	3,332	&	2,052	&	1,642	&	120,292	&	620,805\\
August-1993	&	3,330	&	2,072	&	1,658	&	121,472	&	634,396\\
September-1993	&	3,342	&	2,096	&	1,677	&	120,647	&	630,192\\
October-1993	&	3,352	&	2,137	&	1,710	&	115,020	&	611,060\\
November-1993	&	3,383	&	2,124	&	1,700	&	117,160	&	615,668\\
December-1993	&	3,379	&	2,112	&	1,690	&	122,210	&	651,397\\
January-1994	&	3,379	&	2,163	&	1,731	&	117,728	&	631,120\\
February-1994	&	3,399	&	2,164	&	1,732	&	118,100	&	635,642\\
March-1994	&	3,393	&	2,139	&	1,712	&	117,392	&	610,246\\
April-1994	&	3,394	&	2,117	&	1,694	&	121,911	&	625,030\\
May-1994	&	3,613	&	2,198	&	1,759	&	112,361	&	576,592\\
June-1994	&	3,621	&	2,188	&	1,751	&	112,568	&	577,634\\
July-1994	&	3,634	&	2,205	&	1,764	&	111,037	&	571,058\\
August-1994	&	3,632	&	2,205	&	1,764	&	113,445	&	608,629\\
September-1994	&	3,630	&	2,215	&	1,772	&	114,095	&	595,073\\
October-1994	&	3,630	&	2,214	&	1,772	&	114,985	&	587,746\\
November-1994	&	3,641	&	2,198	&	1,759	&	115,605	&	565,412\\
December-1994	&	3,634	&	2,174	&	1,740	&	117,248	&	592,279\\
\hline

\end{tabular}

\end{center}

\newpage

\begin{center}
{\Large {\bf Table IA-1 -Cont'd.}}\\
{\bf Sample Construction}\\
\end{center}

\baselineskip 12pt
\noindent
Stocks must have 60 months non-missing data on CRSP and a book value in CRSP/Compustat in months $[-18, \; -6]$ to be eligible.  Next a real dollar minimum is applied
(\$110 million in December 2021).  Finally, the smallest 10\% of stocks are removed prior to Nasdaq eligibility, and 20\% after Nasdaq stocks enter the sample (after December,
1977).  This table shows the effect of each filter on the sample size, as well as the smallest and median sized stocks eligible for the ensuing month in \$ millions.

\vskip .1in

\hoffset=-0.85in

\begin{center}

\begin{tabular}{lrrrrr}

      & Eligible & \multicolumn{1}{c}{After} & \multicolumn{1}{c}{Final} & \multicolumn{1}{c}{Minimum} & \multicolumn{1}{c}{Median}\\
 \multicolumn{1}{c}{Month End}      & \multicolumn{1}{c}{Stocks}   & \$ criterion & Sample & Mkt. Cap. & Mkt. Cap. \\
\hline
January-1995	&	3,629	&	2,190	&	1,752	&	116,968	&	577,520\\
February-1995	&	3,627	&	2,209	&	1,768	&	115,998	&	600,214\\
March-1995	&	3,623	&	2,219	&	1,776	&	118,425	&	603,738\\
April-1995	&	3,640	&	2,244	&	1,796	&	119,144	&	603,190\\
May-1995	&	3,643	&	2,264	&	1,812	&	118,035	&	598,975\\
June-1995	&	3,649	&	2,297	&	1,838	&	123,156	&	614,804\\
July-1995	&	3,656	&	2,321	&	1,857	&	125,954	&	620,672\\
August-1995	&	3,656	&	2,348	&	1,879	&	125,100	&	636,805\\
September-1995	&	3,646	&	2,368	&	1,895	&	123,911	&	638,191\\
October-1995	&	3,631	&	2,324	&	1,860	&	121,755	&	621,908\\
November-1995	&	3,616	&	2,312	&	1,850	&	126,588	&	655,911\\
December-1995	&	3,608	&	2,311	&	1,849	&	129,213	&	656,928\\
January-1996	&	3,580	&	2,304	&	1,844	&	127,292	&	656,628\\
February-1996	&	3,562	&	2,301	&	1,841	&	129,587	&	678,347\\
March-1996	&	3,559	&	2,326	&	1,861	&	127,195	&	661,143\\
April-1996	&	3,572	&	2,373	&	1,899	&	127,529	&	666,109\\
May-1996	&	3,580	&	2,421	&	1,937	&	128,628	&	673,644\\
June-1996	&	3,611	&	2,415	&	1,932	&	124,823	&	658,060\\
July-1996	&	3,625	&	2,384	&	1,908	&	124,903	&	637,578\\
August-1996	&	3,631	&	2,403	&	1,923	&	127,721	&	665,696\\
September-1996	&	3,633	&	2,416	&	1,933	&	129,880	&	675,144\\
October-1996	&	3,650	&	2,423	&	1,939	&	133,673	&	667,390\\
November-1996	&	3,670	&	2,444	&	1,956	&	135,281	&	703,688\\
December-1996	&	3,673	&	2,443	&	1,955	&	136,924	&	715,932\\
January-1997    &       3,673   &       2,487   &       1,990   &       135,681 &       712,650\\
February-1997   &       3,678   &       2,465   &       1,972   &       135,401 &       706,475\\
March-1997      &       3,727   &       2,471   &       1,977   &       134,063 &       701,523\\
April-1997      &       3,735   &       2,447   &       1,958   &       132,954 &       719,686\\
May-1997        &       3,743   &       2,520   &       2,016   &       134,228 &       745,841\\
June-1997       &       3,747   &       2,555   &       2,044   &       135,435 &       768,108\\
July-1997       &       3,733   &       2,575   &       2,060   &       136,656 &       784,753\\
August-1997     &       3,729   &       2,592   &       2,074   &       138,231 &       782,523\\
September-1997  &       3,727   &       2,653   &       2,123   &       141,863 &       796,048\\
October-1997    &       3,729   &       2,638   &       2,111   &       143,225 &       784,222\\
November-1997   &       3,728   &       2,602   &       2,082   &       144,167 &       782,661\\
December-1997   &       3,717   &       2,588   &       2,071   &       139,707 &       803,730\\
\hline

\end{tabular}

\end{center}

\newpage

\begin{center}
{\Large {\bf Table IA-1 -Cont'd.}}\\
{\bf Sample Construction}\\
\end{center}

\baselineskip 12pt
\noindent
Stocks must have 60 months non-missing data on CRSP and a book value in CRSP/Compustat in months $[-18, \; -6]$ to be eligible.  Next a real dollar minimum is applied
(\$110 million in December 2021).  Finally, the smallest 10\% of stocks are removed prior to Nasdaq eligibility, and 20\% after Nasdaq stocks enter the sample (after December,
1977).  This table shows the effect of each filter on the sample size, as well as the smallest and median sized stocks eligible for the ensuing month in \$ millions.

\vskip .1in

\hoffset=-0.85in

\begin{center}

\begin{tabular}{lrrrrr}

      & Eligible & \multicolumn{1}{c}{After} & \multicolumn{1}{c}{Final} & \multicolumn{1}{c}{Minimum} & \multicolumn{1}{c}{Median}\\
 \multicolumn{1}{c}{Month End}      & \multicolumn{1}{c}{Stocks}   & \$ criterion & Sample & Mkt. Cap. & Mkt. Cap. \\
\hline
January-1998	&	3,704	&	2,565	&	2,052	&	138,500	&	799,835\\
February-1998	&	3,709	&	2,610	&	2,088	&	140,044	&	826,119\\
March-1998	&	3,715	&	2,657	&	2,126	&	140,649	&	824,276\\
April-1998	&	3,726	&	2,676	&	2,141	&	140,056	&	817,180\\
May-1998	&	3,716	&	2,647	&	2,118	&	140,132	&	798,833\\
June-1998	&	3,721	&	2,610	&	2,088	&	142,238	&	785,022\\
July-1998	&	3,719	&	2,570	&	2,056	&	138,255	&	740,887\\
August-1998	&	3,720	&	2,415	&	1,932	&	132,368	&	691,814\\
September-1998	&	3,724	&	2,443	&	1,955	&	133,194	&	716,236\\
October-1998	&	3,723	&	2,447	&	1,958	&	133,870	&	732,472\\
November-1998	&	3,758	&	2,499	&	2,000	&	135,352	&	731,157\\
December-1998	&	3,779	&	2,505	&	2,004	&	137,484	&	748,786\\
January-1999	&	3,777	&	2,548	&	2,039	&	134,115	&	711,461\\
February-1999	&	3,781	&	2,506	&	2,005	&	132,408	&	678,461\\
March-1999	&	3,784	&	2,455	&	1,964	&	131,195	&	714,592\\
April-1999	&	3,784	&	2,494	&	1,996	&	142,520	&	741,320\\
May-1999	&	3,787	&	2,536	&	2,029	&	138,406	&	743,082\\
June-1999	&	3,774	&	2,541	&	2,033	&	136,391	&	766,721\\
July-1999	&	3,748	&	2,521	&	2,017	&	138,873	&	745,749\\
August-1999	&	3,737	&	2,485	&	1,988	&	136,969	&	745,125\\
September-1999	&	3,712	&	2,447	&	1,958	&	138,574	&	754,150\\
October-1999	&	3,705	&	2,426	&	1,941	&	137,008	&	730,662\\
November-1999	&	3,687	&	2,442	&	1,954	&	138,180	&	753,131\\
December-1999	&	3,681	&	2,479	&	1,984	&	137,286	&	775,016\\
January-2000    &       3,666   &       2,480   &       1,984   &       138,661 &       729,816\\
February-2000   &       3,664   &       2,496   &       1,997   &       147,239 &       761,756\\
March-2000      &       3,651   &       2,487   &       1,990   &       146,980 &       837,646\\
April-2000      &       3,650   &       2,427   &       1,942   &       138,548 &       834,972\\
May-2000        &       3,637   &       2,350   &       1,880   &       140,208 &       858,083\\
June-2000       &       3,646   &       2,386   &       1,909   &       141,869 &       856,196\\
July-2000       &       3,645   &       2,364   &       1,892   &       140,958 &       843,598\\
August-2000     &       3,628   &       2,369   &       1,896   &       147,504 &       914,515\\
September-2000  &       3,605   &       2,330   &       1,864   &       148,376 &       909,018\\
October-2000    &       3,613   &       2,308   &       1,847   &       141,839 &       889,167\\
November-2000   &       3,606   &       2,223   &       1,779   &       145,632 &       897,480\\
December-2000   &       3,608   &       2,234   &       1,788   &       143,353 &       975,016\\
\hline

\end{tabular}

\end{center}

\newpage

\begin{center}
{\Large {\bf Table IA-1 -Cont'd.}}\\
{\bf Sample Construction}\\
\end{center}

\baselineskip 12pt
\noindent
Stocks must have 60 months non-missing data on CRSP and a book value in CRSP/Compustat in months $[-18, \; -6]$ to be eligible.  Next a real dollar minimum is applied
(\$110 million in December 2021).  Finally, the smallest 10\% of stocks are removed prior to Nasdaq eligibility, and 20\% after Nasdaq stocks enter the sample (after December,
1977).  This table shows the effect of each filter on the sample size, as well as the smallest and median sized stocks eligible for the ensuing month in \$ millions.

\vskip .1in

\hoffset=-0.85in

\begin{center}

\begin{tabular}{lrrrrr}

      & Eligible & \multicolumn{1}{c}{After} & \multicolumn{1}{c}{Final} & \multicolumn{1}{c}{Minimum} & \multicolumn{1}{c}{Median}\\
 \multicolumn{1}{c}{Month End}      & \multicolumn{1}{c}{Stocks}   & \$ criterion & Sample & Mkt. Cap. & Mkt. Cap. \\
\hline
January-2001	&	3,583	&	2,304	&	1,844	&	147,903	&	957,634\\
February-2001	&	3,582	&	2,274	&	1,820	&	150,412	&	935,814\\
March-2001	&	3,585	&	2,263	&	1,811	&	149,481	&	903,586\\
April-2001	&	3,585	&	2,273	&	1,819	&	158,111	&	990,158\\
May-2001	&	3,597	&	2,332	&	1,866	&	171,496	&	959,358\\
June-2001	&	3,607	&	2,343	&	1,875	&	172,856	&	989,909\\
July-2001	&	3,609	&	2,352	&	1,882	&	169,456	&	943,538\\
August-2001	&	3,593	&	2,332	&	1,866	&	169,578	&	917,605\\
September-2001	&	3,603	&	2,256	&	1,805	&	159,800	&	838,540\\
October-2001	&	3,621	&	2,304	&	1,844	&	161,772	&	876,072\\
November-2001	&	3,624	&	2,356	&	1,885	&	162,815	&	904,080\\
December-2001	&	3,624	&	2,382	&	1,906	&	169,464	&	955,859\\
January-2002	&	3,624	&	2,408	&	1,927	&	162,954	&	910,238\\
February-2002	&	3,618	&	2,382	&	1,906	&	164,107	&	905,663\\
March-2002	&	3,602	&	2,425	&	1,940	&	169,048	&	958,823\\
April-2002	&	3,601	&	2,428	&	1,943	&	172,130	&	960,196\\
May-2002	&	3,557	&	2,417	&	1,934	&	164,287	&	943,440\\
June-2002	&	3,570	&	2,371	&	1,897	&	178,449	&	927,494\\
July-2002	&	3,582	&	2,305	&	1,844	&	161,529	&	834,056\\
August-2002	&	3,578	&	2,301	&	1,841	&	163,584	&	849,370\\
September-2002	&	3,581	&	2,250	&	1,800	&	159,712	&	811,752\\
October-2002	&	3,588	&	2,295	&	1,836	&	158,529	&	839,902\\
November-2002	&	3,594	&	2,358	&	1,887	&	159,926	&	847,470\\
December-2002	&	3,591	&	2,339	&	1,872	&	157,227	&	823,775\\
January-2003    &       3,588   &       2,334   &       1,868   &       153,143 &       800,345\\
February-2003   &       3,581   &       2,323   &       1,859   &       153,502 &       769,355\\
March-2003      &       3,572   &       2,321   &       1,857   &       157,723 &       790,364\\
April-2003      &       3,577   &       2,386   &       1,909   &       158,653 &       839,884\\
May-2003        &       3,585   &       2,476   &       1,981   &       161,159 &       870,741\\
June-2003       &       3,598   &       2,517   &       2,014   &       166,342 &       848,759\\
July-2003       &       3,612   &       2,574   &       2,060   &       173,601 &       862,115\\
August-2003     &       3,600   &       2,593   &       2,075   &       173,272 &       887,960\\
September-2003  &       3,586   &       2,609   &       2,088   &       170,584 &       837,076\\
October-2003    &       3,569   &       2,650   &       2,120   &       176,368 &       894,627\\
November-2003   &       3,558   &       2,682   &       2,146   &       176,704 &       897,929\\
December-2003   &       3,542   &       2,687   &       2,150   &       179,900 &       916,758\\
\hline

\end{tabular}

\end{center}

\newpage

\begin{center}
{\Large {\bf Table IA-1 -Cont'd.}}\\
{\bf Sample Construction}\\
\end{center}

\baselineskip 12pt
\noindent
Stocks must have 60 months non-missing data on CRSP and a book value in CRSP/Compustat in months $[-18, \; -6]$ to be eligible.  Next a real dollar minimum is applied
(\$110 million in December 2021).  Finally, the smallest 10\% of stocks are removed prior to Nasdaq eligibility, and 20\% after Nasdaq stocks enter the sample (after December,
1977).  This table shows the effect of each filter on the sample size, as well as the smallest and median sized stocks eligible for the ensuing month in \$ millions.

\vskip .1in

\hoffset=-0.85in

\begin{center}

\begin{tabular}{lrrrrr}

      & Eligible & \multicolumn{1}{c}{After} & \multicolumn{1}{c}{Final} & \multicolumn{1}{c}{Minimum} & \multicolumn{1}{c}{Median}\\
 \multicolumn{1}{c}{Month End}      & \multicolumn{1}{c}{Stocks}   & \$ criterion & Sample & Mkt. Cap. & Mkt. Cap. \\
\hline
January-2004	&	3,541	&	2,740	&	2,192	&	179,220	&	923,592\\
February-2004	&	3,546	&	2,749	&	2,200	&	181,157	&	940,672\\
March-2004	&	3,547	&	2,751	&	2,201	&	185,698	&	950,296\\
April-2004	&	3,546	&	2,739	&	2,192	&	180,830	&	937,200\\
May-2004	&	3,545	&	2,745	&	2,196	&	185,336	&	961,805\\
June-2004	&	3,549	&	2,765	&	2,212	&	189,941	&	994,511\\
July-2004	&	3,553	&	2,732	&	2,186	&	181,535	&	955,859\\
August-2004	&	3,553	&	2,747	&	2,198	&	173,728	&	940,951\\
September-2004	&	3,554	&	2,758	&	2,207	&	179,771	&	978,331\\
October-2004	&	3,559	&	2,767	&	2,214	&	181,408	&	995,268\\
November-2004	&	3,564	&	2,800	&	2,240	&	188,966	&	1,071,544\\
December-2004	&	3,562	&	2,843	&	2,275	&	189,464	&	1,063,497\\
January-2005	&	3,546	&	2,822	&	2,258	&	185,772	&	1,030,884\\
February-2005	&	3,555	&	2,829	&	2,264	&	185,685	&	1,061,174\\
March-2005	&	3,560	&	2,821	&	2,257	&	182,520	&	1,059,125\\
April-2005	&	3,580	&	2,794	&	2,236	&	176,961	&	1,017,062\\
May-2005	&	3,566	&	2,790	&	2,232	&	188,530	&	1,069,430\\
June-2005	&	3,570	&	2,819	&	2,256	&	188,423	&	1,110,434\\
July-2005	&	3,580	&	2,856	&	2,285	&	193,418	&	1,134,691\\
August-2005	&	3,587	&	2,853	&	2,283	&	195,411	&	1,115,715\\
September-2005	&	3,587	&	2,844	&	2,276	&	198,252	&	1,139,793\\
October-2005	&	3,583	&	2,826	&	2,261	&	196,487	&	1,106,207\\
November-2005	&	3,575	&	2,837	&	2,270	&	198,837	&	1,127,509\\
December-2005	&	3,551	&	2,828	&	2,263	&	199,279	&	1,110,412\\
January-2006    &       3,526   &       2,839   &       2,272   &       206,133 &       1,206,508\\
February-2006   &       3,517   &       2,851   &       2,281   &       203,219 &       1,195,377\\
March-2006      &       3,502   &       2,862   &       2,290   &       207,362 &       1,235,369\\
April-2006      &       3,487   &       2,857   &       2,286   &       205,734 &       1,236,690\\
May-2006        &       3,463   &       2,814   &       2,252   &       203,810 &       1,194,254\\
June-2006       &       3,454   &       2,792   &       2,234   &       201,826 &       1,196,000\\
July-2006       &       3,452   &       2,771   &       2,217   &       203,666 &       1,169,771\\
August-2006     &       3,440   &       2,766   &       2,213   &       206,333 &       1,215,861\\
September-2006  &       3,429   &       2,771   &       2,217   &       206,667 &       1,215,890\\
October-2006    &       3,413   &       2,774   &       2,220   &       213,259 &       1,271,045\\
November-2006   &       3,400   &       2,775   &       2,220   &       215,115 &       1,306,985\\
December-2006   &       3,382   &       2,765   &       2,212   &       216,046 &       1,309,207\\
\hline

\end{tabular}   

\end{center}

\newpage
     
\begin{center}
{\Large {\bf Table IA-1 -Cont'd.}}\\
{\bf Sample Construction}\\ 
\end{center}

\baselineskip 12pt
\noindent
Stocks must have 60 months non-missing data on CRSP and a book value in CRSP/Compustat in months $[-18, \; -6]$ to be eligible.  Next a real dollar minimum is applied
(\$110 million in December 2021).  Finally, the smallest 10\% of stocks are removed prior to Nasdaq eligibility, and 20\% after Nasdaq stocks enter the sample (after December,
1977).  This table shows the effect of each filter on the sample size, as well as the smallest and median sized stocks eligible for the ensuing month in \$ millions.

\vskip .1in

\hoffset=-0.85in

\begin{center}

\begin{tabular}{lrrrrr}

      & Eligible & \multicolumn{1}{c}{After} & \multicolumn{1}{c}{Final} & \multicolumn{1}{c}{Minimum} & \multicolumn{1}{c}{Median}\\
 \multicolumn{1}{c}{Month End}      & \multicolumn{1}{c}{Stocks}   & \$ criterion & Sample & Mkt. Cap. & Mkt. Cap. \\
\hline
January-2007	&	3,360	&	2,756	&	2,205	&	218,251	&	1,321,867\\
February-2007	&	3,348	&	2,748	&	2,199	&	218,295	&	1,306,914\\
March-2007	&	3,330	&	2,733	&	2,187	&	217,785	&	1,319,870\\
April-2007	&	3,311	&	2,740	&	2,192	&	218,210	&	1,340,112\\
May-2007	&	3,287	&	2,713	&	2,171	&	233,618	&	1,398,424\\
June-2007	&	3,271	&	2,699	&	2,160	&	231,180	&	1,376,766\\
July-2007	&	3,254	&	2,681	&	2,145	&	218,737	&	1,283,589\\
August-2007	&	3,225	&	2,623	&	2,099	&	223,274	&	1,316,534\\
September-2007	&	3,198	&	2,608	&	2,087	&	224,636	&	1,357,727\\
October-2007	&	3,184	&	2,605	&	2,084	&	214,001	&	1,358,045\\
November-2007	&	3,168	&	2,541	&	2,033	&	209,332	&	1,310,035\\
December-2007	&	3,150	&	2,507	&	2,006	&	211,724	&	1,326,367\\
January-2008	&	3,145	&	2,475	&	1,980	&	204,239	&	1,258,913\\
February-2008	&	3,128	&	2,444	&	1,956	&	200,126	&	1,222,064\\
March-2008	&	3,112	&	2,408	&	1,927	&	208,298	&	1,248,193\\
April-2008	&	3,095	&	2,390	&	1,912	&	210,008	&	1,309,163\\
May-2008	&	3,084	&	2,378	&	1,903	&	216,576	&	1,387,309\\
June-2008	&	3,073	&	2,327	&	1,862	&	199,039	&	1,307,111\\
July-2008	&	3,065	&	2,310	&	1,848	&	216,988	&	1,342,931\\
August-2008	&	3,059	&	2,318	&	1,855	&	224,046	&	1,371,713\\
September-2008	&	3,049	&	2,238	&	1,791	&	233,707	&	1,329,074\\
October-2008	&	3,032	&	2,115	&	1,692	&	212,932	&	1,163,097\\
November-2008	&	3,035	&	2,010	&	1,608	&	210,014	&	1,111,979\\
December-2008   &       3,022   &       2,007   &       1,606   &       220,841 &       1,139,157\\
January-2009    &       3,016   &       1,965   &       1,572   &       199,068 &       1,061,132\\
February-2009   &       3,009   &       1,889   &       1,512   &       191,200 &       1,009,453\\
March-2009      &       3,007   &       1,951   &       1,561   &       183,776 &       1,031,365\\
April-2009      &       3,001   &       2,035   &       1,628   &       199,611 &       1,141,281\\
May-2009        &       2,958   &       2,081   &       1,665   &       198,635 &       1,119,698\\
June-2009       &       2,967   &       2,092   &       1,674   &       209,327 &       1,143,122\\
July-2009       &       2,970   &       2,136   &       1,709   &       219,484 &       1,199,247\\
August-2009     &       2,969   &       2,158   &       1,727   &       215,295 &       1,232,221\\
September-2009  &       2,966   &       2,185   &       1,748   &       223,788 &       1,268,425\\
October-2009    &       2,966   &       2,155   &       1,724   &       209,745 &       1,206,184\\
November-2009   &       2,953   &       2,149   &       1,720   &       214,711 &       1,267,479\\
December-2009   &       2,958   &       2,184   &       1,748   &       225,862 &       1,325,731\\
\hline

\end{tabular}   

\end{center}

\newpage
     
\begin{center}
{\Large {\bf Table IA-1 -Cont'd.}}\\
{\bf Sample Construction}\\ 
\end{center}

\baselineskip 12pt
\noindent
Stocks must have 60 months non-missing data on CRSP and a book value in CRSP/Compustat in months $[-18, \; -6]$ to be eligible.  Next a real dollar minimum is applied
(\$110 million in December 2021).  Finally, the smallest 10\% of stocks are removed prior to Nasdaq eligibility, and 20\% after Nasdaq stocks enter the sample (after December,
1977).  This table shows the effect of each filter on the sample size, as well as the smallest and median sized stocks eligible for the ensuing month in \$ millions.

\vskip .1in

\hoffset=-0.85in

\begin{center}

\begin{tabular}{lrrrrr}

      & Eligible & \multicolumn{1}{c}{After} & \multicolumn{1}{c}{Final} & \multicolumn{1}{c}{Minimum} & \multicolumn{1}{c}{Median}\\
 \multicolumn{1}{c}{Month End}      & \multicolumn{1}{c}{Stocks}   & \$ criterion & Sample & Mkt. Cap. & Mkt. Cap. \\
\hline
January-2010	&	2,953	&	2,187	&	1,750	&	214,347	&	1,278,329\\
February-2010	&	2,953	&	2,207	&	1,766	&	225,754	&	1,310,871\\
March-2010	&	2,950	&	2,239	&	1,792	&	229,015	&	1,384,292\\
April-2010	&	2,937	&	2,257	&	1,806	&	241,728	&	1,459,300\\
May-2010	&	2,931	&	2,222	&	1,778	&	234,210	&	1,387,485\\
June-2010	&	2,945	&	2,213	&	1,771	&	222,428	&	1,279,107\\
July-2010	&	2,938	&	2,231	&	1,785	&	232,487	&	1,346,238\\
August-2010	&	2,944	&	2,220	&	1,776	&	213,067	&	1,269,173\\
September-2010	&	2,935	&	2,242	&	1,794	&	224,592	&	1,396,989\\
October-2010	&	2,943	&	2,259	&	1,808	&	223,543	&	1,417,621\\
November-2010	&	2,938	&	2,261	&	1,809	&	234,494	&	1,473,691\\
December-2010	&	2,931	&	2,284	&	1,828	&	249,250	&	1,587,723\\
January-2011    &       2,922   &       2,295   &       1,836   &       239,983 &       1,555,299\\
February-2011   &       2,919   &       2,306   &       1,845   &       245,926 &       1,602,757\\
March-2011      &       2,917   &       2,306   &       1,845   &       253,751 &       1,666,237\\
April-2011      &       2,910   &       2,307   &       1,846   &       254,658 &       1,741,632\\
May-2011        &       2,910   &       2,297   &       1,838   &       252,909 &       1,697,074\\
June-2011       &       2,898   &       2,267   &       1,814   &       253,454 &       1,686,626\\
July-2011       &       2,894   &       2,259   &       1,808   &       244,622 &       1,611,230\\
August-2011     &       2,893   &       2,216   &       1,773   &       234,161 &       1,534,980\\
September-2011  &       2,899   &       2,170   &       1,736   &       221,434 &       1,388,354\\
October-2011    &       2,900   &       2,206   &       1,765   &       245,684 &       1,570,742\\
November-2011   &       2,905   &       2,204   &       1,764   &       242,240 &       1,557,998\\
December-2011   &       2,905   &       2,211   &       1,769   &       244,470 &       1,549,758\\
January-2012    &       2,901   &       2,231   &       1,785   &       254,813 &       1,632,537\\
February-2012   &       2,895   &       2,235   &       1,788   &       253,567 &       1,685,201\\
March-2012      &       2,895   &       2,252   &       1,802   &       257,676 &       1,697,825\\
April-2012      &       2,900   &       2,250   &       1,800   &       250,638 &       1,657,550\\
May-2012        &       2,914   &       2,250   &       1,800   &       244,552 &       1,552,961\\
June-2012       &       2,912   &       2,265   &       1,812   &       258,833 &       1,580,176\\
July-2012       &       2,907   &       2,263   &       1,811   &       245,435 &       1,535,162\\
August-2012     &       2,914   &       2,273   &       1,819   &       255,035 &       1,608,822\\
September-2012  &       2,896   &       2,278   &       1,823   &       263,508 &       1,585,743\\
October-2012    &       2,893   &       2,266   &       1,813   &       253,707 &       1,609,798\\
November-2012   &       2,898   &       2,260   &       1,808   &       258,939 &       1,629,728\\
December-2012   &       2,895   &       2,270   &       1,816   &       263,836 &       1,621,796\\
\hline

\end{tabular}   

\end{center}

\newpage
     
\begin{center}
{\Large {\bf Table IA-1 -Cont'd.}}\\
{\bf Sample Construction}\\ 
\end{center}

\baselineskip 12pt
\noindent
Stocks must have 60 months non-missing data on CRSP and a book value in CRSP/Compustat in months $[-18, \; -6]$ to be eligible.  Next a real dollar minimum is applied
(\$110 million in December 2021).  Finally, the smallest 10\% of stocks are removed prior to Nasdaq eligibility, and 20\% after Nasdaq stocks enter the sample (after December,
1977).  This table shows the effect of each filter on the sample size, as well as the smallest and median sized stocks eligible for the ensuing month in \$ millions.

\vskip .1in

\hoffset=-0.85in

\begin{center}

\begin{tabular}{lrrrrr}

      & Eligible & \multicolumn{1}{c}{After} & \multicolumn{1}{c}{Final} & \multicolumn{1}{c}{Minimum} & \multicolumn{1}{c}{Median}\\
 \multicolumn{1}{c}{Month End}      & \multicolumn{1}{c}{Stocks}   & \$ criterion & Sample & Mkt. Cap. & Mkt. Cap. \\
\hline
January-2013	&	2,892	&	2,293	&	1,835	&	264,818	&	1,705,662\\
February-2013	&	2,887	&	2,296	&	1,837	&	257,386	&	1,731,399\\
March-2013	&	2,883	&	2,297	&	1,838	&	269,967	&	1,819,774\\
April-2013	&	2,877	&	2,296	&	1,837	&	270,702	&	1,779,725\\
May-2013	&	2,854	&	2,298	&	1,839	&	283,240	&	1,853,753\\
June-2013	&	2,838	&	2,283	&	1,827	&	293,563	&	1,838,522\\
July-2013	&	2,830	&	2,288	&	1,831	&	309,792	&	1,948,891\\
August-2013	&	2,823	&	2,293	&	1,835	&	296,906	&	1,883,219\\
September-2013	&	2,815	&	2,309	&	1,848	&	306,836	&	1,918,977\\
October-2013	&	2,805	&	2,315	&	1,852	&	299,331	&	1,996,293\\
November-2013	&	2,790	&	2,310	&	1,848	&	320,760	&	2,057,528\\
December-2013	&	2,777	&	2,314	&	1,852	&	313,811	&	2,082,686\\
January-2014    &       2,766   &       2,307   &       1,846   &       302,315 &       2,010,658\\
February-2014   &       2,753   &       2,320   &       1,856   &       306,613 &       2,081,855\\
March-2014      &       2,749   &       2,304   &       1,844   &       308,877 &       2,118,886\\
April-2014      &       2,741   &       2,288   &       1,831   &       300,389 &       2,079,653\\
May-2014        &       2,737   &       2,274   &       1,820   &       307,235 &       2,103,922\\
June-2014       &       2,732   &       2,278   &       1,823   &       314,848 &       2,186,788\\
July-2014       &       2,725   &       2,267   &       1,814   &       298,267 &       2,081,529\\
August-2014     &       2,717   &       2,270   &       1,816   &       310,069 &       2,155,488\\
September-2014  &       2,716   &       2,250   &       1,800   &       298,624 &       2,065,074\\
October-2014    &       2,711   &       2,243   &       1,795   &       329,103 &       2,193,613\\
November-2014   &       2,717   &       2,248   &       1,799   &       313,943 &       2,187,498\\
December-2014   &       2,711   &       2,237   &       1,790   &       329,349 &       2,245,886\\
January-2015    &       2,703   &       2,230   &       1,784   &       311,089 &       2,124,244\\
February-2015   &       2,693   &       2,230   &       1,784   &       328,474 &       2,237,962\\
March-2015      &       2,690   &       2,221   &       1,777   &       334,813 &       2,286,110\\
April-2015      &       2,687   &       2,225   &       1,780   &       323,457 &       2,254,695\\
May-2015        &       2,686   &       2,217   &       1,774   &       332,846 &       2,319,348\\
June-2015       &       2,680   &       2,200   &       1,760   &       344,643 &       2,334,126\\
July-2015       &       2,675   &       2,186   &       1,749   &       333,666 &       2,322,651\\
August-2015     &       2,670   &       2,176   &       1,741   &       319,244 &       2,221,083\\
September-2015  &       2,659   &       2,151   &       1,721   &       312,979 &       2,161,430\\
October-2015    &       2,643   &       2,149   &       1,720   &       326,646 &       2,267,865\\
November-2015   &       2,645   &       2,156   &       1,725   &       338,271 &       2,309,966\\
December-2015   &       2,639   &       2,145   &       1,716   &       319,739 &       2,163,982\\
\hline

\end{tabular}   

\end{center}

\newpage
     
\begin{center}
{\Large {\bf Table IA-1 -Cont'd.}}\\
{\bf Sample Construction}\\ 
\end{center}

\baselineskip 12pt
\noindent
Stocks must have 60 months non-missing data on CRSP and a book value in CRSP/Compustat in months $[-18, \; -6]$ to be eligible.  Next a real dollar minimum is applied
(\$110 million in December 2021).  Finally, the smallest 10\% of stocks are removed prior to Nasdaq eligibility, and 20\% after Nasdaq stocks enter the sample (after December,
1977).  This table shows the effect of each filter on the sample size, as well as the smallest and median sized stocks eligible for the ensuing month in \$ millions.

\vskip .1in

\hoffset=-0.85in

\begin{center}

\begin{tabular}{lrrrrr}

      & Eligible & \multicolumn{1}{c}{After} & \multicolumn{1}{c}{Final} & \multicolumn{1}{c}{Minimum} & \multicolumn{1}{c}{Median}\\
 \multicolumn{1}{c}{Month End}      & \multicolumn{1}{c}{Stocks}   & \$ criterion & Sample & Mkt. Cap. & Mkt. Cap. \\
\hline
January-2016	&	2,628	&	2,102	&	1,682	&	304,781	&	2,072,964\\
February-2016	&	2,620	&	2,091	&	1,673	&	300,673	&	2,077,084\\
March-2016	&	2,613	&	2,108	&	1,687	&	313,616	&	2,196,329\\
April-2016	&	2,602	&	2,122	&	1,698	&	313,158	&	2,215,682\\
May-2016	&	2,557	&	2,092	&	1,674	&	313,750	&	2,259,515\\
June-2016	&	2,543	&	2,071	&	1,657	&	317,854	&	2,172,081\\
July-2016	&	2,546	&	2,086	&	1,669	&	330,058	&	2,309,892\\
August-2016	&	2,534	&	2,084	&	1,668	&	338,005	&	2,377,730\\
September-2016	&	2,518	&	2,075	&	1,660	&	345,553	&	2,388,112\\
October-2016	&	2,511	&	2,060	&	1,648	&	333,037	&	2,333,828\\
November-2016	&	2,498	&	2,067	&	1,654	&	371,818	&	2,547,856\\
December-2016	&	2,496	&	2,074	&	1,660	&	386,874	&	2,561,973\\
January-2017    &       2,490   &       2,079   &       1,664   &       365,169 &       2,570,078\\
February-2017   &       2,486   &       2,065   &       1,652   &       372,238 &       2,621,476\\
March-2017      &       2,483   &       2,068   &       1,655   &       371,653 &       2,635,251\\
April-2017      &       2,474   &       2,060   &       1,648   &       377,157 &       2,709,580\\
May-2017        &       2,457   &       2,045   &       1,636   &       366,758 &       2,695,521\\
June-2017       &       2,443   &       2,038   &       1,631   &       388,057 &       2,753,834\\
July-2017       &       2,454   &       2,048   &       1,639   &       372,323 &       2,728,318\\
August-2017     &       2,445   &       2,030   &       1,624   &       367,798 &       2,687,630\\
September-2017  &       2,430   &       2,040   &       1,632   &       382,294 &       2,794,427\\
October-2017    &       2,433   &       2,030   &       1,624   &       394,380 &       2,873,361\\
November-2017   &       2,422   &       2,034   &       1,628   &       412,068 &       2,956,042\\
December-2017   &       2,411   &       2,029   &       1,624   &       402,840 &       2,962,668\\
January-2018    &       2,410   &       2,039   &       1,632   &       390,189 &       2,971,092\\
February-2018   &       2,412   &       2,032   &       1,626   &       375,446 &       2,816,824\\
March-2018      &       2,404   &       2,019   &       1,616   &       399,683 &       2,861,387\\
April-2018      &       2,405   &       2,023   &       1,619   &       401,575 &       2,842,501\\
May-2018        &       2,410   &       2,034   &       1,628   &       430,801 &       2,978,070\\
June-2018       &       2,412   &       2,038   &       1,631   &       433,863 &       2,962,780\\
July-2018       &       2,409   &       2,034   &       1,628   &       433,462 &       3,077,618\\
August-2018     &       2,410   &       2,037   &       1,630   &       438,110 &       3,186,325\\
September-2018  &       2,420   &       2,032   &       1,626   &       441,534 &       3,150,733\\
October-2018    &       2,421   &       2,024   &       1,620   &       402,578 &       2,909,988\\
November-2018   &       2,420   &       2,009   &       1,608   &       418,059 &       2,985,492\\
December-2018   &       2,413   &       1,980   &       1,584   &       380,757 &       2,706,263\\
\hline

\end{tabular}   

\end{center}

\newpage
     
\begin{center}
{\Large {\bf Table IA-1 -Cont'd.}}\\
{\bf Sample Construction}\\ 
\end{center}

\baselineskip 12pt
\noindent
Stocks must have 60 months non-missing data on CRSP and a book value in CRSP/Compustat in months $[-18, \; -6]$ to be eligible.  Next a real dollar minimum is applied
(\$110 million in December 2021).  Finally, the smallest 10\% of stocks are removed prior to Nasdaq eligibility, and 20\% after Nasdaq stocks enter the sample (after December,
1977).  This table shows the effect of each filter on the sample size, as well as the smallest and median sized stocks eligible for the ensuing month in \$ millions.

\vskip .1in

\hoffset=-0.85in

\begin{center}

\begin{tabular}{lrrrrr}

      & Eligible & \multicolumn{1}{c}{After} & \multicolumn{1}{c}{Final} & \multicolumn{1}{c}{Minimum} & \multicolumn{1}{c}{Median}\\
 \multicolumn{1}{c}{Month End}      & \multicolumn{1}{c}{Stocks}   & \$ criterion & Sample & Mkt. Cap. & Mkt. Cap. \\
\hline
January-2019	&	2,424	&	2,013	&	1,611	&	393,212	&	2,894,347\\
February-2019	&	2,433	&	2,028	&	1,623	&	402,867	&	2,932,791\\
March-2019	&	2,452	&	2,031	&	1,625	&	394,455	&	2,855,955\\
April-2019	&	2,464	&	2,044	&	1,636	&	400,243	&	2,965,515\\
May-2019	&	2,468	&	2,014	&	1,612	&	389,269	&	2,710,687\\
June-2019	&	2,474	&	2,036	&	1,629	&	399,359	&	2,932,450\\
July-2019	&	2,484	&	2,045	&	1,636	&	385,297	&	2,904,489\\
August-2019	&	2,483	&	2,021	&	1,617	&	379,908	&	2,792,264\\
September-2019	&	2,480	&	2,031	&	1,625	&	373,952	&	2,769,918\\
October-2019	&	2,476	&	2,027	&	1,622	&	389,818	&	2,950,736\\
November-2019	&	2,490	&	2,043	&	1,635	&	399,210	&	2,974,486\\
December-2019	&	2,479	&	2,048	&	1,639	&	412,608	&	3,038,905\\
January-2020	&	2,480	&	2,044	&	1,636	&	374,938	&	2,941,826\\
February-2020	&	2,482	&	2,027	&	1,622	&	353,533	&	2,704,327\\
March-2020	&	2,479	&	1,930	&	1,544	&	309,687	&	2,265,374\\
April-2020	&	2,483	&	1,985	&	1,588	&	329,175	&	2,450,341\\
May-2020	&	2,463	&	1,997	&	1,598	&	322,659	&	2,567,540\\
June-2020	&	2,468	&	2,022	&	1,618	&	324,918	&	2,584,908\\
July-2020	&	2,481	&	2,039	&	1,632	&	323,619	&	2,665,451\\
August-2020	&	2,484	&	2,052	&	1,642	&	340,131	&	2,750,091\\
September-2020	&	2,482	&	2,036	&	1,629	&	348,337	&	2,697,722\\
October-2020	&	2,489	&	2,039	&	1,632	&	347,535	&	2,773,594\\
November-2020	&	2,492	&	2,106	&	1,685	&	372,456	&	3,101,724\\
December-2020	&	2,488	&	2,120	&	1,696	&	377,784	&	3,261,970\\
January-2021	&	2,481	&	2,144	&	1,716	&	380,913	&	3,255,209\\
February-2021	&	2,476	&	2,177	&	1,742	&	378,138	&	3,400,956\\
March-2021	&	2,474	&	2,187	&	1,750	&	395,257	&	3,477,867\\
April-2021	&	2,469	&	2,172	&	1,738	&	408,353	&	3,520,544\\
May-2021	&	2,477	&	2,178	&	1,743	&	410,111	&	3,586,850\\
June-2021	&	2,477	&	2,192	&	1,754	&	400,369	&	3,571,701\\
July-2021	&	2,484	&	2,173	&	1,739	&	398,008	&	3,587,144\\
August-2021	&	2,484	&	2,173	&	1,739	&	401,794	&	3,594,588\\
September-2021	&	2,481	&	2,161	&	1,729	&	411,103	&	3,507,657\\
October-2021	&	2,480	&	2,149	&	1,720	&	443,766	&	3,748,788\\
November-2021	&	2,477	&	2,116	&	1,693	&	437,320	&	3,737,703\\
\hline

\end{tabular}

\end{center}

\newpage 

\begin{center}
{\Large {\bf Table IA-2}}\\
{\bf Ex-post Out-of-Sample Optimal $\gamma^*$ and characteristic set:}\\
{\bf Sampling properties of certainty equivalent returns}\\
{\bf For investor with power utility and coefficient of relative risk aversion,} \boldmath{$\gamma = 5.$}\\
\end{center} 

\baselineskip 12pt
\noindent
Weight tilts ($\theta$) are estimated for 63 characteristic sets under each of 14 values of the loss function curvature ($\gamma^*$), using both
rolling and updating protocols.  Of these 882 cases that with the highest 1\%ile value of the out of sample Certainty Equivalent is reported in basis points per month.
The characteristic symbols are: $\zeta$: momentum, V: book-to-market ratio, S: log size, $\beta$: from lagged 60-month market model,
$\overline{r}$: average same-month return over the previous 5 years, $\sigma_\epsilon$: standard deviation of lagged 60-month market model
residual.
     
\vskip .1in

\hoffset=-0.85in

\begin{tabular}{lccrrrcccrrr} 

     & \multicolumn{5}{c}{Updating Protocol} & & \multicolumn{5}{c}{Rolling Protocol}\\
\cline{2-6} \cline{8-12}  
Next & \multicolumn{2}{c}{Optimal} & \multicolumn{3}{c}{Certainty Equivalent} & &  \multicolumn{2}{c}{Optimal} &  \multicolumn{3}{c}{Certainty Equivalent}\\
\cline{2-3} \cline{8-9}
Year & Chars & $\gamma^*$ & 1\%ile & Mean & Std Dev & & Chars & $\gamma^*$ & 1\%ile & Mean & Std Dev\\
\hline
1990 &                   {\bf VWI} &          &   76.2 & 84.1 & 3.3  & & {\bf VWI}                              &   & 76.2 & 84.1 & 3.3\\
1990 &                   {\bf EWI} &          &   95.0 & 99.5 & 1.9  & & {\bf EWI}                              &   & 95.0 & 99.5 & 1.9\\
\hline
1990 & $\zeta$,V,S,$\overline{r}$,$\sigma_\epsilon$ & 5 & 313.6 & 372.3 & 26.1 & & $\zeta$,V,S,$\overline{r}$,$\sigma_\epsilon$ & 5 & 314.4 & 373.7 & 26.6\\
1991 & $\zeta$,V,S,$\overline{r}$,$\sigma_\epsilon$ & 5 & 290.0 & 345.2 & 24.3 & & $\zeta$,V,S,$\overline{r}$,$\sigma_\epsilon$ & 5 & 285.6 & 342.1 & 25.5\\
1992 & $\zeta$,V,S,$\overline{r}$,$\sigma_\epsilon$ & 5 & 303.8 & 357.5 & 24.1  & & $\zeta$,V,S,$\beta$,$\overline{r}$,$\sigma_\epsilon$ & 5 & 281.5 & 342.3 & 27.7\\
1993 & $\zeta$,V,S,$\overline{r}$,$\sigma_\epsilon$ & 5 & 305.4 & 357.8 & 23.5  & & $\zeta$,V,S,$\beta$,$\overline{r}$,$\sigma_\epsilon$ & 6 & 270.1 & 325.6 & 25.6\\
1994 & $\zeta$,V,S,$\overline{r}$,$\sigma_\epsilon$ & 5 & 307.5 & 359.2 & 23.1  & & $\zeta$,V,S,$\beta$,$\overline{r}$,$\sigma_\epsilon$ & 6 & 271.6 & 327.7 & 25.9\\
1995 & $\zeta$,V,S,$\overline{r}$,$\sigma_\epsilon$ & 5 & 297.4 & 347.5 & 22.4  & & $\zeta$,V,S,$\beta$,$\overline{r}$,$\sigma_\epsilon$ & 6 & 260.4 & 318.9 & 26.8\\
1996 & $\zeta$,V,S,$\beta$,$\overline{r}$,$\sigma_\epsilon$ & 5 & 291.8 & 341.9 & 23.0  & & $\zeta$,V,S,$\overline{r}$,$\sigma_\epsilon$ & 5 & 288.9 & 353.8 & 29.8\\
1997 & $\zeta$,V,S,$\beta$,$\overline{r}$,$\sigma_\epsilon$ & 5 & 286.2 & 334.2 & 22.1  & & $\zeta$,V,S,$\overline{r}$,$\sigma_\epsilon$ & 5 & 291.9 & 358.7 & 30.2\\
1998 & $\zeta$,V,S,$\beta$,$\overline{r}$,$\sigma_\epsilon$ & 5 & 306.4 & 354.8 & 22.5  & & $\zeta$,V,S,$\overline{r}$,$\sigma_\epsilon$ & 5 & 306.2 & 377.8 & 33.1\\
1999 & $\zeta$,V,S,$\beta$,$\overline{r}$,$\sigma_\epsilon$ & 6 & 279.4 & 321.5 & 19.2  & & $\zeta$,V,$\overline{r}$,$\sigma_\epsilon$ & 6 & 251.0 & 319.7 & 30.4\\
2000 & $\zeta$,V,S,$\beta$,$\overline{r}$,$\sigma_\epsilon$ & 5 & 267.8 & 318.7 & 23.3  & & $\zeta$,V,$\overline{r}$,$\sigma_\epsilon$ & 7 & 219.7 & 296.7 & 32.6\\
2001 & $\zeta$,S,$\beta$,$\overline{r}$,$\sigma_\epsilon$ & 7 & 217.5 & 273.8 & 21.5  & & $\zeta$,$\sigma_\epsilon$ & 7 & 110.8 & 141.7 & 13.7\\
2002 & $\zeta$,S,$\beta$,$\overline{r}$,$\sigma_\epsilon$ & 8 & 188.3 & 238.6 & 20.1  & & $\overline{r}$ & 22 & 71.4 & 93.2 & 9.2\\
2003 & $\zeta$,S,$\beta$,$\overline{r}$,$\sigma_\epsilon$ & 8 & 193.1 & 243.9 & 20.3  & & $\zeta$,V & 7 & 83.0 & 114.6 & 14.4\\
2004 & $\zeta$,S,$\beta$,$\overline{r}$,$\sigma_\epsilon$ & 8 & 184.2 & 231.2 & 18.7  & & $\zeta$,V & 8 & 76.3 & 107.3 & 13.8\\
2005 & $\zeta$,S,$\beta$,$\overline{r}$,$\sigma_\epsilon$ & 8 & 183.2 & 230.2 & 18.3  & & $\zeta$,V & 8 & 68.3 & 98.3 & 13.2\\
2006 & $\zeta$,S,$\overline{r}$,$\sigma_\epsilon$ & 9 & 181.1 & 220.5 & 15.6  & & $\zeta$,V,& 8 & 100.9 & 129.8 & 13.1\\
2007 & $\zeta$,S,$\beta$,$\overline{r}$,$\sigma_\epsilon$ & 9 & 177.3 & 215.1 & 15.5  & & $\zeta$,V & 8 & 95.8 & 124.0 & 12.9\\
2008 & $\zeta$,S,$\overline{r}$,$\sigma_\epsilon$ & 9 & 166.3 & 201.8 & 14.0  & & $\zeta$,V & 8 & 65.3 & 90.8 & 11.1\\
2009 & $\zeta$,S,$\overline{r}$,$\sigma_\epsilon$ & 9 & 157.7 & 192.6 & 13.9  & & {\bf EWI} &   &  13.8 & 19.0 & 2.2\\
2010 & $\zeta$,S,$\overline{r}$,$\sigma_\epsilon$ & 11 & 139.9 & 166.7 & 11.4  & & {\bf EWI} &   & 23.4 & 28.8 & 2.3\\
2011 & $\zeta$,S,$\beta$,$\overline{r}$,$\sigma_\epsilon$ & 9 & 139.4 & 173.6 & 14.1  & & {\bf EWI} &   & 16.5 & 21.8 & 2.3\\
2012 & $\zeta$,S,$\beta$,$\overline{r}$,$\sigma_\epsilon$ & 9 & 139.1 & 173.0 & 14.0  & & {\bf EWI} &   & 1.0 & 6.2 & 2.3\\
2013 & $\zeta$,S,$\beta$,$\overline{r}$,$\sigma_\epsilon$ & 9 & 137.9 & 170.6 & 13.6  & & {\bf EWI} &  & -0.4 & 1.3 & 2.3\\
2014 & $\zeta$,S,$\beta$,$\overline{r}$,$\sigma_\epsilon$ & 9 & 140.9 & 173.1 & 13.3  & & {\bf EWI} &  & 19.9 & 25.2 & 2.3\\
2015 & $\zeta$,S,$\overline{r}$,$\sigma_\epsilon$ & 11 & 137.1 & 161.3 & 10.2  & & {\bf EWI} &  & 16.4 & 21.5 & 2.2\\
2016 & $\zeta$,S,$\overline{r}$,$\sigma_\epsilon$ & 11 & 135.7 & 159.5 & 10.0  & & S,$\overline{r}$,$\sigma_\epsilon$ & 11 & 43.9 & 62.5 & 8.4\\
2017 & $\zeta$,S,$\overline{r}$,$\sigma_\epsilon$ & 11 & 134.7 & 157.7 & 9.6  & & S,$\overline{r}$,$\sigma_\epsilon$ & 12 & 42.9 & 56.5 & 6.2\\
2018 & $\zeta$,S,$\overline{r}$,$\sigma_\epsilon$ & 11 & 133.1 & 154.8 & 9.2  & & S,$\sigma_\epsilon$ & 22 & 54.2 & 64.3 & 4.4\\
2019 & $\zeta$,S,$\overline{r}$,$\sigma_\epsilon$ & 11 & 128.8 & 150.2 & 9.1  & & V,S,$\sigma_\epsilon$ & 22 & 41.6 & 55.2 & 5.7\\
2020 & $\zeta$,S,$\overline{r}$,$\sigma_\epsilon$ & 11 & 128.3 & 149.1 & 8.7  & & V,$\sigma_\epsilon$ & 22 & 48.9 & 62.1 & 5.7\\
2021 & $\zeta$,S,$\overline{r}$,$\sigma_\epsilon$ & 11 & 124.9 & 145.1 & 8.5  & & V,S,$\sigma_\epsilon$ & 22 & 45.2 & 61.2 & 6.8\\
\hline

\end{tabular}

\newpage

\begin{center}

\baselineskip 14pt
{\Large {\bf Table IA-3}}\\
{\bf Optimal $\gamma^*$ and each characteristic sets:}\\
{\bf Sampling properties of certainty equivalent returns}\\
{\bf For investor with power utility and coefficient of relative risk aversion,} \boldmath{$\gamma = 8.$}\\
\end{center}

\baselineskip 12pt
\noindent
Weight tilts ($\theta$) are estimated for 63 characteristic sets under each of 14 values of the loss function curvature ($\gamma^*$), using both
rolling and updating protocols.  Of these 882 cases that with the highest 1\%ile value of the out of sample Certainty Equivalent is reported in basis points per month.
The characteristic symbols are: $\zeta$: momentum, V: book-to-market ratio, S: log size, $\beta$: from lagged 60-month market model,
$\overline{r}$: average same-month return over the previous 5 years, $\sigma_\epsilon$: standard deviation of lagged 60-month market model
residual.

\vskip .1in

\hoffset=-0.84in

\begin{tabular}{lccrrrcccrrr}

     & \multicolumn{5}{c}{Updating Protocol} & & \multicolumn{5}{c}{Rolling Protocol}\\
\cline{2-6} \cline{8-12}
Next & \multicolumn{2}{c}{Optimal} & \multicolumn{3}{c}{Certainty Equivalent} & &  \multicolumn{2}{c}{Optimal} &  \multicolumn{3}{c}{Certainty Equivalent}\\
\cline{2-3} \cline{8-9}
Year & Chars & $\gamma^*$ & 1\%ile & Mean & Std Dev & & Chars & $\gamma^*$ & 1\%ile & Mean & Std Dev\\
\hline
1990 & {\bf VWI} & & 32.3                                 &  41.4 &  3.9  &      &  {\bf VWI} &                                        &  32.3 & 41.4 & 3.9\\
1990 & {\bf EWI} & & 30.4                                 &  36.5 &  2.6  &      &  {\bf EWI} &                                        &  30.4 & 36.5 & 2.6\\
1990 & $\zeta$,V,S,$\beta$,$\overline{r}$,$\sigma_\epsilon$ & 8 & 220.7 & 262.7 & 18.5 & & $\zeta$,V,S,$\beta$,$\overline{r}$,$\sigma_\epsilon$ & 8 & 219.8 & 262.2 & 18.6\\
\hline
1990 & $\zeta$,V,S,$\beta$,$\overline{r}$,$\sigma_\epsilon$ & 9 & 222.9 & 260.7 & 16.8 & & $\zeta$,V,S,$\beta$,$\overline{r}$,$\sigma_\epsilon$ & 9 & 220.3 & 258.7 & 17.0\\
1991 & $\zeta$,V,S,$\beta$,$\overline{r}$,$\sigma_\epsilon$ & 9 & 203.0 & 239.2 & 15.9 & & $\zeta$,V,S,$\beta$,$\overline{r}$,$\sigma_\epsilon$ & 9 & 197.5 & 234.2 & 16.6\\
1992 & $\zeta$,V,S,$\overline{r}$,$\sigma_\epsilon$ & 8 & 213.0 & 252.8 & 21.3  & & $\zeta$,V,S,$\beta$,$\overline{r}$,$\sigma_\epsilon$ & 9 & 194.3 & 233.0 & 17.4\\
1993 & $\zeta$,V,S,$\overline{r}$,$\sigma_\epsilon$ & 8 & 214.4 & 253.2 & 17.0  & & $\zeta$,V,S,$\beta$,$\overline{r}$,$\sigma_\epsilon$ & 9 & 183.4 & 225.6 & 18.6\\
1994 & $\zeta$,V,S,$\overline{r}$,$\sigma_\epsilon$ & 8 & 215.2 & 252.8 & 16.6  & & $\zeta$,V,S,$\beta$,$\overline{r}$,$\sigma_\epsilon$ & 9 & 185.8 & 227.7 & 18.6\\
1995 & $\zeta$,V,S,$\overline{r}$,$\sigma_\epsilon$ & 8 & 208.0 & 244.2 & 16.0  & & $\zeta$,V,S,$\beta$,$\overline{r}$,$\sigma_\epsilon$ & 10 & 179.0 & 218.9 & 17.8\\
1996 & $\zeta$,V,S,$\overline{r}$,$\sigma_\epsilon$ & 8 & 207.1 & 241.7 & 15.2  & & $\zeta$,V,S,$\beta$,$\overline{r}$,$\sigma_\epsilon$ & 8 & 204.3 & 250.1 & 20.3\\
1997 & $\zeta$,V,S,$\overline{r}$,$\sigma_\epsilon$ & 8 & 205.9 & 239.0 & 14.6  & & $\zeta$,V,S,$\overline{r}$,$\sigma_\epsilon$ & 8 & 208.4 & 254.5 & 20.5\\
1998 & $\zeta$,V,S,$\overline{r}$,$\sigma_\epsilon$ & 8 & 221.2 & 254.2 & 14.6  & & $\zeta$,V,S,$\beta$,$\overline{r}$,$\sigma_\epsilon$ & 8 & 214.6 & 264.4 & 22.2\\
1999 & $\zeta$,V,S,$\overline{r}$,$\sigma_\epsilon$ & 9 & 206.2 & 235.0 & 12.8  & & $\zeta$,V,S,$\beta$,$\overline{r}$,$\sigma_\epsilon$ & 12 & 173.7 & 211.1 & 16.7\\
2000 & $\zeta$,V,S,$\overline{r}$,$\sigma_\epsilon$ & 9 & 186.2 & 217.7 & 13.4  & & $\zeta$,$\overline{r}$,$\sigma_\epsilon$ & 11 & 125.5 & 166.6 & 17.2\\
2001 & $\zeta$,S,$\sigma_\epsilon$ & 10 & 135.0 & 160.6 & 11.0  & & $\zeta$,$\sigma_\epsilon$ & 8 & 52.6 & 85.6 & 14.0\\
2002 & $\zeta$,S,$\sigma_\epsilon$ & 16 & 76.4 & 105.3 & 11.5  & & {\bf VWI} &    & 9.4 & 21.5 & 5.2\\
2003 & $\zeta$,S,$\sigma_\epsilon$ & 16 & 75.6 & 103.9 & 11.2  & & $\zeta$,V & 9 & 36.3 & 64.4 & 12.2\\
2004 & $\zeta$,S,$\sigma_\epsilon$ & 16 & 75.9 & 102.7 & 10.7  & & $\zeta$,V & 9 & 30.2 & 59.1 & 12.6\\
2005 & $\zeta$,S,$\sigma_\epsilon$ & 16 & 77.8 & 104.0 & 10.4  & & {\bf EWI} &  & 24.0 & 29.2 & 2.3\\
2006 & $\zeta$,S,$\sigma_\epsilon$ & 16 & 77.1 & 102.6 & 10.1  & & $\zeta$,V & 9 & 58.5 & 86.9 & 12.2\\
2007 & $\zeta$,S,$\sigma_\epsilon$ & 16 & 78.7 & 103.5 & 9.8  & & $\zeta$,V & 9 & 56.0 & 84.0 & 12.1\\
2008 & $\zeta$,S,$\sigma_\epsilon$ & 16 & 74.3 & 97.9 & 9.3  & & $\zeta$,V & 9 & 27.3 & 53.1 & 10.8\\
2009 & $\zeta$,S,$\sigma_\epsilon$ & 16 & 63.4 & 87.1 & 9.4  & & {\bf VWI} &   & -29.0 & -15.7 & 5.5\\
2010 & $\zeta$,S,$\sigma_\epsilon$ & 16 & 56.0 & 77.4 & 8.6  & & {\bf VWI} &   & -24.7 & -11.0 & 5.8\\
2011 & $\zeta$,S,$\sigma_\epsilon$ & 16 & 55.9 & 76.9 & 8.4  & & {\bf VWI} &   & -38.4 & -25.0 & 5.8\\
2012 & $\zeta$,S,$\sigma_\epsilon$ & 16 & 55.8 & 76.6 & 8.3  & & {\bf VWI} &   & -51.0 & -37.7 & 5.7\\
2013 & $\zeta$,S,$\sigma_\epsilon$ & 16 & 56.7 & 76.9 & 8.1  & & {\bf VWI} &  & -56.1 & -42.6 & 5.7\\
2014 & $\zeta$,S,$\sigma_\epsilon$ & 16 & 61.0 & 80.6 & 7.9  & & {\bf EWI} &  & -35.3 & -29.0 & 2.7\\
2015 & $\zeta$,S,$\sigma_\epsilon$ & 16 & 61.0 & 80.0 & 7.7  & & {\bf EWI} &  & -38.3 & -32.3 & 2.6\\
2016 & $\zeta$,S,$\sigma_\epsilon$ & 16 & 60.5 & 79.4 & 7.5  & & S,$\beta$,$\overline{r}$ & 13 & 11.8 & 26.7 & 6.5\\
2017 & $\zeta$,S,$\sigma_\epsilon$ & 16 & 62.4 & 80.9 & 7.4  & & V,S,$\sigma_\epsilon$ & 16 & 16.7 & 32.3 & 6.5\\
2018 & $\zeta$,S,$\sigma_\epsilon$ & 16 & 63.6 & 81.8 & 7.2  & & V,S,$\sigma_\epsilon$ & 22 & 30.4 & 45.0 & 6.0\\
2019 & $\zeta$,S,$\sigma_\epsilon$ & 16 & 60.7 & 78.3 & 7.1  & & V,$\sigma_\epsilon$ & 22 & 20.7 & 34.2 & 5.8\\
2020 & $\zeta$,S,$\sigma_\epsilon$ & 16 & 63.2 & 80.5 & 6.9  & & V,$\sigma_\epsilon$ & 22 & 28.9 & 43.1 & 6.1\\
2021 & $\zeta$,S,$\sigma_\epsilon$ & 16 & 59.0 & 75.7 & 6.7  & & V,$\sigma_\epsilon$ & 22 & 21.3 & 37.9 & 7.1\\
\hline

\end{tabular}

\newpage

\begin{landscape}

\voffset=-1.85in

\hoffset=-1in

\begin{center}

\baselineskip 14pt
{\Large {\bf Table IA-4}}\\
{\bf Sampling properties of out-of-sample Portfolio Return Statistics: First Subperiod}\\
\end{center}

\vskip .1in

\begin{tabular}{lrrrrrrrcrrrrrrr}
   
\multicolumn{16}{l}{Sampling properties of dynamic optimal PPPs.  Portfolio characteristic tilts from the best out-of-sampling performer over the relevant preceding period}\\
\multicolumn{16}{l}{(shown in Table 1) each year.  $\cal{CE}$$_k$ is the certainty equivalent return in basis points per month for a power utility investor with coefficient of relative}\\
\multicolumn{16}{l}{risk aversion $(\gamma) = k$.  $E(r)$, $\sigma$, Median, IQR, and MIN are the mean monthly return, the
standard deviation of monthly returns, the median}\\
\multicolumn{16}{l}{monthly return, the interquartile range of monthly returns, and the minimum monthly return--all expressed in basis points per month.}\\
\multicolumn{16}{l}{SKEW and KURT are the return skewness and kurtosis measures, and SR is the Sharpe ratio.}\\
\multicolumn{16}{l}{Results are for the first 9-year out-of-sample subperiod (1990 -- 1998).}\\

    &    &   &   &  &  & & & \hspace{.1in}  &  & & & &  &  &  \\

\multicolumn{16}{l}{{\bf Panel A: Benchmark portfolios}}\\
     & \multicolumn{7}{c}{First 9-year subperiod -- VWI} & & \multicolumn{7}{c}{First 9-year subperiod -- EWI}\\
\cline{2-8} \cline{10-16}
Statistic  & Mean & Std Dev &  2.5\%ile & 25\%ile & Median & 75\%ile & 97.5\%ile & &  Mean & Std Dev &  2.5\%ile & 25\%ile & Median & 75\%ile & 97.5\%ile\\
$\cal{CE}$$_2$ & 128.8 & 4.0 & 121.0 & 126.1 & 128.8 & 131.5 & 136.6 & & 109.4 & 2.3 & 104.8 & 107.9 & 109.4 & 110.9 & 113.9\\
$\cal{CE}$$_5$ & 104.1 & 4.1 & 96.1 & 101.4 & 104.2 & 107.0 & 112.3 & & 77.6 & 2.4 & 72.9 & 75.9 & 77.6 & 79.2 & 82.2\\
$\cal{CE}$$_8$ & 77.2 & 4.5 & 68.3 & 74.1 & 77.2 & 80.3 & 86.1 &  & 41.3 & 2.7 & 36.2 & 39.5 & 41.4 & 43.2 & 46.4\\
$E(r)$ & 144.2 & 4.0 & 136.2 & 141.5 & 144.2 & 146.9 & 152.0 & & 128.6 & 2.3 &  124.1 &  127.1 &   128.6 & 130.1 & 133.1\\
$\sigma$ & 388.6 & 4.6 &  379.6 & 385.5 & 388.6 & 391.7 & 397.7 & & 430.0 &  2.3 &   425.1 &  428.3 &   430.0 &  431.6 &  434.9\\
Median & 167.4 & 11.6 & 144.9 & 159.3 & 167.6 & 175.3 & 189.7 & & 187.5 & 9.8 &  168.1 & 180.0 & 187.6 & 194.1 & 206.5\\
IQR & 469.9 &  22.4 & 426.6 & 454.5 & 469.8 & 484.9 & 514.7 & & 508.7 & 17.5 &     474.6 & 496.7 & 508.7 &  520.5 & 543.3\\
MIN & -1,482.9 & 59.1 & -1,602.5 & -1,521.9 & -1,481.9 & -1,442.2 & -1,368.6 & & -1,747.3 & 28.2 & -1,803.2 & -1,766.3 & -1,747.3 & -1,728.2 & -1,691.9\\
SKEW & -6.0 & 2.9 & -11.6 & -8.0 & -6.0 & -4.0 & -0.4 & & -13.7 & 2.2 & -18.0 & -15.2 & -13.7 &  -12.2 &  -9.3\\
KURT & 26.4 & 4.6 &  17.4 & 23.3 & 26.4 & 29.4 & 35.6 & & 35.2 & 2.5 & 30.3 &  33.5 &  35.2 & 36.9 &  40.2\\
SR & 0.9337 &  0.0364 &  0.8627 &  0.9095 &  0.9337 &  0.9581 &  1.0051 & & 0.7158 &  0.0187 &  0.6787 &  0.7034 &  0.7159 &  0.7284 &  0.7520\\
\hline

   &    &   &   &  &  & & & \hspace{.1in}  &  & & & &  &  &  \\

\end{tabular}

\newpage

\begin{center}

\baselineskip 14pt
{\Large {\bf Table IA-4 --2--}}\\ 
{\bf Sampling properties of out-of-sample Portfolio Return Statistics: First Subperiod  --2--}\\
\end{center}

\baselineskip 12pt

\vskip .1in

\begin{tabular}{lrrrrrrrcrrrrrrr}

\multicolumn{16}{l}{Sampling properties of dynamic optimal PPPs.  Portfolio characteristic tilts from the best out-of-sampling performer over the relevant preceding period}\\
\multicolumn{16}{l}{(shown in Table 1) each year.  $\cal{CE}$$_k$ is the certainty equivalent return in basis points per month for a power utility investor with coefficient of relative}\\
\multicolumn{16}{l}{risk aversion $(\gamma) = k$.  $E(r)$, $\sigma$, Median, IQR, and MIN are the mean monthly return, the
standard deviation of monthly returns, the median}\\
\multicolumn{16}{l}{monthly return, the interquartile range of monthly returns, and the minimum monthly return--all expressed in basis points per month.}\\
\multicolumn{16}{l}{SKEW and KURT are the return skewness and kurtosis measures, and SR is the Sharpe ratio.}\\
\multicolumn{16}{l}{Results are for the first 9-year out-of-sample subperiod (1990 -- 1998).}\\

    &    &   &   &  &  & & & \hspace{.1in}  &  & & & &  &  &  \\

\multicolumn{16}{l}{{\bf Panel B:  \boldmath{$\gamma$} 5 loss function}}\\
     & \multicolumn{7}{c}{First subperiod (1990 - 1998) -- Dyn. Opt. Updating} & & \multicolumn{7}{c}{First subperiod (1990 - 1998) -- Dyn. Opt. Rolling}\\
\cline{2-8} \cline{10-16}
Statistic  & Mean & Std Dev &  2.5\%ile & 25\%ile & Median & 75\%ile & 97.5\%ile & &  Mean & Std Dev &  2.5\%ile & 25\%ile & Median & 75\%ile & 97.5\%ile\\
$\cal{CE}$$_5$ & 250.1 & 36.3 & 173.8 & 228.8 & 251.6 & 274.8 & 317.8 &  &  214.6 & 83.7 & 20.7 & 175.9 & 224.4 & 267.2 & 338.1\\
$E(r)$ & 431.0 & 34.1 & 365.5 & 407.6 & 429.9 & 453.9 & 499.1 & & 580.5 & 50.6 & 485.2 & 545.7 & 579.0 &  613.5 & 682.6\\
$\sigma$ & 831.8 & 53.3 & 732.9 & 794.2 & 829.9 & 867.8 & 938.5 & & 1,185.3 & 74.5 & 1,046.9 & 1,133.3 & 1,182.4 & 1,233.1 & 1,340.5\\
Median & 433.1 & 53.6 & 331.9 & 397.2 & 432.3 & 467.7 & 540.6 & & 522.3 & 74.3 &  378.6 & 472.0 & 521.6 &  571.5 & 671.9\\
IQR & 1,003.6 & 103.3 & 810.3 & 932.1 & 999.8 & 1,072.9 & 1,211.6 & & 1,473.6 & 139.0 & 1,211.8 & 1,378.0 & 1,470.0 & 1,566.2 & 1,757.8\\
MIN & -2,557.6 & 452.3 &  -3,483.6 & -2,852.2 & -2,546.9 & -2,242.3 & -1,710.2 & & -3,175.5 & 727.9 & -4,707.2 & -3,662.6 & -3,143.7 & -2,640.3 & -1,907.7\\
SKEW & -0.2 & 5.2 & -10.3 & -3.7 &  -0.3 & 3.2 & 10.0 & & 4.9 & 5.4 & -6.1 & 1.3 & 5.0 & 8.5 & 15.4\\
KURT & 22.5 & 13.2 & -2.3 & 13.6 & 22.2 & 31.3 & 49.3 & & 16.9 & 14.0 & -9.8 & 7.3 & 16.6 &  26.2 & 45.1\\
SR & 1.6441 &  0.1318 &  1.3880 &  1.5533 &  1.6440 &  1.7320 &  1.9068 & &  1.5950 &  0.1317 &  1.3340 &  1.5057 &  1.5940 &  1.6819 &  1.8545\\
\hline

   &    &   &   &  &  & & & \hspace{.1in}  &  & & & &  &  &  \\

\multicolumn{16}{l}{{\bf Panel C:  \boldmath{$\gamma$} 8 loss function}}\\
     & \multicolumn{7}{c}{First subperiod (1990 - 1998) -- Dyn. Opt. Updating} & & \multicolumn{7}{c}{First subperiod (1990 - 1998) -- Dyn. Opt. Rolling}\\
\cline{2-8} \cline{10-16}
Statistic  & Mean & Std Dev &  2.5\%ile & 25\%ile & Median & 75\%ile & 97.5\%ile & &  Mean & Std Dev &  2.5\%ile & 25\%ile & Median & 75\%ile & 97.5\%ile\\
$\cal{CE}$$_8$ & 182.2 & 25.2 & 129.1 & 166.7 & 183.5 & 199.5 & 227.7 &  &  156.1 & 47.5 & 48.1 & 131.3 & 160.8 & 188.0 & 232.3\\
$E(r)$ & 320.3 &  22.8 & 278.1 & 304.8 & 319.5 & 335.4 & 366.8 & & 400.5 & 33.3 &  338.5 & 377.4 & 399.6 & 422.0 & 469.8\\
$\sigma$ & 576.1 & 37.8 & 506.2 & 550.2 & 574.7 & 600.5 & 653.7 & & 780.3 & 47.9 &  691.6 & 746.9 & 778.4 &     810.6 &     879.9\\
Median & 315.3 & 36.0 & 247.0 & 290.2 & 314.4 & 339.0 & 389.0 & & 354.8 & 48.0 & 264.4 & 322.1 & 353.4 & 386.4 &  452.6\\
IQR & 702.1 & 68.6 & 575.7 & 653.7 & 699.5 & 747.8 & 839.9 & & 964.1 & 87.6 & 797.2 &     904.7 &     963.3 &   1,021.6 & 1,141.6\\
MIN & -1,749.4 & 309.3 & -2,379.6 & -1,950.6 & -1,738.4 &  -1,541.0 & -1,165.9 & & -2,040.7 & 470.7 & -3,046.9 & -2,353.3 & -2,019.0 & -1,694.4 &  -1211.3\\
SKEW & -0.9 & 5.0 & -8.9 & -2.6 & 0.9 & 4.3 & 10.8 & & 5.8 &       5.2 &   -4.6 &      2.3 & 5.8 & 9.4 & 15.8\\
KURT & 21.3 & 11.7 & -1.3 & 13.3 & 21.1 & 29.0 & 45.0 & & 18.8 & 14.1 & -8.0 & 8.4 &  17.3 & 27.1 & 47.1\\
SR & 1.7007 &  0.1266 &  1.4550 &  1.6135 &  1.6991 &  1.7849 &  1.9503 & &  1.6153 &  0.1318 &  1.3634 &  1.5236 &  1.6150 &  1.7026 &  1.8793\\
\hline

\end{tabular}

\newpage 
   
\hoffset=-1in

\begin{center}

\baselineskip 14pt
{\Large {\bf Table IA-5}}\\ 
{\bf Sampling properties of out-of-sample Portfolio Return Statistics: Second Subperiod}\\
\end{center}

\vskip .1in 

\begin{tabular}{lrrrrrrrcrrrrrrr}
   
\multicolumn{16}{l}{Sampling properties of dynamic optimal PPPs.  Portfolio characteristic tilts from the best out-of-sampling performer over the relevant preceding period}\\
\multicolumn{16}{l}{(shown in Table 1) each year.  $\cal{CE}$$_k$ is the certainty equivalent return in basis points per month for a power utility investor with coefficient of relative}\\
\multicolumn{16}{l}{risk aversion $(\gamma) = k$.  $E(r)$, $\sigma$, Median, IQR, and MIN are the mean monthly return, the
standard deviation of monthly returns, the median}\\
\multicolumn{16}{l}{monthly return, the interquartile range of monthly returns, and the minimum monthly return--all expressed in basis points per month.}\\
\multicolumn{16}{l}{SKEW and KURT are the return skewness and kurtosis measures, and SR is the Sharpe ratio.}\\
\multicolumn{16}{l}{Results are for the second 23-year out-of-sample subperiod (1999 -- 2021).}\\
     
    &    &   &   &  &  & & & \hspace{.1in}  &  & & & &  &  &  \\

\multicolumn{16}{l}{{\bf Panel A: Benchmark portfolios}}\\
     & \multicolumn{7}{c}{Second subperiod (1999 - 2021) -- VWI} & & \multicolumn{7}{c}{Second subperiod (1999 - 2021) -- EWI}\\
\cline{2-8} \cline{10-16}
Statistic  & Mean & Std Dev &  2.5\%ile & 25\%ile & Median & 75\%ile & 97.5\%ile & &  Mean & Std Dev &  2.5\%ile & 25\%ile & Median & 75\%ile & 97.5\%ile\\
$\cal{CE}$$_2$ & 60.0 & 3.5 & 53.1 & 57.6 & 60.0 & 62.4 & 66.9 & & 76.6 & 1.7 & 73.4 & 75.4 & 76.6 & 77.7 & 79.8\\
$\cal{CE}$$_5$ & 29.8 & 3.8 & 22.4 & 27.3 & 29.9 & 32.4 & 37.4 &   & 27.8 & 1.8 & 24.3 & 26.6 & 27.8 & 29.0 & 31.2\\
$\cal{CE}$$_8$ & -2.5 & 4.2 & -10.6 & -5.3 & -2.5 & 0.2 & 5.7 &  & -26.9 & 2.2 & -31.2 & -28.4 & -26.9 & -25.3 & -22.61\\
$E(r)$ & 79.2 & 3.5 & 72.4 & 76.8 & 79.2 & 81.6 & 85.9 & & 106.8 & 1.7 & 103.6 & 105.6 & 106.8 & 107.9 & 110.0\\
$\sigma$ & 433.3 & 4.2 & 425.2 & 430.4 & 433.3 & 436.3 & 441.6 & & 542.7 & 2.1 & 538.6 & 541.3 & 542.7 & 544.1 & 546.8\\
Median & 124.0 & 8.0 & 107.9 & 118.6 & 124.1 & 129.4 & 139.4 & & 141.9 & 6.7 & 128.8 & 137.4 & 141.9 & 146.4 & 155.3\\
IQR & 511.2 & 13.9 & 484.6 & 501.8 & 510.9 & 520.6 & 538.8 & & 655.0 & 12.1 & 612.0 & 626.8 &  635.0 &     642.9 &     659.3\\
MIN & -1,667.4 & 78.6 & -1,819.4 & -1,720.6 & -1,667.6 & -1,613.6 & -1,514.3 & & -2119.7 & 42.4 & -2,206.5 & -2,147.9 & -2,118.6 & -2,090.3 & -2,039.7\\
SKEW & -10.3 & 1.8 & -13.9 & -11.5 & -10.4 & -9.1 &  -6.8 & & -6.5 & 1.2 & -8.9 &  -7.3 & -6.5 & -5.6 & -4.1\\
KURT & 27.9 & 3.8 & 20.6 & 25.3 &  27.9 & 30.4 & 35.4 & & 32.5 & 1.5 & 29.5 & 31.5 & 32.5 & 33.5 & 35.5\\
SR & 0.5245 &  0.0286 &  0.4687 &  0.5052 &  0.5246 &  0.5436 &  0.5807 & & 0.5959 &  0.0105 &  0.5756 &  0.5887 &  0.5957 &  0.6029 &  0.6165\\
\hline

\end{tabular}

\newpage

\voffset=-1.85in

\begin{center}

\baselineskip 14pt
{\Large {\bf Table IA-5 --2--}}\\
{\bf Sampling properties of out-of-sample Portfolio Return Statistics: Second Subperiod  --2--}\\
\end{center}

\baselineskip 12pt

\vskip .1in

\begin{tabular}{lrrrrrrrcrrrrrrr}

\multicolumn{16}{l}{{\bf Panel B:  \boldmath{$\gamma$} 5 loss function}}\\
     & \multicolumn{7}{c}{Second subperiod (1990 - 2021) -- Dyn. Opt. Updating} & & \multicolumn{7}{c}{Second subperiod (1990 - 2021) -- Dyn. Opt. Rolling}\\
\cline{2-8} \cline{10-16}
Statistic  & Mean & Std Dev &  2.5\%ile & 25\%ile & Median & 75\%ile & 97.5\%ile & &  Mean & Std Dev &  2.5\%ile & 25\%ile & Median & 75\%ile & 97.5\%ile\\
$\cal{CE}$$_5$ & -27.1 & 89.8 & -212.6 & -38.6 & -8.4 & 12.0 & 41.3 &  &  -334.0 & 668.7 & -1,690.6 & -295.0 & -172.2 & -113.1 & -52.8\\
$E(r)$ & 142.6 & 13.2 & 116.6 & 133.8 & 142.6 & 151.4 & 168.5 & & 104.0 & 13.7 & 78.1 & 94.7 &  103.9 &  113.0 & 131.4\\
$\sigma$ & 717.4 & 35.7 & 652.8 & 692.2 & 714.9 & 739.5 & 794.5 & & 841.1 & 52.3 &  748.0 & 805.0 & 837.5 &  873.6 & 953.9\\
Median & 169.6 &  20.0 &  131.0 &  156.1 &  169.3 & 183.0 & 209.7 & & 122.9 & 15.4 & 93.6 & 112.4 & 122.4 & 133.2 & 154.2\\
IQR & 746.3 & 36.2 & 676.5 & 721.5 & 746.0 & 770.5 & 818.7 & & 688.3 & 27.4 & 635.8 & 669.6 & 688.3 & 706.8 &  742.4\\
MIN & -3,575.9 & 899.6 & -5,689.1 & -4,118.0 & -3,390.7 & -2,900.2 & -2,283.4 & & -5,002.7 & 1,112.2 & -7,586.0 & -5,654.3 & -4,799.2 & -4,161.0 & -3,408.4\\
SKEW & -3.8 & 2.7 & -9.1 & -5.6 &  -3.8 & -2.0 & 1.6 & & -2.3 & 2.8 & -6.2 & -3.6 & -2.2 &  -0.9 & 1.6\\
KURT & 74.0 & 13.0 & 49.4 & 65.3 & 73.7 & 82.6 & 100.4 & & 120.4 & 2.0 & 93.9 &   110.9 &  120.2 &  129.7 & 148.2\\
SR & 0.6273 &  0.0726 &  0.4844 &  0.5785 &  0.6268 &  0.6756 &  0.7692 & & 0.3753 &  0.0611 &  0.2588 &  0.3336 &  0.3750 &  0.4165 &  0.4982\\
\hline

   &    &   &   &  &  & & & \hspace{.1in}  &  & & & &  &  &  \\

\multicolumn{16}{l}{{\bf Panel C:  \boldmath{$\gamma$} 8 loss function}}\\
     & \multicolumn{7}{c}{Second subperiod (1999 - 2021) -- Dyn. Opt. Updating} & & \multicolumn{7}{c}{Second subperiod (1990 - 2021) -- Dyn. Opt. Rolling}\\
\cline{2-8} \cline{10-16}
Statistic  & Mean & Std Dev &  2.5\%ile & 25\%ile & Median & 75\%ile & 97.5\%ile & &  Mean & Std Dev &  2.5\%ile & 25\%ile & Median & 75\%ile & 97.5\%ile\\
$\cal{CE}$$_8$ & -122.5 & 75.0 & -307.2 & -146.7 & -105.8 & -76.4 & -36.4 &  &  -265.8 & 172.2 & -687.0 & -301.3 & -224.7 & -173.0 & -110.4\\
$E(r)$ & 111.7 & 7.8 & 96.5 & 106.4 & 111.7 & 116.9 & 127.1 & & 70.4 & 8.6 & 53.9 & 64.6 & 70.4 & 76.1 & 87.0\\
$\sigma$ & 583.6 & 24.2 & 538.1 & 567.3 & 583.0 & 599.1 & 633.9 & & 653.0 &  30.8 &  597.2 &  631.5 & 651.7 & 672.9 & 718.0\\
Median & 136.7 & 12.1 & 113.4 & 128.5 & 136.5 & 144.9 & 160.7 & & 107.9 & 13.1 & 83.0 & 99.1 & 107.7 & 116.7 & 134.3\\
IQR & 504.9 & 24.9 & 457.0 & 487.9 & 505.1 & 521.6 & 554.0 & & 631.8 & 24.4 & 583.6 & 615.4 & 631.6 & 648.3 & 679.6\\
MIN & -3,317.6 & 497.1 & -4,447.6 & -3,613.3 & -3,254.6 & -2,960.9 & -2,506.6 & & -3,783.4 & 575.4 & -5,105.0 & -4,107.0 & -3,719.0 & -3,383.5 & -2,836.6\\
SKEW & -4.3 & 2.1 & -8.5 & -5.7 & -4.3 & -2.9 & -0.3 & & -5.7 & 208 & -9.7 & -7.1 & -5.7 & -4.4 & -1.8\\
KURT & 108.7 & 11.2 & 87.0 & 101.2 & 108.5 & 116.1 & 131.2 & & 75.7 & 11.1 & 54.9 & 68.1 & 75.4 & 83.1 &  98.5\\
SR & 0.5853 &  0.0524 &  0.4839 &  0.5500 &  0.5848 &  0.6204 &  0.6898 & & 0.3034 &  0.0493 &  0.2087 &  0.2697 &  0.3024 &  0.3363 &  0.4014\\
\hline

\end{tabular}

\newpage

\voffset=-1.95in
\hoffset=-1.1in

\begin{center}

\baselineskip 14pt 
{\Large {\bf Table IA-6}}\\ 
{\bf Sampling properties of out-of-sample Portfolio Performance Statistics}\\
{\bf 384-month out-of-sample period, 1990 -- 2021}\\
\end{center}

\vskip .1in

\begin{tabular}{lrrrrrrrcrrrrrrr}

\multicolumn{16}{l}{Sampling properties of dynamic optimal PPPs.  Portfolio characteristic tilts from the best out-of-sampling performer over the relevant preceding period}\\
\multicolumn{16}{l}{(shown in Table 1) each year.  $\cal{CE}$$_k$ is the certainty equivalent return in basis points per month for a power utility investor with coefficient of relative}\\
\multicolumn{16}{l}{risk aversion $(\gamma) = k$.  $E(r)$, $\sigma$, Median, IQR, and MIN are the mean monthly return, the
standard deviation of monthly returns, the median}\\
\multicolumn{16}{l}{monthly return, the interquartile range of monthly returns, and the minimum monthly return--all expressed in basis points per month.}\\
\multicolumn{16}{l}{SKEW and KURT are the return skewness and kurtosis measures, and SR is the Sharpe ratio.}\\
\multicolumn{16}{l}{Results are for the full 32-year out-of-sample subperiod (1990 -- 2021).}\\
    
    &    &   &   &  &  & & & \hspace{.1in}  &  & & & &  &  &  \\

\multicolumn{16}{l}{{\bf Panel A: Benchmark portfolios}}\\  
     & \multicolumn{7}{c}{VWI} & & \multicolumn{7}{c}{EWI}\\
\cline{2-8} \cline{10-16}
Statistic  & Mean & Std Dev &  2.5\%ile & 25\%ile & Median & 75\%ile & 97.5\%ile & &  Mean & Std Dev &  2.5\%ile & 25\%ile & Median & 75\%ile & 97.5\%ile\\
$\cal{CE}$$_2$ & 79.3 & 2.8 & 73.8 & 77.4 & 79.3 & 81.1 & 84.6 &  &  85.8 & 1.4 & 83.1 & 84.8 & 85.8 & 86.7 & 88.4\\
$\cal{CE}$$_5$ & 50.5 & 2.9 & 44.6 & 48.4 & 50.5 & 52.4 & 56.3 &   & 41.7 & 1.4 & 38.9 & 40.7 & 41.7 & 42.7 & 44.5\\ 
$\cal{CE}$$_8$ & 19.4 & 3.3 & 12.9 & 17.2 & 19.3 & 21.6 & 25.7 &  & -8.1 & 1.8 & -11.5 & -9.3 & -8.1 & -6.9 & -4.6\\
$E(r)$ & 97.4 &  2.8 & 92.0 & 95.6 & 97.4 & 99.3 & 102.8 & & 112.9 & 1.3 &     110.3 &     112.0 &     112.9 &     113.8 &     115.6\\
$\sigma$ & 422.3 & 3.3 & 415.7 & 420.0 &  422.3 & 424.5 & 428.8 & & 513.6 & 1.7 &     510.4 &     512.5 &     513.6 &     514.8 &     517.0\\
Median &  138.3 &       6.5 &     125.7 &     133.9 &     138.3 &     142.7 &     151.1 & &  157.3 & 6.0 & 145.4 & 153.2 & 157.3 &     161.3 &     169.0\\
IQR & 503.7 &      11.9 &     480.2 &     495.8 &     503.7 &     511.7 &     526.7 & & 608.3 &      10.4 &     588.2 &     601.4 &     608.2 &     615.4 &     628.9\\
MIN & -1,669.1 & 77.4 & -1,824.4 & -1,720.9 & -1,666.9 & -1,614.9 & -1,522.1 & &  -2,119.8 &      42.1 &   -2,205.9 &   -2,148.0 &   -2,118.6 &   -2,090.6 &   -2,040.3\\
SKEW &  -31.5 &       4.7 &     -40.5 &     -34.6 &     -31.6 &     -28.5 &     -21.9 & &  -9.9 &       3.1 &     -16.1 &     -12.0 &      -9.9 &      -7.8 &      -4.0\\
KURT &  29.5 &       3.1 &      23.6 &      27.4 &      29.5 &      31.6 &      35.6 & & 33.4 &       1.3 &      30.8 &      32.5 &      33.4 &      34.2 &      36.0\\
SR &  0.6271 &  0.0234 &  0.5810 &  0.6111 &  0.6270 &  0.6429 &  0.6735 & & 0.6196 &  0.0091 &  0.6020 &  0.6134 &  0.6196 &  0.6257 &  0.6375\\
\hline

   &    &   &   &  &  & & & \hspace{.1in}  &  & & & &  &  &  \\

\multicolumn{16}{l}{{\bf Panel B: \boldmath{$\gamma$} 2 loss function}}\\
     & \multicolumn{7}{c}{Dyn. Opt. Updating} & & \multicolumn{7}{c}{Dyn. Opt. Rolling}\\
\cline{2-8} \cline{10-16}
Statistic  & Mean & Std Dev &  2.5\%ile & 25\%ile & Median & 75\%ile & 97.5\%ile & &  Mean & Std Dev &  2.5\%ile & 25\%ile & Median & 75\%ile & 97.5\%ile\\
$\cal{CE}$$_2$ & 114.8 & 168.9 & 33.2 & 96.7 & 120.6 & 142.3 & 166.5 &  &  -7,396.7 & 2,467.7 & -10,000 & -10,000 & -5,395.9 & -4,988.2 & -4,943.1\\
$E(r)$ & 339.8 &      28.9 &     284.1 &     320.3 &     338.7 &     358.9 &     398.4 & & 379.1 & 33.2 & 316.5 & 356.5 &  378.4 & 400.9 & 447.0\\
$\sigma$ & 1,424.0 &      62.9 & 1,306.0 & 1,380.7 & 1,421.0 & 1,465.2 & 1,553.3 & & 1,593.5 & 102.5 & 1,409.7 & 1,522.5 & 1,586.2 & 1,657.3 & 1,812.6\\
Median & 346.7 &      45.3 &     260.6 &     315.6 &     345.9 &     377.0 &     437.5 & & 230.4 & 19.6 & 189.6 & 217.9 &   230.8 & 243.2 &     267.8\\
IQR & 1,670.4 & 94.8 &    1,492.8 &    1,606.0 &    1,667.4 & 1,731.7 & 1,863.4 & & 1,014.0 & 52.5 & 914.8 & 978.6 & 1,012.5 & 1,047.9 & 1,121.1\\
MIN & -5,518.9 & 911.9 & -7,756.8 & -5,968.0 & -5,374.8 & -4,894.4 & -4,179.1 & & -10,086 & 1,857.3 &  -14,310 &  -11,127 & -9,861.4 & -8,789.2 & -7,106.5\\
SKEW &   -0.5 & 2.7 & -5.8 & -2.4 & -0.5 & 1.4 &  4.9 & & 9.3 & 1.9 & 5.7 & 8.1 & 9.3 & 10.6 & 12.9\\
KURT &  35.0 & 8.9 & 18.4 & 28.9 & 34.8 & 40.9 & 52.8 & & 138.8 & 13.3 & 113.6 & 129.6 & 138.6 & 147.5 & 165.8\\
SR &  0.7783 &  0.0656 &  0.6479 &  0.7345 &  0.7778 &  0.8219 &  0.9069  & & 0.7831 &  0.0702 &  0.6449 &  0.7362 &  0.7832 &  0.8302 &  0.9197\\
\hline

\end{tabular}

\newpage  

\begin{center} 

\baselineskip 14pt   
{\Large {\bf Table IA-6  --2--}}\\  
{\bf Sampling properties of out-of-sample Portfolio Performance Statistics  --2--}\\
{\bf 384-month out-of-sample period, 1990 -- 2021}\\
\end{center}

\vskip .1in

\begin{tabular}{lrrrrrrrcrrrrrrr}

\multicolumn{16}{l}{Sampling properties of dynamic optimal PPPs.  Portfolio characteristic tilts from the best out-of-sampling performer over the relevant preceding period}\\
\multicolumn{16}{l}{(shown in Table 1) each year.  $\cal{CE}$$_k$ is the certainty equivalent return in basis points per month for a power utility investor with coefficient of relative}\\
\multicolumn{16}{l}{risk aversion $(\gamma) = k$.  $E(r)$, $\sigma$, Median, IQR, and MIN are the mean monthly return, the 
standard deviation of monthly returns, the median}\\     
\multicolumn{16}{l}{monthly return, the interquartile range of monthly returns, and the minimum monthly return--all expressed in basis points per month.}\\
\multicolumn{16}{l}{SKEW and KURT are the return skewness and kurtosis measures, and SR is the Sharpe ratio.}\\
\multicolumn{16}{l}{Results are for the full 32-year out-of-sample subperiod (1990 -- 2021).}\\

    &    &   &   &  &  & & & \hspace{.1in}  &  & & & &  &  &  \\  

\multicolumn{16}{l}{{\bf Panel C:  \boldmath{$\gamma$} 5 loss function}}\\
     & \multicolumn{7}{c}{Dyn. Opt. Updating} & & \multicolumn{7}{c}{Dyn. Opt. Rolling}\\
\cline{2-8} \cline{10-16}
Statistic  & Mean & Std Dev &  2.5\%ile & 25\%ile & Median & 75\%ile & 97.5\%ile & &  Mean & Std Dev &  2.5\%ile & 25\%ile & Median & 75\%ile & 97.5\%ile\\
$\cal{CE}$$_5$ & 46.7 & 71.2 & -94.3 & 36.7 & 60.2 & 77.1 & 102.3 &  &  -209.0 & 60.6 & -1,325.2 & -166.5 & -73.6 & -26.1 & 28.6\\
$E(r)$ & 223.7 & 13.7 & 197.5 & 214.3 & 223.6 & 232.8 & 250.7 & & 238.0 & 17.3 & 204.8 & 226.2 & 237.8 & 249.5 & 272.6\\
$\sigma$ & 763.0 & 32.8 & 703.9 & 740.0 &  761.3 & 783.8 & 830.9 & & 975.5 & 43.2 & 896.7 & 945.6 & 973.7 &  1,002.6 & 1,066.2\\
Median & 229.1 &      19.1 &     192.1 &     216.1 &     229.1 &     242.0 & 266.4 & & 196.6 & 19.3 & 157.3 & 183.5 & 197.1 & 210.4 & 232.8\\
IQR & 817.2 & 36.8 &     745.6 &     792.2 &     816.4 &     841.2 &     891.9 & & 835.6 & 37.0 & 766.0 & 809.8 &  835.6 & 860.4 & 910.1\\
MIN & -3,625.1 & 857.3 & -5,689.1 & -4,119.4 & -3,421.9 & -2,998.3 & -2,450.4 & & -5,032.0 & 1,089.7 & -7,586.0 & -5,659.1 & -4,825.9 & -4,215.3 & -3,492.5\\
SKEW &  -0.7 & 2.4 & -5.4 &  -2.4 & -0.7 & 0.9 & 4.0 & & 4.2 & 2.1 & 0.2 & 2.9 & 4.2 & 5.6 & 8.3\\
KURT &  57.5 &       9.4 &      39.5 &      51.1 &      57.3 & 63.8 & 76.8 & & 92.5 &       9.8 & 73.7 & 85.9 & 92.4 & 98.9 & 112.4\\
SR &  0.9258 &  0.0646 &  0.8000 &  0.8826 &  0.9254 &  0.9698 &  1.0516 & & 0.7749 &  0.0606 &  0.6558 &  0.7342 &  0.7750 &  0.8158 &  0.8928\\
\hline

   &    &   &   &  &  & & & \hspace{.1in}  &  & & & &  &  &  \\

\multicolumn{16}{l}{{\bf Panel D:  \boldmath{$\gamma$} 8 loss function}}\\
     & \multicolumn{7}{c}{Dyn. Opt. Updating} & & \multicolumn{7}{c}{Dyn. Opt. Rolling}\\
\cline{2-8} \cline{10-16}
Statistic  & Mean & Std Dev &  2.5\%ile & 25\%ile & Median & 75\%ile & 97.5\%ile & &  Mean & Std Dev &  2.5\%ile & 25\%ile & Median & 75\%ile & 97.5\%ile\\
$\cal{CE}$$_8$ & -44.5 & 58.9 & -188.8 & -63.0 & -31.9 & -8.8 & 22.3 &  &  -163.1 & 144.2 & -502.5 & -189.7 & -130.0 & -89.7 & -37.8\\
$E(r)$ & 170.4 & 8.5 & 154.0 & 164.6 & 170.4 & 176.0 & 187.2 & & 163.3 & 11.3 & 141.6 & 155.3 & 163.1 & 170.8 & 186.0\\
$\sigma$ & 589.4 &      20.8 &     550.2 &     575.2 &     588.7 &     602.9 &     632.4 & & 707.5 & 26.6 & 657.9 & 689.0 & 706.9 & 725.0 & 761.2\\
Median & 178.1 &      12.6 &     153.6 &     169.5 &     177.9 &     186.5 &     203.0 & & 163.9 & 15.8 & 133.8 & 153.0 & 163.7 & 174.5 & 195.6\\
IQR & 558.7 &      24.7 &     511.4 &     541.6 &     558.3 &     575.5 &     608.2 & & 697.5 & 26.1 & 647.6 & 679.5 & 696.7 & 715.1 & 749.7\\
MIN & -3317.9 &     496.7 &   -4,447.6 & -3.613.3 & -3,254.7 & -2,961.2 & -2,508.0 & & -3,785.2 & 573.3 & -5,105.0 & -4,107.0 & -3,719.3 & -3,386.5 & -2,847.5\\
SKEW &  -1.3 &       2.1 &      -5.3 &      -2.7 &      -1.3 &       0.1 &       2.7 & & -0.1 & 2.2 &  -4.5 & -1.6 & -0.1 & 1.4 & 4.1\\
KURT &  80.2 & 8.5 & 63.8 & 74.5 & 80.1 & 85.8 & 97.1 & & 64.0 & 8.6 &  47.7 & 58.0 & 63.8 & 69.7 & 81.3\\
SR &  0.8824 &  0.0524 &  0.7807 &  0.8475 &  0.8824 &  0.9181 &  0.9841 & & 0.7001 &  0.0549 &  0.5928 &  0.6627 &  0.6997 &  0.7373 &  0.8082\\
\hline

\end{tabular}

\newpage

\voffset=-1.8in

\begin{center}

{\Large {\bf Table IA-7}}\\
{\bf Out-of-Sample 6-factor Fama-French regressions}\\
$ 
r_{i,t} - r_f = \alpha + \beta_1 \cdot (R_{m,t} - r_f) + \beta_2 \cdot {\rm HML} + \ \beta_3 \cdot {\rm SMB}  + \beta_4 \cdot {\rm MOM} + \beta_5 \cdot {\rm RMW} + \beta_6 \cdot {\rm CMA} + \epsilon_{i,t}
$\\
For power utility investor with coefficient of relative risk aversion, $\gamma = 5.$  Monthly returns; $\alpha$ in basis points per month.\\
\end{center}

\baselineskip 12pt

\vskip .1in

\hoffset=-1in

\begin{tabular}{lrrrrrrrcrrrrrrr}

\multicolumn{16}{l}{{\bf Panel A. 32-year out-of-sample period: 1990 - 2021}}\\

     & \multicolumn{7}{c}{Updating Protocol} & & \multicolumn{7}{c}{Rolling Protocol}\\
\cline{2-8} \cline{10-16}
Coefficient  & Mean & Std Dev &  2.5\%ile & 25\%ile & Median & 75\%ile & 97.5\%ile & &  Mean & Std Dev &  2.5\%ile & 25\%ile & Median & 75\%ile & 97.5\%ile\\
$\alpha$ & 98.44 &  14.92 &  69.39 &  88.53 &  98.20 & 108.47 & 127.98 &  & 133.80 &  18.52 &  98.15 & 121.16 & 133.52 & 146.06 & 170.84\\
Mkt &   0.57 &   0.04 &   0.49 &   0.54 &   0.57 &   0.60 &   0.65 &   & 0.93 &   0.05 &   0.83 &   0.90 &   0.93 &   0.96 &   1.02\\
HML &  1.03 &   0.07 &   0.89 &   0.98 &   1.03 &   1.08 &   1.18 &  & 1.18 &   0.08 &   1.02 &   1.12 &   1.18 &   1.24 &   1.35\\
SMB &  -0.02 &   0.09 &  -0.20 &  -0.08 &  -0.02 &   0.03 &   0.14 &    & -0.98 &   0.10 &  -1.19 &  -1.05 &  -0.98 &  -0.91 &  -0.78\\
MOM &   0.63 &   0.06 &   0.52 &   0.60 &   0.63 &   0.67 &   0.75 &  & 0.43 &   0.07 &   0.30 &   0.39 &   0.43 &   0.48 &   0.57\\
RMW &   0.58 &   0.11 &   0.36 &   0.51 &   0.58 &   0.66 &   0.80 &  &  -0.24 &   0.14 &  -0.51 &  -0.33 &  -0.24 &  -0.14 &   0.04\\
CMA &   0.02 &   0.10 &  -0.19 &  -0.05 &   0.02 &   0.09 &   0.21 &  &   0.21 &   0.12 &  -0.03 &   0.13 &   0.21 &   0.29 &   0.44\\
\hline

   &    &   &   &  &  & & & \hspace{.1in}  &  & & & &  &  &  \\

\multicolumn{16}{l}{{\bf Panel B. Subperiod 1: 1990 - 1998}}\\
     & \multicolumn{7}{c}{Updating protocol} & & \multicolumn{7}{c}{Rolling protocol}\\
\cline{2-8} \cline{10-16}
Coefficient  & Mean & Std Dev &  2.5\%ile & 25\%ile & Median & 75\%ile & 97.5\%ile & &  Mean & Std Dev &  2.5\%ile & 25\%ile & Median & 75\%ile & 97.5\%ile\\
$\alpha$ & 166.16 &  35.41 &  97.74 & 141.43 & 165.73 & 190.19 & 236.61 & & 244.99 &  52.59 & 144.17 & 209.62 & 243.94 & 278.95 & 352.57\\
Mkt & 0.03 &   0.12 &  -0.21 &  -0.05 &   0.03 &   0.11 &   0.25 & & -0.08 &   0.16 &  -0.40 &  -0.19 &  -0.08 &   0.03 &   0.24\\
HML &  2.25 &   0.25 &   1.78 &   2.09 &   2.25 &   2.42 &   2.75 & & 3.49 &   0.38 &   2.76 &   3.22 &   3.48 &   3.74 &   4.27\\
SMB &  0.97 &   0.16 &   0.65 &   0.85 &   0.96 &   1.08 &   1.28 & & 0.45 &   0.26 &  -0.05 &   0.28 &   0.45 &   0.63 &   0.96\\
MOM & 1.69 &   0.16 &   1.40 &   1.59 &   1.69 &   1.80 &   2.01 & & 2.16 &   0.22 &   1.74 &   2.01 &   2.16 &   2.31 &   2.60\\
RMW &  0.71 &   0.27 &   0.19 &   0.53 &   0.71 &   0.89 &   1.25 & & 0.75 &   0.43 &  -0.08 &   0.46 &   0.75 &   1.03 &   1.61\\
CMA & -0.92 &   0.32 &  -1.56 &  -1.13 &  -0.91 &  -0.71 &  -0.31 & & -1.98 &   0.49 &  -2.97 &  -2.31 &  -1.97 &  -1.65 &  -1.04\\
\hline

   &    &   &   &  &  & & & \hspace{.1in}  &  & & & &  &  &  \\

\multicolumn{16}{l}{{\bf Panel C. Subperiod 2: 1999 - 2021}}\\
     & \multicolumn{7}{c}{Updating protocol} & & \multicolumn{7}{c}{Rolling protocol}\\
\cline{2-8} \cline{10-16}
Coefficient  & Mean & Std Dev &  2.5\%ile & 25\%ile & Median & 75\%ile & 97.5\%ile & &  Mean & Std Dev &  2.5\%ile & 25\%ile & Median & 75\%ile & 97.5\%ile\\
$\alpha$ &   34.02 &  14.20 &   5.90 &  24.29 &  33.97 &  43.66 &  61.66 & & 16.61 &  15.22 & -12.55 &   6.20 &  16.33 &  26.73 &  47.19\\
Mkt &  0.75 &   0.04 &   0.67 &   0.72 &   0.75 &   0.78 &   0.83 & & 1.22 &   0.04 &   1.14 &   1.19 &   1.22 &   1.25 &   1.31\\
HML &   0.71 &   0.07 &   0.57 &   0.66 &   0.70 &   0.75 &   0.84 & & 0.63 &   0.07 &   0.49 &   0.58 &   0.63 &   0.67 &   0.76\\
SMB &  -0.08 &   0.10 &  -0.27 &  -0.14 &  -0.08 &  -0.01 &   0.10 & & -1.00 &   0.11 &  -1.23 &  -1.08 &  -1.00 &  -0.92 &  -0.79\\
MOM & 0.54 &   0.06 &   0.41 &   0.49 &   0.54 &   0.58 &   0.66 & & 0.26 &   0.07 &   0.13 &   0.22 &   0.26 &   0.31 &   0.40\\
RMW & 0.84 &   0.11 &   0.62 &   0.77 &   0.84 &   0.92 &   1.07 & & 0.13 &   0.14 &  -0.14 &   0.04 &   0.13 &   0.22 &   0.40\\
CMA &  0.12 &   0.11 &  -0.10 &   0.05 &   0.12 &   0.20 &   0.34 & & 0.46 &   0.12 &   0.23 &   0.38 &   0.46 &   0.54 &   0.69\\
\hline

\end{tabular}

\newpage

\voffset=-1.8in

\begin{center}

{\Large {\bf Table IA-8}}\\
{\bf Out-of-Sample 6-factor Fama-French regressions}\\
$ 
r_{i,t} - r_f = \alpha + \beta_1 \cdot (R_{m,t} - r_f) + \beta_2 \cdot {\rm HML} + \ \beta_3 \cdot {\rm SMB}  + \beta_4 \cdot {\rm MOM} + \beta_5 \cdot {\rm RMW} + \beta_6 \cdot {\rm CMA} + \epsilon_{i,t}   
$\\ 
For power utility investor with coefficient of relative risk aversion, $\gamma = 8.$  Monthly returns; $\alpha$ in basis points per month.\\
\end{center}

\baselineskip 12pt

\vskip .1in

\hoffset=-1in

\begin{tabular}{lrrrrrrrcrrrrrrr}

\multicolumn{16}{l}{{\bf Panel A. 32-year out-of-sample period: 1990 - 2021}}\\
     & \multicolumn{7}{c}{Updating Protocol} & & \multicolumn{7}{c}{Rolling Protocol}\\
\cline{2-8} \cline{10-16}
Coefficient  & Mean & Std Dev &  2.5\%ile & 25\%ile & Median & 75\%ile & 97.5\%ile & &  Mean & Std Dev &  2.5\%ile & 25\%ile & Median & 75\%ile & 97.5\%ile\\
$\alpha$ & 44.88 &   9.20 &  27.04 &  38.62 &  44.86 &  50.98 &  63.26 &  &  53.62 &  12.24 &  30.17 &  45.24 &  53.40 &  62.00 &  77.98\\
Mkt &   0.64 &   0.03 &   0.58 &   0.62 &   0.63 &   0.65 &   0.69 &   & 0.86 &   0.03 &   0.80 &   0.84 &   0.86 &   0.88 &   0.93\\
HML &  0.87 &   0.05 &   0.78 &   0.84 &   0.87 &   0.91 &   0.97 &  & 0.82 &   0.06 &   0.71 &   0.78 &   0.82 &   0.86 &   0.94\\
SMB &   -0.15 &   0.06 &  -0.26 &  -0.19 &  -0.15 &  -0.11 &  -0.04 &    & -0.71 &   0.06 &  -0.83 &  -0.75 &  -0.71 &  -0.67 &  -0.59\\
MOM &   0.43 &   0.04 &   0.36 &   0.40 &   0.43 &   0.45 &   0.50 &  & 0.30 &   0.05 &   0.21 &   0.26 &   0.30 &   0.33 &   0.39\\
RMW &   0.68 &   0.06 &   0.56 &   0.64 &   0.68 &   0.73 &   0.81 &  &   0.15 &   0.08 &  -0.01 &   0.10 &   0.15 &   0.21 &   0.32\\
CMA &   0.26 &   0.06 &   0.13 &   0.21 &   0.26 &   0.30 &   0.38 &  &   0.33 &   0.08 &   0.17 &   0.27 &   0.33 &   0.38 &   0.48\\
\hline

   &    &   &   &  &  & & & \hspace{.1in}  &  & & & &  &  &  \\
\multicolumn{16}{l}{{\bf Panel B. Subperiod 1: 1990 - 1998}}\\
     & \multicolumn{7}{c}{Updating protocol} & & \multicolumn{7}{c}{Rolling protocol}\\
\cline{2-8} \cline{10-16}
Coefficient  & Mean & Std Dev &  2.5\%ile & 25\%ile & Median & 75\%ile & 97.5\%ile & &  Mean & Std Dev &  2.5\%ile & 25\%ile & Median & 75\%ile & 97.5\%ile\\
$\alpha$ &  103.55 &  23.05 &  59.71 &  87.66 & 103.00 & 119.35 & 149.00 & & 144.98 &  33.93 &  81.31 & 122.15 & 144.01 & 166.96 & 214.48\\
Mkt & 0.20 &   0.07 &   0.06 &   0.15 &   0.20 &   0.24 &   0.33 & & 0.14 &   0.11 &  -0.07 &   0.07 &   0.14 &   0.21 &   0.35\\
HML &  1.68 &   0.18 &   1.34 &   1.56 &   1.68 &   1.80 &   2.05 & & 2.29 &   0.25 &   1.82 &   2.12 &   2.28 &   2.45 &   2.79\\
SMB &  0.50 &   0.11 &   0.28 &   0.42 &   0.50 &   0.58 &   0.73 & & 0.15 &   0.15 &  -0.15 &   0.05 &   0.15 &   0.25 &   0.45\\
MOM & 1.12 &   0.11 &   0.91 &   1.05 &   1.11 &   1.19 &   1.33 & & 1.36 &   0.14 &   1.09 &   1.27 &   1.36 &   1.46 &   1.64\\
RMW & 0.53 &   0.18 &   0.19 &   0.41 &   0.53 &   0.65 &   0.88 & & 0.59 &   0.28 &   0.06 &   0.40 &   0.59 &   0.77 &   1.14\\
CMA & -0.63 &   0.22 &  -1.07 &  -0.78 &  -0.63 &  -0.48 &  -0.22 & & -1.29 &   0.32 &  -1.94 &  -1.50 &  -1.28 &  -1.07 &  -0.68\\
\hline

   &    &   &   &  &  & & & \hspace{.1in}  &  & & & &  &  &  \\

\multicolumn{16}{l}{{\bf Panel C. Subperiod 2: 1999 - 2021}}\\
     & \multicolumn{7}{c}{Updating protocol} & & \multicolumn{7}{c}{Rolling protocol}\\
\cline{2-8} \cline{10-16}
Coefficient  & Mean & Std Dev &  2.5\%ile & 25\%ile & Median & 75\%ile & 97.5\%ile & &  Mean & Std Dev &  2.5\%ile & 25\%ile & Median & 75\%ile & 97.5\%ile\\
$\alpha$ &  4.75 &   8.28 & -11.33 &  -0.82 &   4.79 &  10.36 &  21.21 & & -22.81 &  9.58 & -41.57 & -29.19 & -22.89 & -16.37 & -3.78\\
Mkt &  0.77 &   0.03 &   0.71 &   0.75 &   0.77 &   0.79 &   0.82 & & 1.06 &   0.03 &   1.01 &   1.04 &   1.06 &   1.08 &   1.12\\
HML &   0.67 &   0.05 &   0.58 &   0.64 &   0.67 &   0.70 &   0.76 &    & 0.47 &   0.05 &   0.37 &   0.43 &   0.47 &   0.50 &   0.57\\
SMB &  -0.22 &   0.07 &  -0.36 &  -0.27 &  -0.22 &  -0.18 &  -0.09 &  &  -0.73 &   0.07 &  -0.87 &  -0.78 &  -0.73 &  -0.69 &  -0.60\\
MOM &   0.37 &   0.04 &   0.30 &   0.35 &   0.37 &   0.40 &   0.45 &  & 0.20 &   0.05 &   0.10 &   0.16 &   0.20 &   0.23 &   0.29\\
RMW &   0.84 &   0.06 &   0.72 &   0.80 &   0.84 &   0.88 &   0.97 &  & 0.39 &   0.08 &   0.24 &   0.34 &   0.39 &   0.44 &   0.55\\
CMA &   0.38 &   0.07 &   0.25 &   0.33 &   0.38 &   0.42 &   0.51 &  &  -0.20 &   0.07 &  -0.34 &  -0.24 &  -0.19 &  -0.15 &  -0.06\\
\hline

\end{tabular}

\newpage

\voffset=-1.8in

\begin{center}

{\Large {\bf Table IA-9}}\\
{\bf Out-of-Sample 6-factor Fama-French regressions}\\
$
r_{i,t} - r_f = \alpha + \beta_1 \cdot (R_{m,t} - r_f) + \beta_2 \cdot {\rm HML} + \ \beta_3 \cdot {\rm SMB}  + \beta_4 \cdot {\rm MOM} + \beta_5 \cdot {\rm RMW} + \beta_6 \cdot {\rm CMA} + \epsilon_{i,t}
$\\
Full out-of-sample period: 1990 - 2021\\
For power utility investor with coefficient of relative risk aversion, $\gamma = 2.$  Monthly returns; $\alpha$ in basis points per month.\\
\end{center}

\baselineskip 12pt

\vskip .1in

\hoffset=-1in

\begin{tabular}{lrrrrrrrcrrrrrrr}

     & \multicolumn{7}{c}{Updating Protocol} & & \multicolumn{7}{c}{Rolling Protocol}\\
\cline{2-8} \cline{10-16}
Coefficient  & Mean & Std Dev &  2.5\%ile & 25\%ile & Median & 75\%ile & 97.5\%ile & &  Mean & Std Dev &  2.5\%ile & 25\%ile & Median & 75\%ile & 97.5\%ile\\
$\alpha$ & 166.03 &  31.59 & 105.70 & 144.62 & 165.20 & 186.97 & 229.37 &  & 285.47 &  35.29 & 218.45 & 260.98 & 285.07 & 309.16 & 355.75\\
Mkt &   0.62 &   0.09 &   0.45 &   0.56 &   0.62 &   0.68 &   0.79 &   & 0.89 &   0.09 &   0.72 &   0.83 &   0.89 &   0.95 &   1.07\\
HML &   1.80 &   0.16 &   1.50 &   1.70 &   1.80 &   1.90 &   2.11 &  & 1.72 &   0.16 &   1.41 &   1.61 &   1.72 &   1.83 &   2.05\\
SMB &  0.63 &   0.19 &   0.24 &   0.50 &   0.63 &   0.76 &   1.00 &    & -1.28 &   0.22 &  -1.74 &  -1.42 &  -1.27 &  -1.13 &  -0.87\\
MOM &   1.14 &   0.14 &   0.87 &   1.05 &   1.14 &   1.23 &   1.41 &  & 0.84 &   0.13 &   0.57 &   0.75 &   0.84 &   0.93 &   1.11\\
RMW &   0.87 &   0.20 &   0.49 &   0.74 &   0.87 &   1.00 &   1.27 &  &  -0.62 &   0.30 &  -1.23 &  -0.83 &  -0.62 &  -0.42 &  -0.03\\
CMA &  -0.32 &   0.22 &  -0.75 &  -0.47 &  -0.32 &  -0.17 &   0.12 &  &  -0.50 &   0.24 &  -0.97 &  -0.66 &  -0.50 &  -0.34 &  -0.03\\
\hline

\end{tabular}

\newpage

\voffset=-1.8in

\begin{center}

{\Large {\bf Table IA-10}}\\
{\bf Out-of-Sample 6-factor Fama-French regressions}\\
{\bf  Subperiod 1: 1990 - 1998   Updating Protocol}\\
{\bf Portfolio mean and variance decompositions}\\
$ 
r_{i,t} - r_f = \alpha + \beta_1 \cdot (R_{m,t} - r_f) + \beta_2 \cdot {\rm HML} + \ \beta_3 \cdot {\rm SMB}  + \beta_4 \cdot {\rm MOM} + \beta_5 \cdot {\rm RMW} + \beta_6 \cdot {\rm CMA} + \epsilon_{i,t}   
$\\  
For power utility investor with coefficient of relative risk aversion, $\gamma = 8.$  Monthly returns; $\alpha$ in basis points per month.\\
\end{center}

\baselineskip 12pt

\vskip .1in

\hoffset=-1in

\begin{tabular}{lrrrrrrrcrrrrrrr}

\multicolumn{16}{l}{{\bf Panel A. Power utility investor with \boldmath{$\gamma = 5$}}}\\
     & \multicolumn{7}{c}{Mean Return Decomposition} & & \multicolumn{7}{c}{Variance Decomposition}\\
\cline{2-8} \cline{10-16}
Coefficient  & Mean & Std Dev &  2.5\%ile & 25\%ile & Median & 75\%ile & 97.5\%ile & &  Mean & Std Dev &  2.5\%ile & 25\%ile & Median & 75\%ile & 97.5\%ile\\
$\alpha$ / orthog. & 42.21 &   6.52 &  28.64 &  38.00 &  42.45 &  46.75 &  54.38 & & 52.83 &   4.02 &  45.03 &  50.14 &  52.79 &  55.51 &  60.72\\
Mkt &   0.77 &   2.99 &  -5.29 &  -1.21 &   0.79 &   2.80 &   6.50 & & 0.33 &   0.46 &   0.00 &   0.03 &   0.15 &   0.44 &   1.63\\
HML &  4.28 &   1.82 &  10.93 &  13.03 &  14.20 &  15.47 &  18.06 & & 43.77 &   7.79 &  29.18 &  38.34 &  43.55 &  48.89 &  59.85\\
SMB &   -6.35 &   1.23 &  -8.91 &  -7.16 &  -6.31 &  -5.49 &  -4.09 & & 10.13 &   3.34 &   4.41 &   7.71 &   9.88 &  12.28 &  17.42\\
MOM &   43.58 &   4.57 &  35.22 &  40.42 &  43.41 &  46.54 &  53.06 & & 32.48 &   4.35 &  24.07 &  29.55 &  32.46 &  35.44 &  41.00\\
RMW &   7.87 &   3.07 &   2.05 &   5.79 &   7.81 &   9.89 &  14.09 & & 1.57 &   1.07 &   0.10 &   0.77 &   1.37 &   2.17 &   4.16\\
CMA &   -2.37 &   0.83 &  -4.04 &  -2.93 &  -2.36 &  -1.82 &  -0.78 & & 4.69 &   2.88 &   0.49 &   2.58 &   4.24 &   6.34 &  11.48\\
\hline

   &    &   &   &  &  & & & \hspace{.1in}  &  & & & &  &  &  \\
\multicolumn{16}{l}{{\bf Panel B. Power utility investor with \boldmath{$\gamma = 8$}}}\\
     & \multicolumn{7}{c}{Mean Return Decomposition} & & \multicolumn{7}{c}{Variance Decomposition}\\
\cline{2-8} \cline{10-16}
Coefficient  & Mean & Std Dev &  2.5\%ile & 25\%ile & Median & 75\%ile & 97.5\%ile & &  Mean & Std Dev &  2.5\%ile & 25\%ile & Median & 75\%ile & 97.5\%ile\\
$\alpha$ / orthog. &  36.69 &   6.12 &  23.90 &  32.72 &  36.96 &  40.96 &  47.83 & & 52.38 &   3.91 &  44.87 &  49.68 &  52.32 &  55.04 &  60.10\\
Mkt & 7.14 &   2.68 &   1.96 &   5.31 &   7.10 &   8.91 &  12.51 & & 2.16 &   1.54 &   0.13 &   1.03 &   1.86 &   2.95 &   6.02\\
HML &  14.82 &   1.67 &  11.71 &  13.69 &  14.74 &  15.89 &  18.32 & & 50.43 &   7.80 &  35.55 &  45.24 &  50.33 &  55.53 &  66.24\\
SMB & -4.61 &   1.12 &  -6.90 &  -5.35 &  -4.58 &  -3.84 &  -2.50 & & 5.84 &   2.49 &   1.76 &   4.02 &   5.57 &   7.34 &  11.44\\
MOM & 40.04 &   4.13 &  32.39 &  37.15 &  39.89 &  42.73 &  48.60 & & 29.39 &   4.08 &  21.60 &  26.60 &  29.33 &  32.10 &  37.51\\
RMW & 8.19 &   2.80 &   2.95 &   6.27 &   8.12 &  10.05 &  13.89 & & 1.77 &   1.08 &   0.22 &   0.96 &   1.60 &   2.39 &   4.32\\
CMA & -2.26 &   0.77 &  -3.82 &  -2.77 &  -2.24 &  -1.74 &  -0.81 & & 4.56 &   2.76 &   0.54 &   2.50 &   4.11 &   6.17 &  11.02\\
\hline

\end{tabular}

\end{landscape}

\end{document}